\documentclass[aps,twocolumn,superscriptaddress,floatfix,longbibliography]{revtex4-2}
\usepackage{graphicx}
\usepackage{appendix}
\usepackage{lipsum}
\usepackage{physics}
\usepackage{bbold}
\usepackage{hyperref}
\usepackage{natbib}
\usepackage{amsmath,mathtools}
\usepackage[normalem]{ulem}
\hypersetup{colorlinks=true,linkcolor=blue,citecolor=blue,urlcolor=blue}

\newcommand{\beq}{\begin{equation}}
\newcommand{\eeq}{\end{equation}}
\newcommand{\beqar}{\begin{eqnarray}}
\newcommand{\eeqar}{\end{eqnarray}}
\newcommand{\bea}{\begin{eqnarray}}
\newcommand{\eea}{\end{eqnarray}}
\newcommand{\bcen}{\begin{center}}
\newcommand{\ecen}{\end{center}}

\begin{document}
\title{Control of open quantum systems via dynamical invariants }
\author{Loris M. Cangemi}
\altaffiliation{Present address: Dept. of Electrical Engineering and Information Technology,
Università degli Studi di Napoli Federico II, via Claudio 21, Napoli, 80125, Italy}

\email{lorismaria.cangemi@unina.it}
\affiliation{Department of Chemistry; Institute of Nanotechnology and Advanced Materials; Center for Quantum
Entanglement Science and Technology, Bar-Ilan University, Ramat-Gan, 52900 Israel}
\author{Hilario Espin\'{o}s}
\affiliation{Department of Physics, Universidad Carlos III de Madrid, Avda. de la Universidad 30, Leganés, 28911 Madrid, Spain}
\author{Ricardo Puebla}
\affiliation{Department of Physics, Universidad Carlos III de Madrid, Avda. de la Universidad 30, Leganés, 28911 Madrid, Spain}
\author{Erik Torrontegui}
\affiliation{Department of Physics, Universidad Carlos III de Madrid, Avda. de la Universidad 30, Leganés, 28911 Madrid, Spain}
\author{Amikam Levy}
\email{amikam.levy@biu.ac.il}
\affiliation{Department of Chemistry; Institute of Nanotechnology and Advanced Materials; Center for Quantum
Entanglement Science and Technology, Bar-Ilan University, Ramat-Gan, 52900 Israel}

\begin{abstract}

In this study, we address the challenge of controlling quantum systems under environmental influences using the theory of dynamical invariants. We employ a reverse engineering approach to develop control protocols designed to be robust against environmental noise and dissipation. This technique offers significant improvements over traditional quantum control methods by accounting for the time-dependent dissipation factor in the master equation, which results from modulating the system’s Hamiltonian (the control fields). Additionally, our method obviates the need for iterative propagation of the system state, a resource-intensive process. 
The method can be applied to any open system dynamics that can be described using a time-dependent Master equation.
We demonstrate the effectiveness and practicality of our approach through applications to two fundamental models: a two-level quantum system and a quantum harmonic oscillator, both interacting with a thermal bath.

\end{abstract}
\maketitle

\section{Introduction}

Quantum technologies are the focus of intense ongoing research~\cite{acin2018quantum}. As quantum information and computation devices have now crossed the border of the NISQ era~\cite{preskill2018quantum}, the issue of optimization of quantum devices' performance against noise and dissipation has become pivotal~\cite{koch2022quantum}. 

In the last four decades, Quantum Control (QC) theories~\cite{d2007introduction,glaser2015training,koch2022quantum} have been developed to devise suitable quantum protocols to steer a quantum system from a  predetermined initial state to a desired target state. These theoretical efforts permitted the achievement of a high level of control in electromagnetic pulse shaping, so that they found applications in the field of Nuclear Magnetic Resonance (NMR)\cite{skinner2003application,khaneja2005optimal} and laser-cooling \cite{koch2004stabilization,Reich_2013}. Moreover, they were employed in the design of quantum information devices \cite{nielsenchuang2010,schulte2005optimal}, such as optimal quantum gates based on Nitrogen-Vacancy (NV) centers \cite{Waldherr_2014_NV}, trapped ions  \cite{garcia2003speed,garcia2005coherent,manning2014optimal}, and cold atoms platforms \cite{leibfried2005creation,goerz2014robustness,omran2019generation}. 
However, the target state could either be the outcome of a quantum state preparation protocol~\cite{rojan_stateprep_2014,omran2019generation}, or the result of a quantum algorithm \cite{moll2018quantum,choquette2021quantum}, and it could encode the solution to a complex computational task~\cite{albash2018adiabatic,hauke2020perspectives}. As a consequence, more recently, quantum control theories have found applications in novel experimental settings within the realm of quantum technologies, ranging from cQED~\cite{blais2021circuit,goerz2017charting}, to superconducting qubits~\cite{setiawan_2021, werninghaus2021leakage}.  

QC theory tackles several fundamental issues. For instance, controllability, i.e., the study of the set of quantum states that can actually be employed as target states along each quantum protocol. Furthermore, Quantum Optimal Control (QOC)~\cite{doria2011optimal}, based on established results in optimal control theory for classical systems~\cite{d2007introduction}, provides necessary conditions for the existence of optimal control fields that minimize a given cost function. In the case of finite-dimensional closed quantum systems, QOC provides rigorous conditions to controllability~\cite{d2007introduction}. Moreover, powerful numerical algorithms such as GRAPE~\cite{khaneja2005optimal}, CRAB/DCRAB~\cite{caneva2011chopped,muller2022one} have been devised to tackle the control of closed quantum systems.

In the last two decades, alternative routes to QC have been put forward by means of the so-called Shortcuts To Adiabaticity (STA) methods~\cite{torrontegui2013shortcuts,guery2019shortcuts}. STA protocols for closed quantum systems are devised to achieve the final result of an adiabatic evolution in a finite amount of time. In order to implement STA protocols, a wide variety of theoretical techniques have been proposed, namely, transitionless quantum driving ~\cite{demirplak2003adiabatic,demirplak2005assisted, berry2009transitionless,del2013shortcuts,an2016shortcuts}, dynamical invariants~\cite{chen2011lewis}, fast forward approach \cite{masuda2008fast,masuda2010fast}, or Gauge potentials~\cite{kolodrubetz2017geometry}, to name a few. The main advantage brought in by these methods is that, under suitable constraints, the target state of the protocol can be reached with ideal fidelity in a finite amount of time. They thus contrast with adiabatic protocols, such as STIRAP \cite{vitanov2017stimulated}, which are more prone to the detrimental effects of noise and decoherence due to their long time duration as compared with the characteristic timescale of the driven quantum system \cite{Du_fastSTIRAP_2016,Vespsalsuper_2019}. Although the effectiveness of STA protocols in  controlling driven many-body quantum systems may require sophisticated theoretical approaches ~\cite{del2012assisted,sels2017minimizing,claeys2019floquet,vcepaite2023counterdiabatic,sun2022optimizing,chen2022digitized,espinós2023invariantbased}, these protocols have been tested experimentally using NMR~\cite{Zhoulocalcounter2020}, quantum computing platforms~\cite{hegade2021shortcuts}, and cQED architectures \cite{cardenas2023shortcuts,yin2022shortcuts}.  

One of the main issues is to assess to what extent control protocols, such as STA, can be useful when dealing with open quantum systems \cite{koch2016controlling,levy2017action,levy2018noise,kallush2022controlling,venuti2021optimal},~e.g., quantum systems evolving in the presence of several sources of noise. By themselves, open quantum systems cannot meet the controllability condition, and for generic open quantum systems, a full understanding of the issues posed by controllability is yet to come~\cite{koch2016controlling,koch2022quantum}. However, the effectiveness of STA protocols under the influence of noise and dissipation has been investigated in several works \cite{vacanti2014transitionless,villazon2019swift}. More recent approaches focused on balanced gain and loss \cite{alipour2020shortcuts} as well as on the definition of general bounds to the performance of CD driving in the presence of the environment \cite{funo2021general}.        

In this work, we employ the theory of dynamical invariants~\cite{lewis69,Dodonov79,Dodonov89} to devise control protocols of quantum systems, such as STA, that can be robust against the detrimental effects of noise and dissipation.
\textit{The main challenge we are facing is that the noise model depends on the control Hamiltonian itself. That is, different control protocols are subject to different dissipation and decoherence processes. While this can be a major difficulty, it may also provide an opportunity for finding a control protocol that is minimally sensitive to a certain type of noise.
The clear advantage of our approach is that we find such a control protocol without the need to solve the dynamics iteratively, as is often done in optimal control techniques.}

The effect of noise is modeled by means of a time-dependent Markovian quantum master equation, which is valid in the nonadiabatic driving regime~\cite{dann2018time}.
We show that, by means of a reverse engineering approach~\cite{chen2011lewis,levy2018noise,Espinós_2023}, a set of invariant operators for the closed system can be found such that the effect of noise on the protocol can be minimized, leading to improvements in its success.
We focus on paradigmatic models of driven quantum systems, i.e., a two-level system (TLS) and a driven harmonic oscillator. The former can be employed as a simple model of single-qubit gate, while the latter can be used to simulate harmonic traps~\cite{chen2010fast,uzdin2013effects}.

The work is organized as follows: in Sec.~\ref{sec:STA}, we recap the main theoretical properties of dynamical invariants and how they can be used to achieve STA in the closed setting. In Sec.~\ref{sec:noise}, we outline how the physical scenario changes when the quantum system is coupled to its environment, restricting our analysis to the Markovian setting and inertial approximation~\cite{dann2021inertial}. We also explain how a reverse-engineering approach can be set up to find the invariant of the closed system that leads to optimal protocol fidelity. Eventually, in Sec.~\ref{sec:qubit} and Sec.~\ref{sec:oscillator}, we show how it can be useful in the instance of the single quantum systems of interest.          
\section{STA via dynamical invariant}\label{sec:STA}
\begin{figure*}
\includegraphics[scale=0.78]{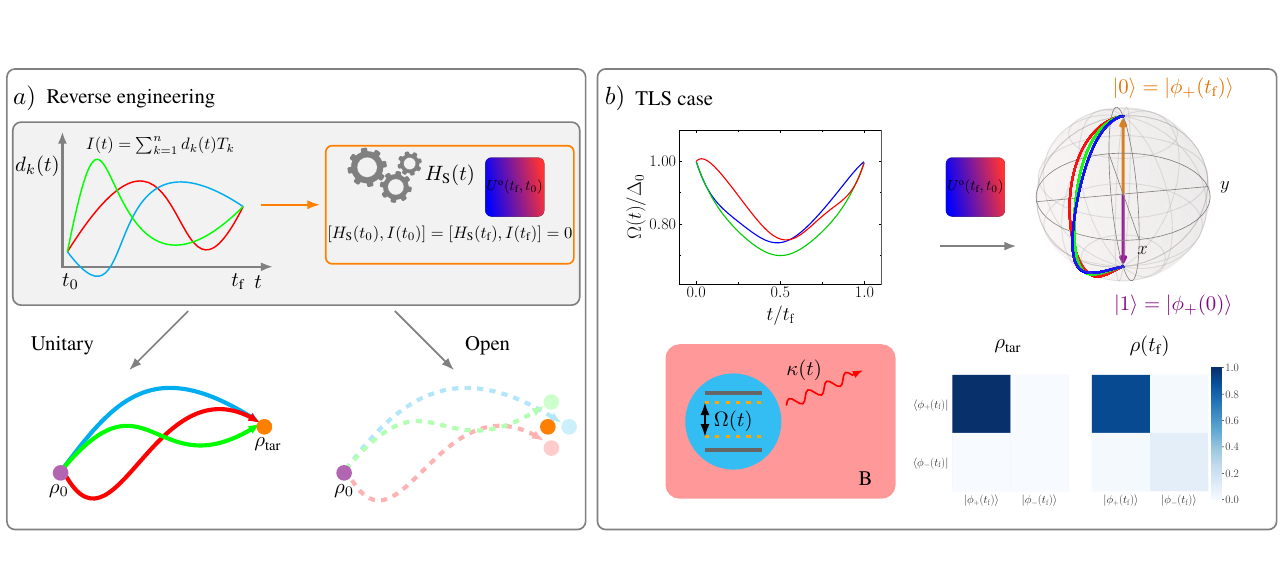}
\caption{Schematic diagram of the reverse-engineering method based on the Lewis-Riesenfeld invariant. $a)$  The invariant operator $I(t)$ belonging to a fixed Lie algebra is designed to share common eigenstates with the Hamiltonian $H_{\text{S}}(t)$ at the start and at the end of the protocol, i.e., $\comm{I(t_{\text{a}})}{H(t_{\text{a}})}=0$, $t_{\text{a}}=t_{0},t_{\text{f}}$. From Eqs.~\eqref{eq:invariant1} and \eqref{eq:pop}, an infinite set of protocols parametrized by means of the functions ${\bf d}(t)=(d_{1}(t),\dots, d_{n}(t))$,  can be defined which exactly transfer the system between pairs of states $\rho_{0},\rho_{\text{tar}}$ preserving their populations in the invariant eigenstate basis  $\{\ket{\phi_{n}(t)}\}$. The freedom in choosing the operators $H_{\text{S}}(t)$ at intermediate times can be exploited to reduce the protocol errors due to the presence of noise (see Sec.~\ref{sec:noise}). The choice of the endpoint states $(\rho_{0},\rho_{\text{tar}})$ is not limited to pure states.   $b)$ Sketch of different instances of the reverse-engineered protocols in the case of a driven TLS. Here, the energy gap $\Omega(t)$ is changed in time by means of time-dependent driving fields along the $x$ and $z$ directions (see Sec.~\ref{sec:qubit}).
The unitary dynamics linked to the reverse-engineered Hamiltonians $H_{\text{S}}(t)$ are designed to connect a pair of pure states that are also eigenstates of $H_{\text{S}}(t)$ and $I(t)$ at $t_{a}=0,t_{\text{f}}$. 
In the course of the evolution, the system is subject to the influence of its environment, and the dissipative effects also depend on time (see Sec.~\ref{sec:noise}). As a result, population leakages from the invariant eigenstate $\ket{\phi_{+}(t)}$ occur, hindering the success of the reverse-engineered protocols.  
%$c)$ Similar diagram in the case of a driven harmonic oscillator, where the frequency is modulated in time from $\omega(0)=\omega_{\text{in}}$ to $\omega(t_{\text{f}})=\omega_{\text{fin}}$ and the oscillator is prepared in a coherent superposition of eigenstates of $H(0)$. 
}
\label{fig:Protocolsketch}
\end{figure*}

We consider the instance of a driven quantum system, whose Hilbert space $\mathcal{H}$ has dimension $\text{dim}(\mathcal{H})$. In the Schr{\"o}dinger picture, it is described by a time-dependent Hamiltonian operator $H(t)$. The evolution of the state operator $\rho(t)$ is ruled by the Von Neumann equation (hereafter we assume $\hbar=1$), 
\beq\label{eq:VonNeu}
    \frac{\mathrm d}{\mathrm d t}{\rho}(t)=-i\comm{H(t)}{\rho(t)}.
\eeq
The state $\rho(t)$ can also be written as 
\beq\label{eq:statet}
    \rho(t)=U(t,t_{0})\rho(t_{0})U^{\dagger}(t,t_{0}),
\eeq
where $U(t,t_{0})=\mathcal{T}\exp[-i\int_{t_{0}}^{t} H(t^{\prime}) \mathrm{d} t^{\prime} ]$ is the time evolution operator, $\mathcal{T}$ denotes the time-ordering operator and $t_{0}$ is a fixed initial time.  

\noindent In the Heisenberg picture, the equation of motion of a generic, possibly time-dependent operator $O(t)$ reads
\beq\label{eq:Heisenberg}
    \frac{\mathrm d}{\mathrm d t}{O^{\mathrm H}}(t)=i\comm{H^{\mathrm H}(t)}{O^{\mathrm H}(t)} + \frac{\partial}{\partial t}O^{\mathrm H}(t),
\eeq
with $O^{\mathrm H}(t)=U^{\dagger}(t,t_{0})O(t)U(t,t_{0})$ and $
(\partial/\partial t) O^{\mathrm H}(t)=U^{\dagger}(t,t_{0})\qty((\partial/\partial t)O(t))U(t,t_{0}).$ 
A Lewis-Riesenfeld (Dodonov-Man'ko) invariant operator \cite{lewis69,Dodonov79,Dodonov89} $I(t)$ is defined as a Hermitian, time-dependent operator in the Schr{\"o}dinger picture, such that
\beq\label{eq:invariant1}
    \frac{\partial}{\partial t}I^{\mathrm H}(t)+i\comm{H^{\mathrm H}(t)}{I^{\mathrm H}(t)}  =0.
\eeq
 A relation analogous to Eq.~\eqref{eq:invariant1} is valid for the operators $(I(t),H(t))$ written in the Schr{\"o}dinger picture. From the Hermiticity property and  Eq.~\eqref{eq:invariant1}, it follows that, at each point in time, the operator is diagonalizable and its eigenvalues $\{\lambda_{j},j=1,\dots,\text{dim}(\mathcal{H})\}$ are constant in time. Therefore, a basis of instantaneous eigenvectors $\ket{\phi_{j}(t)}$ exists such that $I(t)=\sum_{j}\lambda_{j}\ket{\phi_{j}(t)}\bra{\phi_{j}(t)}$. The remarkable property of this reference basis is that, denoting with $\rho_{jk}(t)=\bra{\phi_{j}(t)}\rho(t)\ket{\phi_{k}(t)}$ the expectation values of the state at any point in time, from Eqs.~\eqref{eq:VonNeu} and \eqref{eq:invariant1} it follows     
\beq\label{eq:pop}
    \frac{\mathrm d}{\mathrm d t}\rho_{kk}(t)=0, 
\eeq    
i.e., the populations do not evolve in time. On the contrary, the off-diagonal elements of the state in the basis evolve according to
%\begin{multline}\label{eq:coher}
%    \frac{\mathrm d}{\mathrm d t}\rho_{kl}(t)=-i\sum_{m}\bra{\phi_{m}(t)}\qty(i\frac{\partial}{\partial t}-H(t))\ket{\phi_{l}(t)}\rho_{km}(t)+\\
%    +i\sum_{m}\bra{\phi_{m}(t)}\qty(i\frac{\partial}{\partial t}-H(t))\ket{\phi_{k}(t)}^{*}\rho_{ml}(t). 
%\end{multline}
\beq\label{eq:coher}
    \frac{\mathrm d}{\mathrm d t}\rho_{kl}(t)=-i(\dot{\varphi}_{l}(t)-\dot{\varphi}_{k}(t))\rho_{kl}(t), 
\eeq
where $\dot{\varphi}_{k}(t)=\bra{\phi_{k}(t)}\qty(i\partial/\partial t-H(t))\ket{\phi_{k}(t)}$. As first formalized in~\cite{chen2011lewis}, in the same spirit of transitionless quantum driving~\cite{berry2009transitionless}, Eqs.~\eqref{eq:invariant1} and~\eqref{eq:pop} allow to set up a reverse engineering procedure~\cite{chen2010fast} of the Hamiltonian $H(t)$ such that the system is exactly transferred from an initial to a target state within a finite time~$t_{\text f}$. Such a tailored protocol thus provides a speed-up with respect to its adiabatic counterparts working between the same pair of states.

A typical choice for the initial and target states are the eigenstates of the system Hamiltonian at $t=0,t_{\text f}$.~A suitable condition is thus that $I(t)$ shares common eigenvectors with $H(t)$ both at the start and end of the protocol, i.e., $\comm{H(t_{a})}{I(t_{a})}=0$, for $t_{a}=t_{0},t_{\text f}$. If the latter equations (also known as the frictionless conditions) are fulfilled (see also App.~\ref{sec:appA}), then from Eqs.~\eqref{eq:pop} and \eqref{eq:coher} it follows that the engineered protocol transfers the quantum state from an arbitrary superposition of Hamiltonian eigenstates at $t=t_{0}$ to a superposition of eigenstates at~$t=t_\text{f}$, preserving the initial state populations. The choice of the initial and target states can be also extended to mixed states. 

In Fig.~\ref{fig:Protocolsketch}a, we provide a pictorial representation of our reverse-engineering approach. Following~\cite{levy2018noise}, our aim is to exploit the high degree of freedom in the choice of the invariant  operator $I(t)$ at the intermediate times to engineer the corresponding operator $H(t)$. Among the infinite solutions to Eq.~\eqref{eq:invariant1}, it is possible to choose the ones that obey desired physical constraints, e.g., the driving protocol that minimizes the detrimental effect of the environment, as we will explain in Sec.~\ref{sec:noise}. 
However, in order to set up our reverse engineering procedure from Eq.~\eqref{eq:invariant1}, more information is needed on the properties of the operator $I(t)$. In this regard, we assume the operators $(H(t),I(t))$ belong to the same Lie algebra~\cite{gilmore_2008}.~We can then consider the representation of the algebra related to the Lie group of order $n$ by means of its generators $\{T_{a},a=1,\dots n\}$ and the structure factors $f_{abc}$ such that $\comm{T_a}{T_b}=i \sum_{c=1}^{n}f_{abc} T_c $ (see App.~\ref{sec:appA}).~Due to the closure of the Lie Algebra, any operator belonging to it can be written in terms of the generators $T_a$.~Thus, under the previous assumption, we can represent the Hamiltonian and the invariant operator in the Schr{\"o}dinger picture as $H(t)=\sum_{k=1}^{n} h_{k}(t) T_k$ and $I(t)=\sum_{k=1}^{n} d_{k}(t) T_k$, where $h_{k}(t)$ are the control fields and $d_{k}(t)$ are arbitrary functions of time. The advantages of such a representation become evident as we insert the previous expressions into Eq.~\eqref{eq:invariant1} to find 
\beq\label{eq:diffEq}
\dot{\bf{d}}(t)=\mathcal{M}({\bf h}(t)) {\bf d}(t), 
\eeq
where ${\bf d}(t)=(d_1(t),\dots,d_n(t))^\text{T}$ and $\mathcal{M}_{kj}(t)=\sum_{m} f_{kmj} h_{m}(t)$.~It follows that, under the closure condition, if the Hamiltonian $H(t)$ is known at any point in time, then, employing the frictionless conditions, the first order ODE in Eq.~\eqref{eq:diffEq} can be solved and the operator $I(t)$ can be fully determined. 

From a reverse engineering perspective, we can consider Eq.~\eqref{eq:diffEq} as a set of algebraic equations in the vector ${\bf h}(t)$. While it is not generally possible to invert Eq.~\eqref{eq:diffEq}, i.e., to express ${\bf h}(t)=\mathcal{N}({\bf d}(t))$, we can adopt a suitable parametrization of the functions $d_{k}(t)$ and express the $h_k(t)$ in terms of the latter.Following this route, we have access to an infinite parameter space for the operator $I(t)$, and each point in this space corresponds to a single STA protocol for the closed system connecting the desired initial and final states. However, some caveats are in order. First, the parametrized invariant needs to obey suitable constraints at the start and the end of the protocol related to frictionless conditions, and second, the algebraic relations need to be well-posed, so that no divergencies in the expressions for the driving fields occur.

%Namely, it is possible to tailor suitable Hamiltonian controls

\section{Reverse engineering in the presence of the environment}\label{sec:noise} The approach outlined in Sec.~\ref{sec:STA} allows in principle to find infinite invariant operators $I(t)$ such that, in the closed system scenario, it is possible to achieve STA with fixed initial and target states. The success of each of these protocols can be measured by means of the fidelity~\cite{nielsenchuang2010} with respect to the target state $\rho_{\text{tar}}$      
 \beq\label{eq:fides}
    \mathcal{F}(\rho(t_{\rm f}),\rho_{\text{tar}})=\tr \sqrt{\sqrt{\rho(t_{\rm f})} \rho_{\text{tar}} \sqrt{\rho(t_{\rm f})} }, 
 \eeq
%{\color{red} [I think the Fidelity is usually defined as the square of the trace.]}
where $\rho(t_{\rm f})$ is the state of the system at the end of the protocol. In the absence of external influences, each reverse-engineered protocol performs the task with ideal fidelity.
However, as sketched in Fig.~\ref{fig:Protocolsketch}a, due to the unavoidable influence of the environment, the success of the STA protocol is expected to decrease. Indeed, the actual system dynamics will deviate from the predictions of Eq.~\eqref{eq:VonNeu} and, as a consequence, Eq.~\eqref{eq:invariant1} will not hold. 

\subsection{Models of dissipation}

To account for the presence of the environment, which acts as a source of noise and induces dissipation and decoherence, we adopt the model of an open quantum system~\cite{weissbook,breuer,rivas2012open}, where the driven system is in contact with a thermal bath. In this setting, many rigorous approaches are available to treat the combined effects of system-bath interactions and the driving fields~\cite{GRABERT1988115,grifoni1998driven}. Herein the dynamics is modeled as a unitary dynamics of system + bath, and the bath is usually described as a set of quantum harmonic oscillators whose frequencies follow a spectral distribution function~\cite{weissbook}. The advantage of these approaches is that they allow to explore even non-Markovian and strong coupling effects \cite{prior2010efficient,strathearn2018efficient} on the dynamics. 

The starting point is the conventional formulation of the composite system Hamiltonian as
\beq\label{eq:sysbath}
     H_{\text{tot} }(t)= H_{\text S}(t)\otimes \mathbb{1}_{\text B} + \mathbb{1}_{\text S}\otimes H_{ \text B} + H_{ \text {I}},
 \eeq
where $H_{\text{S}}(t)$, $H_{\text{B}}$, $H_{\text{I}}$ are the Hamiltonian of the system (S), the bath (B) and the interaction (I), respectively. Here, we assume that the only control is on the system Hamiltonian and that the coupling to the bath cannot be turned off, such that $H_{\text{I}}$ is independent on time and reads 
\beq\label{eq:Hint}
     H_{\text{I}}= \sum_{k} c_{k} A_{k}\otimes B_{k},
\eeq
where $A_{k}(B_{k})$ are operators acting on the Hilbert space of S(B). The state of the whole S+B system $\rho_{\text{SB}}(t)$ evolves as written in Eq.~\eqref{eq:statet}, where the unitary operator is replaced by $U_{\text{tot}}(t,t_0)=\mathcal{T}\exp[-i\int_{t_{0}}^{t} H_{\text{tot}}(t^{\prime}) \mathrm{d} t^{\prime} ]$. From Eq.~\eqref{eq:sysbath}, it follows that S undergoes a time-dependent protocol while it is continuously interacting with the bath. Moreover, the interaction Hamiltonian in Eq.~\eqref{eq:Hint} may not belong to the Lie algebra of operators of the system S. 
From Eqs.~\eqref{eq:Heisenberg},~\eqref{eq:invariant1}, and~\eqref{eq:sysbath}, and from the fact that $I(t)$ is an invariant operator of S in the Schr\"odinger picture, we can readily show that its derivative under the system-bath interaction reads
\beq\label{eq:derinvopen}
    \frac{\mathrm d}{\mathrm d t}{I^{\mathrm H}}(t)=i\sum_{k}c_{k}U_{\text{tot}}^{\dagger}(t,t_{0})\qty(\comm{A_{k}}{I(t)}\otimes B_{k}) U_{\text{tot}}(t,t_{0}).
\eeq
Equation~\eqref{eq:derinvopen} shows that the operator $I(t)$ is not an invariant of the open system. Furthermore, Eq.~\eqref{eq:derinvopen} describes an involved operator acting on the composite Hilbert space of S + B. It also depends on the actual unitary dynamics, which is not closed under the same Lie algebra of S. Additional issues are linked to the equation for the reduced state $\rho_{\text{S}}(t)$ under Hamiltonian Eq.~\eqref{eq:sysbath}. Indeed, depending on the chosen model of S and B, it might not be accessible in closed form, and memory effects due to the system-bath interaction may be present~\cite{weissbook}.           

In order to achieve a simplified description of the open system dynamics, we confine our analysis to the Markovian setting, i.e., we model the dissipative dynamics of the reduced system by means of a time-dependent quantum master equation (QME) of the Lindblad-Gorini-Kossakowski-Sudarshan (LGKS) kind~\cite{breuer,alicki2007quantum,rivas2012open}. Similar equations have been derived both in the adiabatic~\cite{albash2012quantum} and nonadiabatic~\cite{dann2018time,mozgunov2020completely} regimes, by using different sets of approximations \cite{Wudriven2022} and numerical techniques \cite{di2023time}.  Moreover, it has been shown that they can be directly simulated on quantum computing processors~\cite{watad2023variational}. However, several theoretical issues may arise, as those related to their thermodynamic consistency~\cite{kosloff2013quantum,dann2021open}, linked to the local and global properties of the dissipator~\cite{levy2014local,de2018reconciliation}, as well as their inaccuracy beyond the weak coupling regime~\cite{cangemi2019dissipative}.   
    
In order to set up a QME, we assume $\rho_{\text{SB}}(t)$ to be a product state at each point in time $t$ (Born approximation), i.e., $ 
     \rho_{\text{SB}}(t)\approx\rho_{\text S}(t)\otimes \rho_{\text B},$
where the state of the bath is a thermal state  $\rho_{\text B}=(1/Z)\exp(-\beta H_{\text B}),$ with $ Z=\tr[\exp(-\beta H_{\text B})]$. This is a good approximation when working in the weak system-bath coupling limit. 
%From the Von Neumann equation for the state $\rho_{\text{SB}}(t)$, written 
By denoting with $\tilde{\rho}_{\text{SB}}(t)$ the density matrix of S+B in the interaction picture,  %$\dot{\tilde{\rho}}_{\text{SB}}(t)=-i\comm{\tilde{H}_{\text{I}}(t)}{\tilde{\rho}_{\text{SB}}(t)}$, 
after taking the trace with respect to B, we obtain in second-order
\beq\label{eq:BM1}
     \frac{\mathrm{d}}{\mathrm{d} t}\tilde{\rho}_{\text S}(t)=-\int_{t_{0}}^{t}\mathrm{d}s \tr_{\text B}\comm{\tilde{H}_{\text{I}}(t)}{\comm{\tilde{H}_{\text{I}}(s)}{\tilde{\rho}_{\text S}(s)\otimes \tilde{\rho}_{\text B}}}.
\eeq
where $\tilde{H}_{\text{I}}(t)=U^{(0)\dagger}_{\text{SB}}(t,t_{0})H_{\text{I}}U^{(0)}_{\text{SB}}(t,t_{0})$ and $U^{(0)}_{\text{SB}}(t,t_{0})$ is the evolution operator of the noninteracting S+B system. Equation~\eqref{eq:BM1} is the starting point to perform the Markov approximation, which consists in neglecting memory effects in the dynamics of the reduced state $\tilde{\rho}_{\text S}$ and exploiting the condition of fast decay of bath correlations with respect to the characteristic timescale of the dynamics~\cite{breuer}.    

However, as we are dealing with a driven quantum system, the derivation of the time-dependent QME from the Hamiltonian in Eq.~\eqref{eq:sysbath} poses additional challenges. The degree of adiabaticity of the time-dependent protocol ruled by $H_{\text S}(t)$ is conventionally quantified by means of the parameter $\mu(t)$,
\beq\label{eq:mupar}
\mu(t)=\sum_{n\neq m} \frac{\bra{\epsilon_{n}(t)}\dot{H}_{\text{S}}(t)\ket{\epsilon_{m}(t)}}{(\epsilon_{n}(t)-\epsilon_{m}(t))^2}, 
\eeq
where $\ket{\epsilon_{m}(t)}$ are the instantaneous eigenstates of $H_{\text{S}}(t)$ with energy $\epsilon_{m}(t)$. Due to the arbitrary time dependence of the reverse-engineered fields in Eq.~\eqref{eq:diffEq}, the resulting Hamiltonian $H_{\text{S}}(t)$ may not change adiabatically over time, i.e., $\mu(t)$ can be different from zero, as well as its first derivative. 
%\noindent Eq.~\eqref{eq:BMfinal} describes the evolution of the state when the system is subject to the influence of a bath. As a consequence, the dynamical equations for any operator acting on the Hilbert space of S change. Let's focus on the invariant operator of the system S. Due to the influence of the environment, this operator does not obey Eq.~\eqref{eq:invariant1}.~It can be understood by rewriting Eq.~\eqref{eq:BMfinal} in the Schr{\"o}dinger picture

Below, we assume a time-local QME of the LGKS form to exist, which allows for the description of the open system in the presence of driving fields evolving nonadiabatically in the course of time. Although in Sec.~\ref{sec:apps} we will employ a detailed microscopic derivation of the nonadiabatic QME (see also App.~\ref{sec:appC}), here we only refer to its LGKS form. Namely, we assume that, after a given set of approximations, from Eq.~\eqref{eq:BM1}, a dynamical equation in the Schr\"{o}dinger picture can be obtained that reads
\beq\label{eq:BMSchro}
     \frac{\mathrm{d}}{\mathrm{d} t}\rho_{\text S}(t)=  \mathcal{L}_{t}(\rho_{\text S}(t)),
\eeq     
where $\mathcal{L}_{t}$ is the time-dependent generator of the open system dynamics, i.e., 
\beq\label{eq:generator}
  \mathcal{L}_{t}=-i\comm{H_{\text{S}}(t) + H_{\text{LS}}(t)}{\circ} +\mathcal{D}_{t}.
\eeq
Here, the $H_{\text{LS}}(t)$ is the Lamb shift term, which provides a renormalization of the closed system Hamiltonian $H_{\text S}(t)$. $\mathcal{D}_{t}$ is the dissipation superoperator that, due to the LGKS property, reads 
\beq\label{eq:Dissip}
  \mathcal{D}_{t}= \sum_{l} \kappa_{l}(t)\qty(F_{l,t}\circ F^{\dagger}_{l,t}-\frac{1}{2}\qty{F^{\dagger}_{l,t}F_{l,t},\circ}),
\eeq
where $F_{l,t}$ denote the time-dependent LGKS operators and $\kappa_{l}(t)\geq 0$ the corresponding rates. Its detailed form depends on the bath equilibrium correlation function (see for instance App.~\ref{sec:appC}). Then, the reduced system state at each time $t$ can be obtained from the initial time as follows
\beq\label{eq:twopoint}
  \rho_{\text S}(t)= V(t,t_{0})\rho_{\text S}(t_{0}),
\eeq   
where we introduced the two-point dynamical map $V(t,t_{0})= \mathcal{T}e^{\int_{t_{0}}^{t}\mathcal{L}_{t^{\prime}}\mathrm{d}t^{\prime}}$~\cite{rivas2012open,chruscinski2022dynamical}.~Given the form of Eq.~\eqref{eq:generator}, the dynamical equation for a generic operator that shows an explicit time dependence in the Schr{\"o}dinger picture reads 
\beq\label{eq:adjointeq}
\frac{\mathrm{d}}{\mathrm{d}t} O^{\text{H}}(t)=V^{\dagger}(t,t_{0})\qty[\mathcal{L}^{\dagger}_{t}O(t) +\frac{\partial}{\partial t} O(t)],
\eeq  
where $\mathcal{L}^{\dagger}_{t}$ is the adjoint generator at any time $t$ and we denoted with $V^{\dagger}(t,t_{0})$ the adjoint map.~It is evident from Eq.~\eqref{eq:adjointeq} that, contrary to the case of one-parameter dynamical semigroup, the adjoint map does not commute with $\mathcal{L}^{\dagger}_{t}$. Moreover, following Eq.~\eqref{eq:adjointeq}, under the action of the two-point map, the derivative of $I(t)$ in Eq.~\eqref{eq:derinvopen} can be replaced by the following expression
\beq\label{eq:adjointeqINV}
\frac{\mathrm{d}}{\mathrm{d}t} I^{\text{H}}(t)=V^{\dagger}(t,t_{0})\qty[i\comm{H_{\text{LS}}(t)}{\circ} + \mathcal{D}^{\dagger}_{t}]I(t). 
\eeq  
On the physical side, the main consequence of Eq.~\eqref{eq:adjointeqINV} is that the simple Eq.~\eqref{eq:pop} does not hold anymore,~i.e., the instantaneous populations in the basis $\ket{\phi_{m}(t)}$ change over time ruled by the time-local dissipator $\mathcal{D}_{t}$, and the STA protocol as devised in Sec.~\ref{sec:STA} cannot be successful.  

\subsection{Reverse-engineering against noise}\label{sec:reverse} 

%We aim at finding a strategy to minimize the change in the populations of $\rho_{\text{S}}(t)$ in the invariant eigenbasis that would not require the full-time propagation of the invariant operator under Eq.~\eqref{eq:adjointeqINV} or any iterative expensive computation.
\textit{Our aim is to devise a strategy to maximize the fidelity in Eq.~\eqref{eq:fides} that would not require the full-time propagation of the invariant operator under Eq.~\eqref{eq:adjointeqINV} or any iterative expensive computation that requires solving the open system dynamics.}
Moreover, as explained in Sec.~\ref{sec:STA}, we look for an open-loop control protocol, i.e., we avoid using any measurement-feedback scheme \cite{wiseman_2009,roy2020measurement} in the course of the dynamics. Next, we consider a generic invariant operator for the closed system, as it results from Eq.~\eqref{eq:diffEq}, and we write the expression for the derivative of the populations in its eigenbasis under the influence of noise.~From Eqs.~\eqref{eq:invariant1} and \eqref{eq:BMSchro}, it follows
\beq\label{eq:newpop}
\dot{\rho}_{kk}(t)=\tr\qty[\ket{\phi_{k}(t)}\bra{\phi_{k}(t)}(-i\comm{H_{\text{LS}}(t)}{\circ} +\mathcal{D}_{t})\rho_{\text S}(t)].
\eeq
Hereafter, we assume to safely neglect the Lamb shift contribution, as it just provides a small renormalization of the closed system Hamiltonian $H_{\text S}(t)$. By denoting with $\rho_{mn}(t)$ the coefficient of the state in the instantaneous eigenbasis, i.e., $\rho_{\text S}(t)=\sum_{nm}\rho_{nm}(t)\ket{\phi_{n}(t)}\bra{\phi_{m}(t)}$, we find that
\begin{align}
\label{eq:rates}
\dot{\rho}_{kk}(t)&=\sum_{mn}\mathcal{R}^{(kk)}_{nm}(t)\rho_{mn}(t), \;\; \; {\rm with},
\\ \nonumber
\mathcal{R}^{(kk)}_{nm}(t)&=\tr\qty[\ket{\phi_{k}(t)}\bra{\phi_{k}(t)}(\mathcal{D}_{t}\ket{\phi_{m}(t)}\bra{\phi_{n}(t)})].
\end{align}
The matrix $\mathcal{R}^{(kk)}_{nm}(t)$ describes the rates of change over time of the $k$-th population due to each element of the state. Notice that, due to the LGKS form, the matrix $\mathcal{R}^{(kk)}_{nm}$ can be computed once the action of the dissipator on the projectors of the instantaneous eigenbasis is known, while the full solution of the equation depends on the dynamical map $V(t,t_{0})$. It follows that, if $\mathcal{D}_{t}\ket{\phi_{m}(t)}\bra{\phi_{n}(t)}=0\mbox{ } \forall \mbox{} m, n$, the populations in the invariant eigenbasis are preserved.~Alternatively, one may require the $\ket{\phi_{k}(t)}\bra{\phi_{k}(t)}$ to be a left eigenoperator of the dissipator with null eigenvalue. These requirements are akin to finding an instantaneous decoherence-free subspace (DFS)~\cite{lidar1998decoherence,knill2000theory,lidar2003decoherence}. However, as we allow for the LGKS operators to change over time, the required conditions for the existence of a DFS could not be met at every point in time in the course of the protocol.

%Our approach is thus focused on finding the invariant of the closed system that, on average, minimizes the detrimental action of dissipation and decoherence on the populations of $\rho_{\text S}(t)$ in the eigenbasis $\ket{\phi_{n}(t)}$.%
Our approach is thus focused on finding the invariant of the closed system that, on average, minimizes the detrimental action of dissipation and decoherence on the protocol fidelity at the final time. A rather intuitive way to achieve this is to require the time-averaged derivatives of the populations in Eq.~\eqref{eq:rates} to be minimized. However, due to our reverse-engineering approach, the target state can be a rather generic superposition of invariant eigenstates at final times, so that both populations and coherences of $\rho_{\text S}(t)$ in the invariant eigenbasis will contribute to the fidelity. For a given choice of the target state, the quantity in  Eq.~\eqref{eq:fides} depends on the open-system map $V(t_{\text{f}},t_0)$. Different approaches have been devised to evaluate similar quantities, exploiting the properties of the instantaneous attractor \cite{cavina2017slow,Scandi2019thermodynamiclength}, as well as exact factorization of the map \cite{scopa2019exact}.~Eventually, Eq.~\eqref{eq:fides} could in principle be evaluated by considering a Dyson series of the dynamical map that, for a time-local LGKS generator as in Eq.~\eqref{eq:BMSchro}, guarantees the property of Completely Positive (CP) divisibility  \cite{chruscinski2022dynamical}.

In what follows, we propose to simplify the problem by choosing the target state as a pure state,~i.e., $\rho_{\text{tar}}=\ket{\psi_{\text{tar}}}\bra{\psi_{\text{tar}}}$, so that Eq.~\eqref{eq:fides} reduces to $\mathcal{F}=\sqrt{\bra{\psi_{\text{tar}}}\rho_{\text{S}}(t_{\text{f}})\ket{\psi_{\text{tar}}}}$. We stress that this is not strictly required, as our approach can be extended to mixed states allowed by the reverse-engineered protocol \cite{Liang_2019}. Moreover, we express the formal solution of the dissipative map $V(t,t_{0})$ in terms of the closed system one, i.e., $V^{0}(t,t_{0})= \mathcal{T} e^{-i\int_{t_{0}}^{t}\comm{H_{\text{S}}(t^{\prime})}{\circ}\mathrm{d}t^{\prime}}$, that reads
\beq\label{eq:dyson1}
V(t,t_{0})=V^{0}(t,t_{0}) + \int_{t_{0}}^{t} V^{0}(t,s)\mathcal{D}_{s}V(s,t_{0})\mathrm{d}s.
\eeq
Equation~\eqref{eq:dyson1} allows for a Dyson expansion of the full map in terms of $V^{0}(t^{\prime},t^{\prime\prime})$,where $t^{\prime\prime}<t^{\prime}$ are intermediate times, and of increasing powers of the dissipator $\mathcal{D}_{t}$. The advantage brought in by Eq.~\eqref{eq:dyson1} is evident by looking at the key properties of the invariant explained in Sec.~\ref{sec:STA}. Indeed, when written in the invariant eigenbasis $\{\ket{\phi_{n}(t)}\}$, the map $V^{0}(t^{\prime},t^{\prime\prime})$ can be expressed in closed analytic form  so that Eq.~\eqref{eq:dyson1} can be evaluated at any order in $\mathcal{D}_{t}$. At first order in $\mathcal{D}_{t}$, we can define a suitable functional of the control fields and of the closed system state as follows
\beq\label{eq:functional3}
\mathcal{Z}=\bra{\psi_{\text{tar}}}\int_{t_{0}}^{t_{\text{f}}}V^{0}(t_{\text{f}},s)\mathcal{D}_{s}(\rho^{0}_{\text{S}}(s))\mathrm{d}s\ket{\psi_{\text{tar}}},
\eeq
with $ \rho^{0}_{\text{S}}(s)=V^{0}(s,t_{0})(\rho_{\text{S}}(t_{0}))$.
It is easy to prove that, by minimizing $\abs{\mathcal{Z}}$ the protocol fidelity can be maximized (see App.~\ref{sec:appD} for details). An intuitive explanation for the choice of the functional, Eq.~\eqref{eq:functional3}, stems from the fact that if all the rates $\mathcal{R}^{(kk)}_{nm}(t)$ were to be zero, as for the closed quantum system case, then fidelity one would be reached.
Including the information about the initial state in the functional, i.e., the term $\rho^{(0)}_{mn}(t)$,  accounts for the different weights of the rates and becomes important when the initial state is a superposition state, as studied in  Sec.~\ref{sec:oscillator}.   
In principle, higher-order corrections involving additional contributions linked to the dissipator may be taken into account. However, in the next sections, we will show that even the simple form of Eq.~\eqref{eq:functional3} is enough to find a set of dynamical invariants that minimize the influence of decoherence and dissipation. The optimization can be further simplified by using the control fields parametrization $d_{k}(t)$ introduced in Sec.~\ref{sec:STA}, which also takes into account the additional constraints related to the open system setting. The details of the parametrization depend on the properties of the Lie algebra of the system S (see also App.~\ref{sec:appA}). 

\section{Examples}\label{sec:apps}
Below, we demonstrate the feasibility of our reverse-engineering approach by addressing two paradigmatic instances of driven quantum systems, i.e., a two-level system and a parametric harmonic oscillator. As our model of QME, we adopt the recently developed Non-Adiabatic quantum Master Equation (NAME) \cite{dann2018time} in the inertial approximation \cite{dann2020fast,dann2021inertial,turyansky2023inertial}, which holds valid for sufficiently slow driving acceleration.
The basic idea underlying the NAME approach is to find a system of dynamical left eigenoperators of the propagator in the Liouville-Hilbert space of the system S, which can be used as a reference basis to expand at each time $t$ the  interaction operator $\tilde{H}_{\text{I}}(t)$ (see App.~\ref{sec:appC}). The advantages brought in by this set of operators is that, under suitable conditions and for $\mu(t)=const$, they lead to an expansion of $\tilde{H}_{\text{I}}(t)$ in terms of operators computed at $t=0$. As a consequence, the secular approximation can be performed using this set, and from Eq.~\eqref{eq:BM1}, a time-local LGKS dissipator can be derived. As shown in \cite{dann2020fast,dann2021inertial} (see also App.~\ref{sec:appC}), in the inertial approximation limit, this strategy can be employed even in the instance of a time-dependent adiabaticity parameter $\mu(t)$, provided that $(\dot{\mu}(t)/\Omega(t))^2\ll 1$, where $\Omega(t)$ is the energy gap of the system $\text{S}$. The latter constraint is motivated by the need for sufficiently small acceleration of the driving fields.
We report the microscopic derivation of NAME equation in App.~\ref{sec:appC}. 

\subsection{The driven TLS}\label{sec:qubit}
We consider the instance of a driven TLS described by the Hamiltonian operator
\beq\label{eq:Hqubit}
H_{\text S}(t)= \frac{\Delta(t)}{2}\sigma_z + \frac{\epsilon(t)}{2}\sigma_x, 
\eeq
where $\sigma_a$ are Pauli operators and $(\Delta(t),\epsilon(t))$ are the driving fields.~As sketched in Fig.~\ref{fig:Protocolsketch}b, the fields in Eq.~\eqref{eq:Hqubit} modulate the instantaneous energy gap $\Omega(t)=\sqrt{\Delta^2(t) +\epsilon^2(t)}$. The Hamiltonian in Eq.~\eqref{eq:Hqubit} belongs to the $\text{su}(2)$ Lie algebra.~The invariant operator $I(t)$ can also be written as a combination of the three Pauli operators, which are the generators of the same Lie algebra (see also App.~\ref{sec:appA}),~i.e.,
$T_{1}=
\frac{1}{2}\sigma_x$, $T_{2}=\frac{1}{2}\sigma_y$, $T_{3}=\frac{1}{2}\sigma_z\mbox{, with } \comm{T_a}{T_b}=i\sum_{c=1}^3\varepsilon_{abc}T_c.$
The structure constants, $\varepsilon_{abc}$, are the elements of the Levi-Civita tensor.
It can be shown that, from Eq.~\eqref{eq:diffEq}  (see also App.~\ref{sec:appA}), a representation of the invariant operator $I(t)$ reads~\cite{levy2018noise}
\beq\label{eq:su2inv}
I(t)=\Omega(0)( r_x(t)\sigma_x + r_y(t)\sigma_y +r_z(t)\sigma_z ), 
\eeq
with $\vec{r}(t)=(\sin G(t)\cos B(t),\sin G(t)\sin B(t),\cos G(t))$, and $G(t)$ and $B(t)$ are a couple of auxiliary fields that are the solution of the following differential equations 
\beq\label{eq:diffeqQUBIT}
\Delta(t)=\dot{B}(t) -\frac{\dot{G}(t)}{\tan G(t) \tan B(t) }, \epsilon(t)=-\frac{\dot{G}(t)}{\sin B(t)}.
\eeq
From Eq.~\eqref{eq:su2inv}, it also follows that the instantaneous eigenvalues and eigenstates of $I(t)$ read
\beq\label{eq:eigensu2}
\lambda_{\pm}=\pm \Omega(0), \ket{\phi_{\pm}(t)}=f_{\pm}(t)\ket{0}\pm f_{\mp}(t)e^{iB(t)}\ket{1},
\eeq
where $\{\ket{0},\ket{1}\}$ are the eigenstates of $\sigma_{z}$ with eigenvalues $\{+1,-1\}$, $f_{+}(t)=\sin G(t)/\sqrt{2(1-\cos G(t))}$ and $f_{-}(t)=\sqrt{(1-\cos G(t))/2}$.~Moreover, it is easy to show from Eqs.~\eqref{eq:mupar} and~\eqref{eq:Hqubit} that the adiabaticity parameter reads
\beq\label{eq:muparqubit}
\mu(t)=\frac{\dot{\Delta}(t)\epsilon(t)-\Delta(t)\dot{\epsilon}(t)}{\Omega^3(t)}.
\eeq

\noindent The dissipative dynamics of the TLS governed by Eq.~\eqref{eq:Hqubit} can be easily studied in the context of the NAME QME \cite{dann2020fast}. Indeed, let us assume the system is in contact with a bath, modeled with a set of harmonic oscillators. The TLS is linearly coupled to the position of each oscillator of the bath by means of the $\sigma_{y}$ operator. We can thus replace the bath and interaction operator in Eq.~\eqref{eq:sysbath} with
\beq\label{eq:Hamiltotqubit}
H_{\text{B}} = \sum_{k}\omega_{k} a^{\dagger}_{k}a_{k} \;\; {\rm and} \;\; H_{ \text {I}}=\frac{\sigma_{y}}{2}\sum_{k}\nu_{k}(a_{k} + a^{\dagger}_{k}),
\eeq
where $a_{k}(a^{\dagger}_{k})$ are conventional bosonic annihilation (creation) operators of the $k$-th mode of the bath and $\nu_{k}$ has the dimension of a frequency. Moreover, the bath is described by means of a Ohmic spectral density function \cite{caldeira1983,weissbook}, 
\beq\label{eq:spectral}
\tilde{J}(\omega)= \sum_{k}\nu^2_{k} \delta(\omega -\omega_{k})=2\tilde{\gamma}\omega e^{-\omega/\omega_{c}},
\eeq
where $\tilde{\gamma}$ is a dimensionless parameter describing the strength of the dissipation and $\omega_{c}$ is the cutoff frequency. As we show in App.~\ref{sec:appCqubit}, the peculiar choice of the coupling operator in Eq.~\eqref{eq:Hamiltotqubit} is motivated by its simplicity, while the method allows for an arbitrary system-bath coupling operator.  

The resulting form of the LGKS dissipator (see App.~\ref{sec:appCqubit}) reads 
\beq\label{eq:dissipatorqubit}
\mathcal{D}_{t}=\sum_{l=1}^{3}\kappa_{l}(t)\qty(F_{l,t}\circ F^{\dagger}_{l,t} - \frac{1}{2}\{F^{\dagger}_{l,t}F_{l,t},\circ \}),
\eeq
where, employing a peculiar set of operators in the Schr\"{o}dinger picture $\qty(H_{\text{S}}(t),L(t),D(t))=\{H_{\text{S}}(t),(\epsilon(t)/2) \sigma_{z} -(\Delta(t)/2)\sigma_{x},(\Omega(t)/2)\sigma_{y}\}$ and setting $k(t)=\sqrt{1+\mu^2(t)}$, the LGKS operators read
\begin{align}\label{eq:jumpqubit}
F_{1,t}&=\frac{1}{k(t)}H_{\text{S}}(0) +\frac{\mu(t)}{k(t)}D(0),\nonumber\\
F_{2,t}&=\frac{\mu(t)}{\sqrt{2}k(t)}H_{\text{S}}(0) +\frac{i}{\sqrt{2}}L(0) -\frac{1}{\sqrt{2}k(t)}D(0),\nonumber\\
F_{3,t}&= \frac{\mu(t)}{\sqrt{2}k(t)}H_{\text{S}}(0) -\frac{i}{\sqrt{2}}L(0) -\frac{1}{\sqrt{2}k(t)}D(0).
\end{align}
\noindent As follows from the inertial and secular approximation (see App.~\ref{sec:appC}), the rates $\kappa_{l}(t)$ can be computed in terms of the equilibrium correlation function $\Gamma_{kk^{\prime}}(\omega)=\int_{0}^{+\infty} e^{i\omega s}\tr_{\text B}[\tilde{B}_{k}(t)\tilde{B}_{k^{\prime}}(t-s)\rho_{\text B}]\mathrm{d}s =\frac{1}{2}\gamma_{kk^{\prime}}(\omega) + iS_{kk^{\prime}}(\omega)$, where $\tilde{B}_k(t)=\sqrt{1/(2 m_{k}\omega_{k})}(\tilde{a}_{k}(t)+\tilde{a}^{\dagger}_{k}(t))$ denote the bath operators  in the interaction picture, as it also follows from Eq.~\eqref{eq:BM1}.~The rates read $\kappa_{1}(t)=\gamma(0)\mu^2(t)/(\Omega^2(0) k^{2}(t)),\:\kappa_{2}(t)=\gamma(\dot{\Lambda}_{2}(t))/(2\Omega^2(0) k^{2}(t)),\:\kappa_{3}(t)=\gamma(\dot{\Lambda}_{3}(t))/(2\Omega^2(0) k^{2}(t))$, with $\dot{\Lambda}_{2}(t)=\Omega(t)k(t),\:\dot{\Lambda}_{3}(t)=-\Omega(t)k(t)$ and $\gamma(\omega)=\sum_{k}\gamma_{kk}(\omega)$. Notice that, as a consequence of the NAME approach and the inertial approximation, each operator $F_{l,t}$ is written in terms of a linear combination of $\text{su}(2)$ generators computed at time $t=0$ with time-dependent coefficients. Once the driving fields in Eq.~\eqref{eq:Hqubit} are known, the dissipator in Eq.~\eqref{eq:Dissip} is analytically computed.

%{\color{red} [Let's think if we want all these details in the main text, it shifts the attention to the details of the NAME derivation instead of the control problem. Maybe we can move some of the details to the appendix and just provide the Master equation. We can talk about it.]  }

\subsubsection{The protocol}

Without loss of generality, we focus on a family of STA protocols to steer the TLS from the ground state of $H_{\text S}(0)$ to the ground state of the Hamiltonian $H_{\text S}(t_{\text{f}})$, in a finite amount of time $t_{\text{f}}$. These protocols provide a speed-up with respect to any adiabatic protocol working between the same couple of states. We also impose the frictionless conditions at the start and at the end of the protocol,~i.e., $\comm{H_{\text S}(t_{a})}{I(t_{a})}=0,$ for $ t_{a}=0,  t_{\text f}$.~It follows that at the boundaries of the protocol, the Hamiltonian and the invariant operator share common eigenvectors. If the state of the system is prepared in the ground state of $H_{\text{S}}(0)$, in the absence of degeneracies this state coincides with an eigenstate of $I(0)$.~As a direct consequence of Eqs.~\eqref{eq:pop} and \eqref{eq:coher}, under the unitary dynamics, the population of this instantaneous state does not change in time.~Moreover, if at $t=t_{\text f}$ the target state $\rho_{\text{tar}}$ is the density matrix corresponding to the previous eigenstate of the invariant, then the protocol achieves ideal fidelity. 

The properties of the Hamiltonian fields at the boundary need to be fixed. We choose $\Delta(0)=\Delta_0>0,\Delta(t_{\text f})=-\Delta_0$, and $\epsilon(0)=\epsilon(t_{\text f})=0$, thus the TLS state is transferred from $\rho(0)=\ket{1}\bra{1}$ to $\rho_{\text{tar}}=\ket{0}\bra{0}$ (see also Fig.~\ref{fig:Protocolsketch}b).~Taking into account the frictionless conditions, along with the values of the fields at the boundaries, it follows that the values of $G(t)$ and of its first derivative at the boundaries remain fixed,~i.e., $G(0)=\pi, G(t_{\text f})=0,$ and $ \dot{G}(0)=\dot{G}(t_{\text f})=0$.~Additional constraints on $B(t)$ arise from the requirement of absence of divergencies in Eq.~\eqref{eq:diffeqQUBIT}, which translates to $B(t_a)=-\pi/2$ and $\dot{B}(t_a)\neq 0$ for $t_a=\{0,t_{\text f}\}$ (see App. \ref{sec:appA} for details). As a consequence, the form of the invariant operator at the boundary remains fixed as well, so that from Eqs.~\eqref{eq:su2inv} and \eqref{eq:eigensu2} $\rho_{\text S}(0)=\ket{\phi_{+}(0)}\bra{\phi_{+}(0)} \text{and } \rho_{\text {tar}}=\ket{\phi_{+}(t_{\text f})}\bra{\phi_{+}(t_{\text f})}$. Although the control fields have well-defined values at the boundaries, at the intermediate times they can have arbitrary shapes. This freedom can be exploited to reverse-engineer the control fields as explained below.  

\subsubsection{Results of reverse engineering}

In the same spirit of \cite{levy2018noise}, we adopt a parametrization of the angles $(G(t),B(t))$ in terms of simple polynomia, 
\beq\label{eq:paramet}
G(t)=\sum_{n=0}^{n_{g}}g_{n} \qty(\frac{t}{t_{\text{f}}})^n,B(t)=\sum_{n=0}^{n_{b}}b_{n} \qty(\frac{t}{t_{\text{f}}})^n,
\eeq
where the parameters $(g_n,b_n)$ are real numbers that can be chosen to fulfill the above constraints.  Indeed, by inserting Eq.~\eqref{eq:paramet} into Eq.~\eqref{eq:diffeqQUBIT}, it is possible to parametrize the fields $(\Delta(t),\epsilon(t))$. 
%i.e., to reverse-engineer the STA protocol avoiding the solution of Eq.~\eqref{eq:diffeqQUBIT}.
For instance, in the case of unitary evolution, the control fields are fully determined to obey frictionless conditions and boundary values by fixing the minimum number of parameters as $n_{g}=3$ and $n_{b}=2$ (see App.~\ref{sec:appA} for details). 

We can thus tackle the problem of the STA protocol in the presence of the environment by simply adding more parameters to Eq.~\eqref{eq:paramet}. We  choose the STA protocol that is less prone to noise and dissipation in the enlarged parameter region allowed by Eqs.~\eqref{eq:diffeqQUBIT}. Indeed, by using the actual form of the $\rho_{\text{tar}}$ described in the previous section, for $t_{0}=0$ Eq.~\eqref{eq:functional3} reduces to     \beq\label{eq:functionalqubit}
\mathcal{Z}=\int_{0}^{t_{\text f}}R^{(++)}_{++}(t^{\prime})\mathrm{d}t^{\prime}.
\eeq

\noindent Notice that Eq.~\eqref{eq:functionalqubit} is proportional to the time average of the decay rate linked to the populations in the state $\ket{\phi_{+}(t)}$, so that minimizing $\abs{\mathcal{Z}}$ is equivalent to requiring the average decay of the population of this state to be reduced over each protocol. 
\begin{figure}
\includegraphics[scale=0.9]{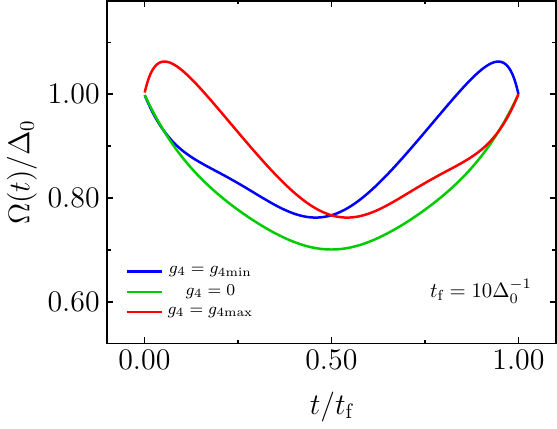}
\caption{The TLS gap plotted against time for three reverse-engineered protocol, as ruled by Eq.~\eqref{eq:diffeqQUBIT}.~Each protocol is obtained by changing the values of the parameter $g_{4}$ in the closed interval $[g_{\text{min}},g_{\text{max}}]=[-8.65,8.65]$. The green curve corresponds to the minimal choice of parameters in the closed case ($g_{4}=0$).~The protocol duration is $t_{\text{f}}=10\Delta^{-1}_{0}$.}
\label{fig:gapfields}
\end{figure}
\begin{figure}
\includegraphics[scale=0.9]{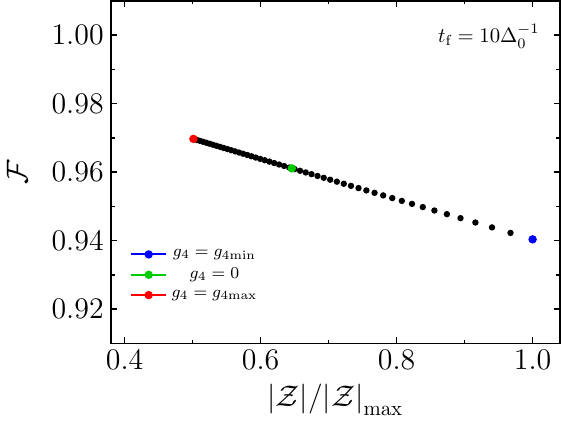}
\caption{The fidelity, as computed numerically from Eqs.~\eqref{eq:fides},\eqref{eq:BMSchro}, plotted versus the normalized functional $\mathcal{Z}$, for fixed dissipation strength $\tilde{\gamma}=5\cdot 10^{-3}$, inverse temperature $\beta=100\Delta^{-1}_{0}$ and protocol duration $t_{\text{f}}=10\Delta^{-1}_{0}$.~The colored dots denote the values of the fidelity corresponding to the reverse-engineered protocols plotted in Fig.~\ref{fig:gapfields}.~The best protocol (red dot) can be found as the one corresponding to minimal $\abs{\mathcal{Z}}$.}
\label{fig:fidelityqubit}
\end{figure}

In Figs.~\ref{fig:gapfields} and \ref{fig:fidelityqubit}, we present the results of our reverse engineering approach. 
In Fig.~\ref{fig:gapfields}, different instances of the STA protocol are shown, corresponding to the choice $n_{g}=4$ and $n_{b}=3$.~The modulation of the qubit gap $\Omega(t)$ is performed in a finite amount of time $t_{\text f}$, and the maximum gap is achieved at different points in time during the protocol. From Eqs.~\eqref{eq:diffeqQUBIT} and \eqref{eq:paramet}, it also follows that the maximum value of the gap increases with the protocol duration $t_{\text f}$. Figure~\ref{fig:fidelityqubit} shows the protocol fidelity,~i.e., Eq.~\eqref{eq:fides} plotted against the values of $\abs{\mathcal{Z}}$, for fixed $t_{\text f}$ and a set of STA protocols chosen in the parameter range allowed by Eq.~\eqref{eq:diffeqQUBIT}.~The protocol fidelity is a monotonically decreasing function of $\abs{\mathcal{Z}}$, showing that a robust STA protocol can always be found by minimizing the latter in a given region of the parameter space. Notice also that, under the effect of noise and dissipation, the STA protocol corresponding to the minimal choice of parameters,~i.e.~$n_{g}=3,n_{b}=2$, achieves a lower fidelity as compared to the optimal one.~Moreover, it can be shown that the choice of Eq.~\eqref{eq:functionalqubit} successfully reproduces the expected behavior of the fidelity w.r.t. the average of the total derivative, i.e., $\mathcal{F}=\sqrt{1-(1/2)\int_{0}^{t_{\text f}}\abs{\ev{\mathrm{d}I^{\text H}/\mathrm{d}t}}\mathrm{d} t}$. However, this easy functional relation between the fidelity and the average of $\ev{\mathrm{d}I/\mathrm{d}t}$ does not hold true if the target state is a superposition of different basis states $\ket{\phi_{n}(t_{\text f})}$, as explained in Sec.~\ref{sec:oscillator}.  

\begin{figure}
\includegraphics[scale=0.9]{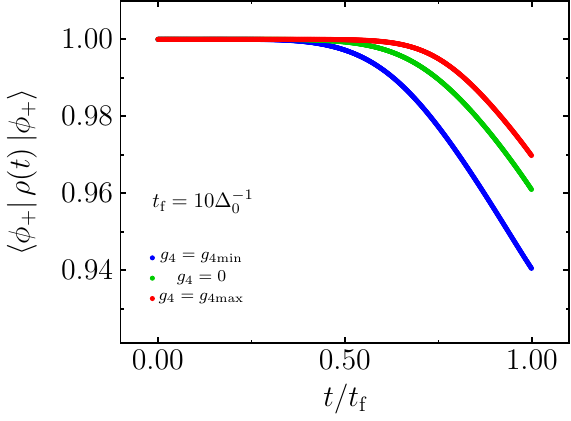}
\caption{Dynamics of the instantaneous populations in the invariant eigenstate $\ket{\phi_{+}(t)}$, numerically computed for $\tilde{\gamma}=5\cdot 10^{-3}$, $\beta=100\Delta^{-1}_{0}$ and protocol duration $t_{\text{f}}=10\Delta^{-1}_{0}$. The reverse-engineered protocols are the same as those reported in Fig.~\ref{fig:gapfields}.  In the label, the time dependence is omitted for brevity.}
\label{fig:rhopp}
\end{figure}

%\begin{figure}
%\includegraphics[scale=1.30]{rhopp_cont.pdf}
%\caption{fig}
%\label{fig:rhocoher}
%\end{figure}

Given our choice of the target state, the optimal STA protocol minimizes the loss in the populations from the instantaneous eigenstate $\ket{\phi_{+}(t_{\text f})}$. It is shown in Fig.~\ref{fig:rhopp}, where the dynamics of the instantaneous populations is plotted for different protocols as a function of time. The best STA protocol is the one for which the decay of the instantaneous populations can be successfully delayed in time. Actually, this dynamical feature is the result of the interplay between the time-dependent dissipation and the driving protocol. Indeed, in the absence of dissipation, the STA protocol is tailored to steer the state from $\ket{\phi_{+}(0)}=\ket{1}$ to  $\ket{\phi_{+}(t_{\text f})}=\ket{0}$. Then, the populations of $\rho_{\text{S}}(t)$ in these two eigenstates of $\sigma_{z}$ are required to flip at some point in time in the course of the protocol. If the flip is performed too early, then the dissipation acts to decrease the populations in the desired target state, leading to a noticeable decrease in the fidelity. 

For increasing $t_{\text{f}}$, the fidelity of the reverse-engineered protocols is expected to decrease as a result of the prolonged interaction with the environment. On the other hand, there are several limitations to the lowest value of $t_{\text f}$. In the first place, the Quantum Speed Limit (QSL) \cite{deffner2017quantum} sets a fundamental lower bound to any driven quantum system, and different formulations have been presented in the closed as well as open system setting \cite{taddei2013quantum,del2013quantum,deffner2013quantum,pires2016generalized}. Furthermore, another limitation may come from the same validity of the NAME equation. Indeed, from Eqs.~\eqref{eq:diffeqQUBIT} and ~\eqref{eq:muparqubit}, and from the frictionless conditions, it follows that by decreasing $t_{\text f}$, $\dot{\mu}(t)$ can exceed the constraints set by the inertial approximation, thus requiring a numerically-exact approach to the dissipative dynamics.  

\subsection{The driven harmonic oscillator}\label{sec:oscillator}

\begin{figure}
\includegraphics[scale=0.80]{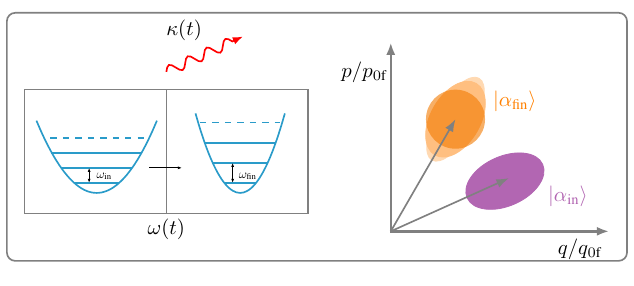}
\caption{Schematic diagram of the reverse-engineering method based on the Lewis-Riesenfeld invariant in the case of a driven harmonic oscillator. The oscillator frequency is modulated in time from $\omega(0)=\omega_{\text{in}}$ to $\omega(t_{\text{f}})=\omega_{\text{fin}}$ and $q_{0\text{f}}=1/\sqrt{2\omega_{\text{fin}}}$, $p_{0\text{f}}=\sqrt{\omega_{\text{fin}}/2}$. Here $\rho_{0}=\ket{\alpha}\bra{\alpha}$, i.e, a coherent state in the basis of eigenstates of $H_{\text{S}}(0)$. In contrast to Sec.~\ref{sec:qubit}, this protocol can transfer the system from $\rho_{0}$ to an infinite set of coherent states in the basis of eigenstates of $H(t_{\text{f}})$, one for each choice of the reverse-engineered control.}
\label{fig:newfig}
\end{figure}
As a second example of our reverse-engineering approach, we study the instance of a driven harmonic oscillator,
sketched in Fig.~\ref{fig:newfig}, ruled by the Hamiltonian
\beq\label{eq:Hamoscillator}
H_{\text S}(t)= \frac{p^2}{2m} + \frac{m\omega^2(t)}{2}q^2.
\eeq
Although the modulation of the oscillator frequency can take arbitrary shapes, we assume the initial and final values to be held fixed,~i.e.,~$\omega(0)=\omega_0$, $\omega(t_{\text f})=\omega_{\text f}$.~The Hamiltonian operator in Eq.~\eqref{eq:Hamoscillator} is written in terms of the operator basis $\{q^2/2,\, p^2/2,\, (qp+pq)/2\}$, which is closed under the action of the commutator between any pair of its operators. Indeed, it is possible to show that the operator basis $\{H_{\text S}(t),\,p^2/2 - \omega^2(t)q^2/2,\,\omega(t)(qp+pq)/2\}$ spans a Lie algebra \cite{rezek2006irreversible,uzdin2013effects}, whose structure constants are similar to the ones of the non-compact Lie group $\text{SU}(1,1)$ \cite{bargmann47,novaes2004some, gilmore_2008}. However, for our purpose, we adopt as a basis of generators   \cite{chen2011lewis,torrontegui2011fast,levy2018noise} the former one, i.e., $T_{1}=p^2/2$, $T_{2}=q^2/2$, $T_{3}=(qp + pq)/2$
%and $ \comm{T_a}{T_b}=i\sum_{c=1}^3 f_{abc}T_c$, where the matrix $ f_{abc} $ is determined by the following commutator
which satisfies the commutation relations
$\comm{T_{1}}{T_{2}}=-iT_3,\comm{T_{1}}{T_{3}}=-2iT_1$, and $\comm{T_{2}}{T_{3}}=+2i T_2$.

By expressing $(H_{\text S}(t),I(t))$ in terms of the previous basis, and inserting them into Eq.~\eqref{eq:invariant1}, we can achieve the differential equations in Eq.~\eqref{eq:diffEq} linking the control fields of the invariant to the time-dependent frequency $\omega(t)$  (see App.~\ref{sec:appA} for details). The final form of the operator $I(t)$ can thus be derived and reads
\beq\label{eq:geneoscomm}
I(t)= \frac{\mathcal{P}^2_{t}}{2m} + \frac{1}{2}m\omega_0^2 \mathcal{Q}^2_{t},
\eeq
where we introduced a couple of time-dependent operators $\mathcal{Q}_{t}=q/\zeta(t),\mathcal{P}_{t}=\zeta(t)p -m\dot{\zeta}(t)q$, and $\zeta(t)$ is a solution of the Ermakov equation \cite{lewis69}
\beq\label{eq:Ermakov}
\ddot{\zeta} + \omega^2(t)\zeta-\frac{\omega^2_0}{\zeta^3}=0.
\eeq
Notice that the invariant operator can be computed once the Ermakov parameter $\zeta(t)$ is known. Moreover, the operator $I(t)$ is written as a quadratic operator, and the operators $(\mathcal{Q}_{t},\mathcal{P}_{t})$ preserve the canonical commutation relations,~i.e.~$\comm{\mathcal{Q}_{t}}{\mathcal{P}_{t}}=i$. It follows that it can be diagonalized analytically at any point in time by introducing a pair of instantaneous creation and annihilation operators $(a^{\dagger}_t,a_t)$. The ground state of the instantaneous eigenbasis $\ket{\phi_{0}(t)}$ can be defined such that $a_t\ket{\phi_0(t)}=0$. Once the expansion of $\ket{\phi_0(t)}$ is known in a time-independent basis, e.g., the Fock basis $\ket{n}$ of $H_{\text S}(0)$, then each element of the infinite set of eigenstates can be defined in the usual way as $\ket{\phi_{n}(t)}=(a^{\dagger}_t)^n/\sqrt{n!}\ket{\phi_0(t)}$ (see App.~\ref{sec:appA}).

 The NAME equation can be devised in an analogous way to Sec.~\ref{sec:qubit}. In this case, we consider the driven harmonic oscillator to be linearly coupled to a harmonic bath by means of the position operator $q$, i.e., 
\beq\label{eq:Hamiltotosc}
H_{\text{B}} = \sum_{k}\frac{P^2_{k}}{2 m_{k}} +\frac{1}{2}m_{k}\omega^{2}_{k}Q^{2}_{k} \;\; {\rm and} \;\; H_{ \text {I}}=q\sum_{k}c_{k}Q_{k}.
\eeq
We assume the bath to be described by the spectral function 
\beq\label{eq:spectralosc}
J(\omega)=\frac{\pi}{2}\sum_{k} \frac{c^2_{k}}{m_{k}\omega_{k}}\delta(\omega -\omega_{k})=2\pi m\gamma_{s}\omega e^{-\omega/\omega_{c}},
\eeq
where $\gamma_{s}$ has dimensions of a frequency. Notice that Eq.~\eqref{eq:Hamiltotosc} shares some similarities with the  Caldeira-Leggett model \cite{caldeira1983,GRABERT1988115,weissbook}. However, as we work in the weak-coupling limit and we want to reduce the dynamics to a QME of the form written in Eq.~\eqref{eq:BMSchro}, we neglect the renormalization of the driving frequency $\omega(t)$ arising from the system-bath coupling as well as all the memory effects and system-bath correlations.   

In order to derive the QME, it is enough to consider the dynamics of a vector of operators $\vec{v}(t)=(\hat{q}(t),\hat{p}(t))$, where $\hat{q}(t)=\sqrt{m\omega(t)}q,\hat{p}(t)= p/(\sqrt{m\omega(t)})$. We denote $l(t)=\sqrt{4-\mu^2(t)}$ and assume it to be real. Then, the LGKS dissipator in the inertial approximation can be found as (see App.~\ref{sec:appCosc})
\begin{multline}\label{eq:Dissosc}
\mathcal{D}_{t}=\kappa_{+}(t)\qty(F_{t,+}\circ F_{t,-} -\frac{1}{2}\{F_{t,-}F_{t,+},\circ\}) +\\ + \kappa_{-}(t)\qty(F_{t,-}\circ F_{t,+} -\frac{1}{2}\{F_{t,+}F_{t,-},\circ\}),
\end{multline}
where $F_{t,\pm}$ are time-dependent LGKS operators that read
%\begin{align}\label{eq:jumposc}
%F_{t,\pm}&=\pm \qty(\frac{i}{l(t)\sqrt{\omega(0)}})u_{\mp}(t),\nonumber \\
%u_{\pm}(t)&=\frac{1}{2}(\mu(t)\pm i l(t))\hat{q}(0) +\hat{p}(0). 
%\end{align}
\begin{align}\label{eq:jumposc}
F_{t,\pm}&=\pm \frac{i}{l(t)}u_{\mp}(t),\nonumber \\
u_{\pm}(t)&=\frac{1}{2}(\mu(t)\pm i l(t))\hat{q}(0) +\hat{p}(0). 
\end{align}
Assuming the inertial approximation to hold, the operators $u_{\pm}(t)$ can be derived as the instantaneous left-eigenoperators of the time propagator in the Liouville-Hilbert space  (see App.~\ref{sec:appCosc}), while the time-dependent rates 
read $\kappa_{\pm}(t)=q^2_{0}(\omega_{0}/\omega(t))\gamma(\pm \alpha(t))$, $\alpha(t)=\dot{\theta}^{\text{in}
}_{-}(t)=(l(t)/2)\omega(t)$, where $q_{0}=\sqrt{1/(m\omega_{0})}$ and $\gamma(\omega)$ is computed from the equilibrium bath correlation function (see App.~\ref{sec:appCosc}).

\subsubsection{The protocol}

Without loss of generality we consider as example a compression where the driving frequency of the oscillator changes from $\omega(0)=\omega_0$ to $\omega(t_{\text f})=\omega_{\text f}$ such $\omega_{\text f}\geq\omega_{\text 0}$. We keep on imposing the frictionless conditions as in Sec.~\ref{sec:qubit} at $t=0,t_{\text f}$, so that, at these times, the invariant eigenstates coincide with the Fock states of the oscillator with frequencies $\omega_{0}$ and $\omega_{\text f}$ respectively.~As a consequence, invariant-based protocols can be devised to transfer the state $\rho_{\text{S}}(t)$ between a couple of Fock states of different frequencies. 
As in Sec.~\ref{sec:qubit}, Eq.~\eqref{eq:invariant1} can be combined with the frictionless conditions to find the values of Ermakov parameter and its derivatives at the boundaries (see App.~\ref{sec:appA}),  
\begin{align}\label{eq:Ermakovcond}
\zeta(0)&=1,\; \zeta(t_{\text{f}})=\sqrt{\omega_{0}/\omega_{\text{f}}},\nonumber\\
\dot{\zeta}(0)&=\dot{\zeta}(t_{\text f})=\ddot{\zeta}(0)= \ddot{\zeta}(t_{\text f})=0.
\end{align}
~Any function $\zeta(t)$ allowed by Eq.~\eqref{eq:Ermakov} and fulfilling the conditions written in Eq.~\eqref{eq:Ermakovcond} can be used to directly compute the driving frequency $\omega(t)$ linked to a STA protocol such that $\omega(0)=\omega_{0}$ and $\omega(t_{\text{f}})=\omega_{\text{f}}$.

Below, rather than looking at single eigenstates, we study the transfer between a couple of Gaussian states, which have been of interest in the field of quantum information protocols \cite{weedbrook2012gaussian}. We focus on one of the simplest instances, namely, at $t=0$ we assume our system to be prepared in a coherent state $\ket{\alpha}=\sum_{n}(\alpha^{n}/\sqrt{n!})e^{-\abs{\alpha}^2/2}\ket{n}$, i.e., $\rho_{\text{S}}(0)=\ket{\alpha}\bra{\alpha}$.~From Eqs.~\eqref{eq:pop} and ~\eqref{eq:coher}, it follows that the dynamics preserves the initial state populations and  develops time-dependent relative phases $\varphi_n(t_{\text f})-\varphi_n(0)=-(n+1/2)\omega_{0}\int_0^{t_{\text f}}\mathrm{d}t^{\prime}/\zeta^2(t^{\prime})$,~i.e., the target state reads $\rho_{\text{tar}}=\ket{\tilde{\alpha}}\bra{\tilde{\alpha}}$ with $\tilde{\alpha}=\alpha e^{-i\omega_{0} \int_0^{t_{\text f}}\mathrm{d}t^{\prime}/\zeta^2(t^{\prime})}$.~It should be noticed that, contrary to the previous example, the target state now changes with the Ermakov parameter $\zeta(t)$. 
\subsubsection{Results of the reverse engineering}

We choose to parametrize $\zeta(t)$ as follows 
\beq\label{eq:parametri2}
\zeta(t)=\qty[\sum_{n=0}^{n_{\text max}}g_{n} \qty(\frac{t}{t_{\text f}})^n]^{-1/2}.
\eeq
In order to obey Eqs.~\eqref{eq:Ermakovcond}, in the absence of interactions with the environment, it is enough to set $n_{\text max}=5$. Moreover, to find optimal invariants under the influence of the environment, we include additional parameters,~i.e, $(g_6,g_7)$, and select the STA protocol for the closed system that minimizes Eq.~\eqref{eq:functional3}. Indeed, for a pure target state that is a linear combination of invariant eigenstates at $t=t_{\text{f}}$, after setting $t_{0}=0$, Eq.~\eqref{eq:functional3} reduces to
\begin{multline}\label{eq:functional3osc}
\mathcal{Z}=\sum_{nmjk}\bra{\psi_{\text{tar}}}\ket{\phi_{n}(t_{\text{f}})} \bra{\phi_{m}(t_{\text{f}})}\ket{\psi_{\text{tar}}}\cdot\\\cdot
\int_{0}^{t_{\text{f}}}\mathrm{d}s e^{i(\varphi_{n}(t_{\text{f}})-\varphi_{n}(s))-i(\varphi_{m}(t_{\text{f}})-\varphi_{m}(s))}R^{(nm)}_{jk}(s)\rho^{0}_{\text{S} jk}(s).
\end{multline}
From Eq.~\eqref{eq:functional3osc}, it is evident that populations as well as coherences of the $\rho_{\text S}(t)$ in the invariant eigenbasis contribute to $\mathcal{Z}$,~i.e., the latter depends on the Ermakov parameter via the elements of dissipator $R^{(nm)}_{jk}(s)$ as well as the dynamical phases $\varphi_{n}(t)$. Equation~\eqref{eq:functional3osc} thus provides quite an easy way to take into account the effects of the derivative of the state in Eq.~\eqref{eq:rates}, for a sufficiently weak dissipation strength. Thanks to the special features of Eq.~\eqref{eq:Dissip}, the analytical expression of $R^{(nm)}_{jk}(s)$ can be easily derived by writing the LGKS operators $F_{l,t}$ in terms of the operators $(a_t,a^{\dagger}_t)$ at each point in time. 

\begin{figure}
\includegraphics[scale=0.9]{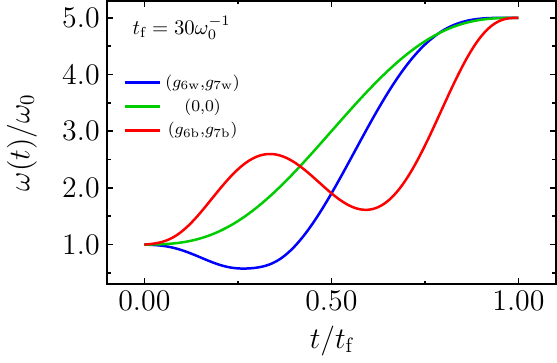}
\caption{Plot of the reverse-engineered protocols for the oscillator frequency $\omega(t)$. We set $\omega_{\text{f}}=5\omega_{0}$, $t_{\text{f}}=30\omega^{-1}_{0}$ and choose a parametrization with $n_{\text{max}}=7$. The green curve corresponds to the minimal choice of parameters for the closed system,~i.e.,~$n_{\text{max}}=5$. The red and blue curves are computed by choosing the parameters $(g_{6},g_{7})$ along the line $g_{7}=-(2/7)g_{6} +g^{*}$, with $g^*=20$. For the best protocol along this line, $g_{6\text{b}}=-2857$ (red curve), and for the worst, $g_{6\text{w}}=928$ (blue curve). For each choice of the parameter $g^*$, the target state at time $t=t_{\text{f}}$ remains fixed.}
\label{fig:drivingosc}
\end{figure}

\begin{figure}
\includegraphics[scale=0.90]{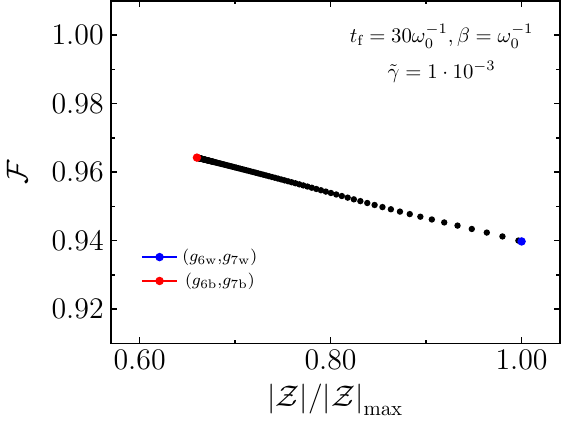}
\caption{Fidelity of a set of reverse-engineered protocols corresponding to different values of $(g_{6},g_{7})$, plotted against the normalized $\abs{\mathcal{Z}}$. The colored dots denote the protocols sketched in Fig.~\ref{fig:drivingosc}. Here we set $g^{*}=20,\gamma_{s}=10^{-3} \omega_{0}, \beta=1.0\omega^{-1}_{0}$. The coherent state parameter $\alpha$ is chosen such that $\Re\alpha =\Im\alpha=0.30$.}
\label{fig:fidelityosc}
\end{figure}

In Fig.~\ref{fig:drivingosc}, we show several instances of the reverse-engineered fields as a function of time, for fixed values of protocol duration and frequency ratio $\omega_0/\omega_{\text f}$, obtained by setting $n_{\text{max}}=7$. Furthermore, in Fig.~\ref{fig:fidelityosc}, we plot the fidelity against the normalized $\abs{\mathcal{Z}}$ for different protocols set by choosing $(g_{6},g_{7})$ to span a given line of the parameters' space, which corresponds to a specific target state $\rho_{\text{tar}}$. Using Eq.~\eqref{eq:parametri2}, these lines can be found analytically.  

\begin{figure*}[t!]
\centering
\includegraphics[scale=0.40]{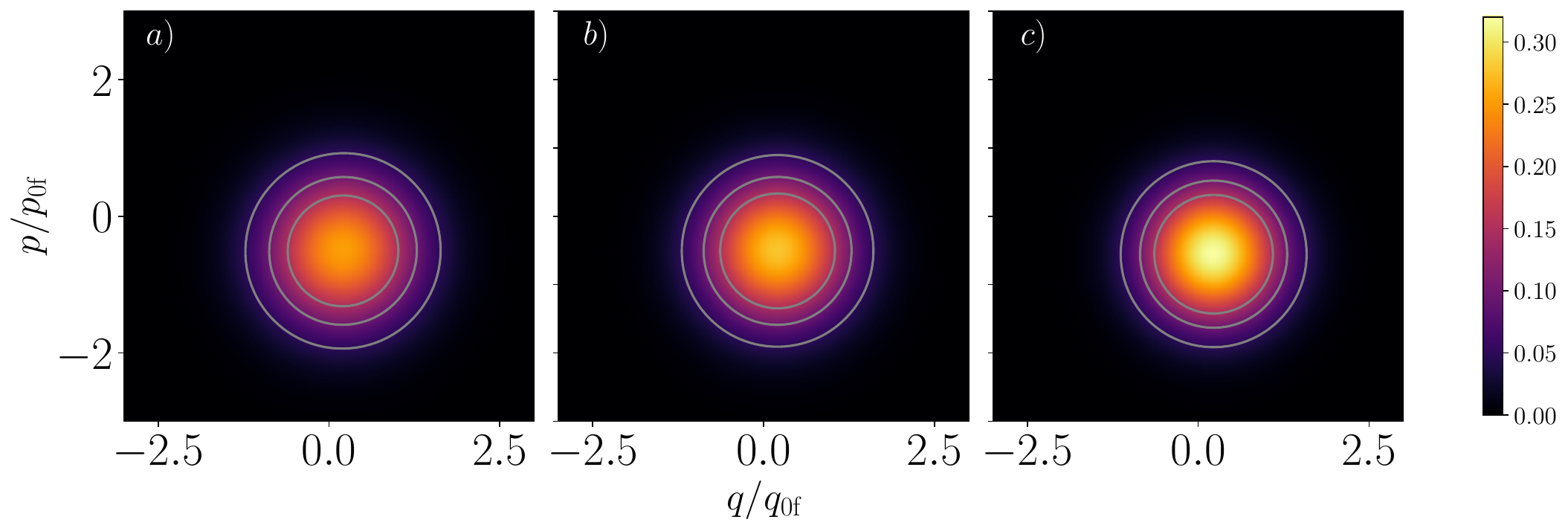}
\caption{Contour plot of the Wigner quasi-probability distribution, Eq.~\eqref{eq:wigner} for the reduced system state $\rho(t_{\text{f}})$ that corresponds to the blue and red protocols in Figs.~\ref{fig:drivingosc} and~\ref{fig:fidelityosc}   are shown in panels a) and b) respectively, whereas the target state $\rho_{\text{tar}}$ is shown in panel c).  Here $(q,p)$ are reported in units of the final frequency $\omega_{\text{f}}$, $\Re \alpha =\Im \alpha=0.30$, $\beta=1.0\omega^{-1}_{0}$, $\gamma_s=10^{-3} \omega_{0}$ .}
\label{fig:Wigner}
\end{figure*}

It is evident from Fig.~\ref{fig:fidelityosc} that, for fixed values of dissipation strength and inverse temperature, the actual fidelity decreases monotonically as a function of $\abs{\mathcal{Z}}$. Moreover, in the chosen parameters region, the field optimizing $\abs{\mathcal{Z}}$ exhibits a non-monotonic behavior as a function of time, i.e., the oscillator frequency is first increased, then reduced before reaching the target value at $t=t_{\text{f}}$. 
Adding degrees of freedom in the invariant for minimizing Eq.~\eqref{eq:functional3} is thus crucial for the success of finding the least noise-sensitive protocol.
%Moreover, \eras{ as shown in Sec.~\ref{sec:qubit}} \textcolor{red}{similar to Sec.~\ref{sec:qubit}, it can be shown that} the optimal invariant under the action of dissipation is different from the one corresponding to the minimal choice of parameters,~i.e., $(g_{6},g_{7})=(0,0)$\textcolor{red}{(green curve in Fig.~\ref{fig:drivingosc})}. As a consequence, the addition of the constraint to minimize Eq.~\eqref{eq:functional3} also provides an improvement in the fidelity w.r.t. the minimal choice of parameters. 

%Under the assumption of regularity of the functions involved, the position of local maxima of the fidelity can be found within a good degree of approximation by looking at the stationary points of \eqref{eq:functional3},i.e., 
%\beq
%\nabla_{g_m}\mathcal{\bar{R}}^{(k)}=0.
%\eeq
%\sout{By adding more parameters to Eq.~\eqref{eq:parametri2}, we can also expect that the effectiveness of the method does not change, as the fidelity is maximized in the regions where Eq.~\eqref{eq:functional3osc} is minimized.}

The advantages of the optimal fields can be further understood by comparing the representation of the target state with the actual state at $t=t_{\text{f}}$. We thus employ the phase space representation of the oscillator's state, namely, the Wigner-Weyl quasi-probability distribution \cite{HILLERY1984121,LEE1995147},
\beq\label{eq:wigner}
W(q,p)=\frac{1}{\pi}\int_{-\infty}^{+\infty}\mathrm{d} y e^{-2 i y p}\bra{q-y}\rho\ket{q+y},
\eeq
where $(q,p)$ denotes a point in the oscillator phase space. 

In Fig.~\ref{fig:Wigner}, we show the contour plots of $W(q,p)$ for the actual state $\rho_{\text{S}}(t_{\text{f}})$ ( panels a and b), under two different reverse-engineered protocols corresponding to maximum (red point) and minimum (blue point) fidelity, respectively~\footnote{We emphasize that all these protocols would produce fidelity one in case the dynamics is assumed closed. Maximum and minimum refer to optimization with respect to the parameter $g_6,g_{7}$ in Eq.~(\ref{eq:parametri2}). Better or worse results can be obtained by considering higher-order expansion. }, of Fig.~\ref{fig:fidelityosc}, and compare them with the corresponding target state $\rho_{\text{tar}}$ (panel  c). It follows that the peak of the quasi-probability differs noticeably from the target one, the actual difference depending on the value of the inverse temperature $\beta$ as compared with $\omega_{0}$. However, the driving fields corresponding to the optimal invariant reduce the difference in the peak, as well as the asymmetry in the whole profile of $W(q,p)$. Furthermore, the peak of the distribution tends to coincide with the centroid of the target distribution for the maximum fidelity invariant.   

\section{Conclusions}\label{sec:Conc}
In our work, we address the control of prototypical examples of open quantum systems, where the action of noise and dissipation explicitly depends on the driving fields. We propose to reverse-engineer optimal STA-like protocols that can minimize losses due to the detrimental influence of dissipation and decoherence on the driven system. Our approach holds valid for weak system-environment coupling and works in the Markovian limit. Compared to conventional QOC approaches, it does not require the full propagation of the dynamical equation and it explicitly takes into account the instance of time-dependent dissipative effects. Due to its similarity to DFS theory, it can, in principle, be formulated for arbitrarily driven systems and it can be also extended to different kinds of Markovian QMEs. Challenging generalizations of the work might involve coupled physical systems, e.g. models of driven qubit-resonators devices, as well as engineered quantum many-body systems that can be driven by tuning a minimal set of Hamiltonian parameters. A further challenge would be to tackle the control of driven quantum systems beyond the inertial approximation, possibly evolving under non-Markovian dissipative dynamics, where the present approach is not expected to be working anymore.      

\section{Acknoledgements}
We acknowledge financial support from the Israel Science
Foundation (Grant No. 1364/21), the Spanish Government via the projects PID2021-126694NA-C22 (MCIU/AEI/FEDER, EU) and TSI-069100-2023-8 (Perte Chip-NextGenerationEU). H. E. acknowledges the Spanish
Ministry of Science, Innovation and Universities for funding through the FPU program (FPU20/03409). E. T. acknowledges the Ram{\'o}n y Cajal (RYC2020-030060-I) research fellowship. 

\appendix

\section{Reverse-engineering of the invariant fields}\label{sec:appA}

The definition of the invariant operator in Eq.~\eqref{eq:invariant1}, along with the frictionless conditions, allows to obtain a system of differential equations linking the Hamiltonian driving fields $h_{i}(t)$ to the invariant ones $d_{i}(t)$.~After imposing a number of constraints, this system can be inverted. Therefore, by making an arbitrary choice of the invariant fields, an infinite set of driving Hamiltonians can be devised such that the system dynamics preserve the populations of the state in the invariant eigenbasis. Below, we discuss in detail the derivation of equations used in the main text, Sec. \ref{sec:apps}, e.g., Eqs.~\eqref{eq:diffeqQUBIT} and ~\eqref{eq:Ermakov} and we also set out additional properties of the invariant eigenstates.   

\subsection{The driven TLS}

Here we assume the system to be isolated from its environment and to evolve under the Hamiltonian $H_{\text{S}}(t)$. Furthermore, $H_{\text{S}}(t)$ and the invariant $I(t)$  belong to the Lie Algebra linked to the $\text{su}(2)$ group,~i.e.,
\beq\label{eq:HIsu2}
H_{\text{S}}(t)=\sum_{i=1}^{3}h_{i}(t)T_{i},\quad I(t)=\sum_{i=1}^{3}d_{i}(t)T_{i}, 
\eeq
with $T_{1}=\sigma_x/2$, $T_{2}=\sigma_y/2$, $T_{3}=\sigma_z/2,\mbox{and } \comm{T_{i}}{T_{j}}=i\varepsilon_{ijk}T_{k}$.~The main equation of the dynamical invariant in the Scr\"{o}dinger picture sets the basic link between the two operators and reads ($\hbar=1$)  
\beq\label{eq:SchrodInv}
    i\comm{H_{\text{S}}(t)}{I(t)} + \frac{\partial}{\partial t}I(t)=0.
\eeq
To perform the reverse-engineering procedure as set forth in Sec.~\ref{sec:STA}, we also impose the frictionless conditions,~i.e., we require the two operators to commute at the initial and final time
\beq\label{eq:frictionless}
    \comm{H_{\text{S}}(t_{a})}{I(t_{a})}=0 \mbox{, with } t_{a}=0,t_{\text{f}}. 
\eeq
Inserting Eq.~\eqref{eq:HIsu2} into Eq.~\eqref{eq:SchrodInv} it follows
\beq\label{eq:linearsys1}
 \dot{d}_{k}(t)=\sum_{ln}\varepsilon_{lnk}h_{l}(t)d_{n}(t).  
\eeq
Equation~\eqref{eq:linearsys1} is a system of differential equations as written in Eq.~\eqref{eq:diffEq}. It allows us to find the auxiliary fields of the invariant $d_{k}(t)$ once the Hamiltonian is known. However, for each time $t$ the system in Eq.~\eqref{eq:linearsys1} can be recast as an algebraic linear system linking the derivatives to the Hamiltonian fields $h_{l}(t)$.~After dropping the dependence on time $t$ and vectorizing   the functions $\dot{{\bf d}}=(\dot{d}_{1},\dot{d}_{2},\dot{d}_{3})^{\text{T}}$, ${\bf h}=(h_{1},h_{2},h_{3})^{\text{T}}$, Eq.~\eqref{eq:linearsys1} reads
\beq\label{eq:linearsys2}
\bf\dot{d}=\mathcal{A}({\bf d}){\bf h}, \mbox{ with }
\mathcal{A}=
\begin{pmatrix}
0 & d_{3} & -d_{2}\\
-d_{3} & 0 &d_{1} \\
d_{2}& -d_{1} &0
\end{pmatrix}.
\eeq
The algebraic system in Eq.~\eqref{eq:linearsys2} is not invertible. However, from Eq.~\eqref{eq:linearsys1} it follows that 
\beq\label{eq:degk}
\sum_{k=1}^{3}d_{k}
\dot{d}_{k}=0,
\eeq
i.e.,~$\sum_{k=1}^{3}d^2_{k}=c$, where $c$ is a real and nonnegative constant. If we allow $c > 0$, from Eq.~\eqref{eq:linearsys1}, using the contraction of the Levi-Civita tensor $\sum_{k}\varepsilon_{i j k}\varepsilon_{l m k}=\delta_{il}\delta_{jm}-\delta_{im}\delta_{jl}$, we can find at least one index $k$ such that Eq.~\eqref{eq:linearsys2} allows us to write
\beq\label{eq:linearsys3}
h_{i}=-\sum_{j}\varepsilon_{ijk}\frac{\dot{d}_{j}}{d_{k}} + h_{k}\frac{d_{i}}{d_{k}}.   
\eeq
Furthermore, Eq.~\eqref{eq:frictionless} translates to additional conditions on the functions for $t_{a}=0,t_{\text{f}}$  that read 
\beq\label{eq:linearsys4}
h_{l}(t_{a})d_{m}(t_{a})-h_{m}(t_{a})d_{l}(t_{a})=0.  
\eeq
Equations ~\eqref{eq:linearsys3} and ~\eqref{eq:linearsys4} allow us to find an explicit form linking the driving fields of the Hamiltonian to the invariant ones. Indeed, from Eq.~\eqref{eq:Hqubit}, we can restrict to the case where $h_{2}(t)=0$ at any point in time.~Furthermore, in order to achieve the protocol described in Sec.~\ref{sec:qubit}, we also impose  
\beq\label{eq:boundary}
h_{3}(0)=\Delta_{0},\ h_{3}(t_{\text{f}})=-\Delta_{0},\ h_{1}(t_{a})=0.
\eeq
Thus, from Eqs.~\eqref{eq:linearsys4} and \eqref{eq:boundary} it readily follows $d_{1}(t_{a})=d_{2}(t_{a})=0$, so that from Eq.~\eqref{eq:degk} $d^2_{3}(t_{a})=c$, $\dot{d}_{3}(t_{a})=0$. Moreover, from Eq.~\eqref{eq:linearsys3}, setting $k=3$, $i=1,2$ we also find that $\dot{d}_{1}(t_{a})=\dot{d}_{2}(t_{a})=0$. Eventually, combining Eq.~\eqref{eq:degk} with Eq.~\eqref{eq:linearsys3} we achieve
\beq\label{eq:finalderieq1}
h_{1}=\frac{\dot{d}_{3}}{d_{2}},\; h_{3}=-\frac{\dot{d}_{1}}{d_{2}}, 
\eeq
We can further make Eq.~\eqref{eq:finalderieq1} explicit by setting $c=1$, i.e., the vector ${\bf d}$ can be interpreted as a vector on the su(2) unit sphere. Then, we can put $d_{1}(t)=\sin G(t)\cos B(t)$, $d_{2}(t)=\sin G(t)\sin B(t)$, $d_{3}(t)=\cos G(t)$, where the angles $(G(t),B(t))$ now parametrize the invariant operator at any point in time $t$. From Eq.~\eqref{eq:finalderieq1}, we thus find 
\beq\label{eq:diffeqQUBITapp}
\Delta(t)=\dot{B}(t) -\frac{\dot{G}(t)}{\tan G(t) \tan B(t) },\  \epsilon(t)=-\frac{\dot{G}(t)}{\sin B(t)}.
\eeq

\begin{figure}[h!]
\centering
\includegraphics[scale=0.9]{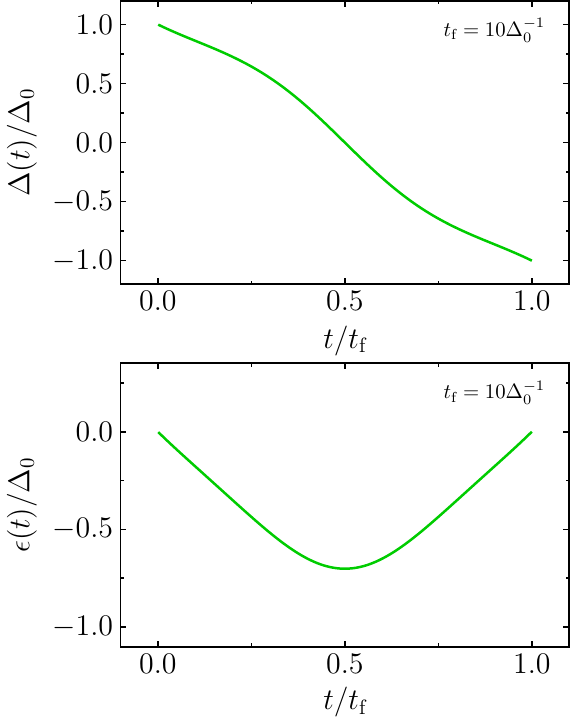}
\caption{Plot of the reverse-engineered driving fields $(\Delta(s),\epsilon(s))\mbox{ , } s=t/t_{\text{f}}$ for the closed system dynamics and $t_{\text{f}}=10\Delta^{-1}_{0}$, adopting the minimal number of parameters, as ruled by Eq.~\eqref{eq:invfieldslast}.}
\label{fig:fieldsapp}
\end{figure}

Equations~\eqref{eq:diffeqQUBITapp} easily link the invariant auxiliary fields to the Hamiltonian drivings. They can be used to reverse-engineer the Hamiltonian fields $(\Delta(t),\epsilon(t))$ by introducing a parametrization of the functions $(G(t),B(t))$. Although several choices of parametrization are feasible, as in Eq.~\eqref{eq:paramet}, main text, we adopt a polynomial shape of the auxiliary fields as follows  
\beq\label{eq:paramet2}
G(t)=\sum_{n=0}^{n_{g}}g_{n} \qty(\frac{t}{t_{\text{f}}})^n,\ B(t)=\sum_{n=0}^{n_{b}}b_{n} \qty(\frac{t}{t_{\text{f}}})^n.
\eeq
In principle, an arbitrary number of parameters can be used to devise invariant operators of the closed system leading to population-preserving driving protocols. However, they can be fixed by requiring the protocol to obey a given number of constraints. In the instance of the closed system dynamics, we can easily derive the minimum number of parameters required for the protocol to obey boundary conditions as well as Eqs.~\eqref{eq:frictionless}. Indeed, from Eqs.~\eqref{eq:boundary} and the derivatives at the boundary, we find that Eqs.~\eqref{eq:diffeqQUBITapp} are well-posed at any time $t\in [0,t_{\text{f}}]$ provided that we choose the fields to fulfill the following 
\begin{align}\label{eq:condGB}
G(0)&=\pi,\ G(t_{\text{f}})=0,\ \dot{G}(0)=\dot{G}(t_{\text{f}})=0,\nonumber  \\
B(0)&=B(t_{\text{f}})=-\frac{\pi}{2},\ \dot{B}(t_{a})\neq 0.
\end{align}  
Combining Eqs.~\eqref{eq:diffeqQUBITapp},~\eqref{eq:paramet2}, and~\eqref{eq:condGB}, we find the minimum number of parameter to be $n_{g}=3,n_{b}=2$, so that the final form of the invariant fields reads    
\begin{align}\label{eq:invfieldslast}
G(s)&=\pi -3\pi s^2 + 2\pi s^3 ,\nonumber \\ 
B(s)&=-\frac{\pi}{2} + \frac{\Delta_{0} t_{\text{f}}}{3}s -\frac{\Delta_{0} t_{\text{f}}}{3}s^2 .
\end{align} 
with $s=t/t_{\text{f}}$ and thus $s\in [0,1]$. In Fig.~\ref{fig:fieldsapp}, we plot the Hamiltonian drivings $(\Delta(t),\epsilon(t))$ as obtained from Eq.~\eqref{eq:invfieldslast}.   
Starting from Eq.~\eqref{eq:paramet2},  additional parameters can be employed to define infinite invariant operators obeying the previous conditions. In this enlarged parameter space, we can choose the optimal invariant according to the procedure described in Sec.~\ref{sec:STA}, main text.  
\subsection{The driven harmonic oscillator}
In the instance of the driven harmonic oscillator, as anticipated in the main text the Hamiltonian operator can be written in terms of su$(1,1)$-like generators,i.e., unbounded operators of the harmonic oscillator that read
\beq\label{eq:geneosc}
T_{1}=\frac{p^2}{2},\ T_{2}=\frac{q^2}{2},\ T_{3}=\frac{qp + pq}{2}. 
\eeq
From the canonical commutation relations $[q,p]=i$, it is easy to find the commutation relations, $\comm{T_a}{T_b}=i\sum_{c=1}^3 f_{abc}T_c$, with the structure factors $f_{abc}$ that fulfill $f_{123}=-1$, $f_{131}=-2$, $f_{232}=2$. Following Eqs.~\eqref{eq:Hamiltotosc} and \eqref{eq:HIsu2}, the Hamiltonian fields read $h_{1}(t)=1/m,\ h_{2}(t)=m\omega^2(t),\ h_{3}(t)=0$. By working out Eq.~\eqref{eq:SchrodInv} in analogous way to the su(2) instance, we find a system of differential equations that read
\beq\label{eq:linearsysosc}
{\bf\dot{d}}=\mathcal{A}({\bf d}){\bf h}, \mbox{ with }
\mathcal{A}=
\begin{pmatrix}
-2d_{3} & 0 & 2d_{1}\\
0 & 2 d_{3} &-2d_{2} \\
-d_{2}& d_{1} &0
\end{pmatrix}.
\eeq
From Eqs.~\eqref{eq:linearsysosc}, it follows that the fields and their first derivatives fulfill the following equality
\beq\label{eq:degosc}
2d_{3}\dot{d}_{3} -d_{1}\dot{d}_{2} -d_{2}\dot{d}_{1}=0.
\eeq
As in the previous case, Eq.~\eqref{eq:degosc} can be used to express the Hamiltonian fields in terms of the auxiliary ones. Since we focus on the instance of a single driving field,~i.e., $h_{2}(t)$, the task can be accomplished easily by writing down a differential equation for a single auxiliary field depending on $h_{2}$.~In the following, we choose to derive the equation for the field $d_{1}(t)$. From \eqref{eq:degosc}, there exist a constant $c$ such that $d_{2}=(1/d_{1})(d^2_{3} + c)$. Combining this with  Eqs.~\eqref{eq:linearsysosc}, from the form of the Hamiltonian a nonlinear differential equation can be derived for $d_{1}(t)$   
\begin{align}\label{eq:inverosc}
\ddot{d}_{1} +2h_{1} h_{2} d_{1} - \frac{\dot{d}^2_{1}}{2d_{1}} -2 c \frac{h^2_{1}}{d_{1}} =0. 
\end{align}  
We also need to impose frictionless conditions at the boundaries as in Eq.~\eqref{eq:frictionless}. It follows that 
\beq\label{eq:frictosc}
\sum_{nm}f_{nma} h_{n}(t_{a})d_{m}(t_{a})=0,\ \mbox{any }a.
\eeq
Equation~\eqref{eq:frictosc}, along with Eqs.~\eqref{eq:linearsysosc} and \eqref{eq:inverosc}, imply the following set of conditions at the boundaries
\begin{align}\label{eq:oscboundary}
d_{3}(t_{a})&=0,\; d_{1}(t_{a})=\sqrt{c\frac{ h_{1}(t_{a})}{h_{2}(t_{a})}},\nonumber\\ 
\dot{d}_{1}(t_{a})&=\dot{d}_{2}(t_{a})=\dot{d}_{3}(t_{a})=0.
\end{align}
To recast Eq.~\eqref{eq:inverosc} in a more convenient form, we choose $c= \omega^2_{0} $, and we set $d_{1}(t)=\zeta^2(t)/m$, where $\zeta(t)$, i.e., the Ermakov parameter, is a positive function of time. From the structure of the Hamiltonian Eq.~\eqref{eq:inverosc} can be written as
\beq\label{eq:Ermakovapp}
\ddot{\zeta} + \omega^2(t)\zeta-\frac{\omega^2_0}{\zeta^3}=0.
\eeq
Equation~\eqref{eq:Ermakovapp} is the Ermakov equation used in the main text. From Eq.~\eqref{eq:oscboundary}, the boundary conditions for $\zeta(t)$ follow as $\zeta(0)=1,\ \zeta(t_{\text{f}})=\sqrt{\omega_{0}/\omega_{\text{f}}},\ \dot{\zeta}(0)=\dot{\zeta}(t_{\text{f}})= \ddot{\zeta}(0)=\ddot{\zeta}(t_{\text{f}})=0$. 
We can also write down the expression of the invariant operator in terms of $\zeta(t)$. Indeed, from Eqs.~\eqref{eq:linearsysosc} we find $d_{2}(t)=m(\dot{\zeta}^2(t) + \omega^2_{0}/\zeta^2(t))$, $d_{3}(t)=-\dot{\zeta}(t)\zeta(t)$. As a result, from Eqs.~\eqref{eq:HIsu2} and \eqref{eq:geneosc}, we can write the invariant operator as follows 
\beq\label{eq:Invaroscapp}
I(t)=\frac{\mathcal{P}^2_{t}}{2m} + \frac{1}{2}m\omega^2_{0}\mathcal{Q}^2_{t},
\eeq
where $(\mathcal{Q}_{t},\mathcal{P}_{t})$ are a couple of time-dependent operators that read 
\beq\label{eq:QPtimedep}
\mathcal{Q}_{t}=\frac{q}{\zeta(t)},\quad\mathcal{P}_{t}=\zeta(t)p -m\dot{\zeta}(t)q. 
\eeq
Notice that the time dependence only results from the coefficients of these operators,~i.e., from the Ermakov parameter $\zeta(t)$.~Moreover, the canonical commutation relations are preserved,~i.e., $\comm{\mathcal{Q}_{t}}{\mathcal{P}_{t}}=i$. This property allows the definition of instantaneous ladder operators that diagonalize Eq.~\eqref{eq:Invaroscapp} at any time $t$. It reads
\beq\label{eq:creant}
a_{t}=\sqrt{\frac{m \omega_{0}}{2}} \mathcal{Q}_{t} +\frac{i}{\sqrt{2 m\omega_{0}}}\mathcal{P}_{t}.
\eeq
The operator can thus be diagonalized as $I(t)=\omega_{0}(a^{\dagger}_{t}a_{t} + 1/2)$, i.e., its eigenvalues coincide with the eigenvalues of the undriven oscillator of frequency $\omega_{0}$ at each point in time t, however, the instantaneous eigenvectors $\ket{\phi_{n}(t)}$ explicitly depend on $t$. In order to construct them, we can proceed in an analogous way as done in standard quantum mechanics. The ground state of Eq.~\eqref{eq:Invaroscapp} is defined as $a_{t}\ket{\phi_{0}(t)}=0$. Expanding the ground state in terms of Fock states of the oscillator with frequency $\omega_{0}$,~i.e, $\ket{\phi_{0}(t)}=\sum_{n}c_{nt}\ket{n}$, the action of $a_{t}$ allows to find a recursion relation for the same coefficients. After the normalization, the instantaneous ground state of the invariant reads
\beq\label{eq:groundket}
\ket{\phi_{0}(t)}= c_{0t}\qty(\ket{0} +\sum_{n=1}^{\infty} (-1)^{n} \sqrt{\frac{(2n-1)!!}{2^{n} n!}} z_{t}^{n}\ket{2n} )
\eeq
with $z_{t}=(1 -\zeta^2(t) -i \dot{\zeta}(t)\zeta(t)/\omega_{0})/(1 +\zeta^2(t) -i \dot{\zeta}(t)\zeta(t)/\omega_{0})$ and $c_{0t}=(1-\abs{z_{t}}^2)^{1/4}$. All the excited states can be computed from Eq.~\eqref{eq:groundket} as $\ket{\phi_{n}(t)}=(a^{\dagger}_{t})^{n}\ket{\phi_{0}(t)}/\sqrt{n!}$. Alternatively, the wave functions corresponding to each $\ket{\phi_{n}(t)}$ can be computed in the basis of the eigenstates of $q$ as
\beq\label{eq:groundwave}
\bra{q}\ket{\phi_{n}(t)}=\frac{1}{\sqrt{\zeta(t)}}\Phi_{n}\qty(\frac{\bar{q}}{\zeta(t)})\exp[\frac{i}{2\omega_{0}}\dot{\zeta}(t)\zeta(t) \qty(\frac{\bar{q}}{\zeta(t)})^2]
\eeq
where $\bar{q}=\sqrt{m\omega_{0}}q$ and  $\Phi_{n}(\xi)$ are the conventional eigenfunctions of the harmonic oscillator defined in terms of the Hermite polynomials $H_{n}(\xi)$,~i.e., $\Phi_{n}(\xi)=(1/\sqrt{2^n n!})(m\omega_{0}/\pi)^{1/4}e^{-\xi^2/2}H_{n}(\xi)$.

\section{Derivation of NAME equation}\label{sec:appC}
In this section, we outline the derivation of the Non-Adiabatic quantum Master Equation in the inertial approximation limit as developed in \cite{dann2018time,dann2020fast,dann2021inertial}. Hereafter we set $t_{0}=0$.
 As with many microscopic derivations of QME \cite{breuer}, the starting point for all the approximations involved is the Born-Markov equation \cite{breuer} that can be obtained from Eq.~\eqref{eq:BM1} by replacing $s\rightarrow t-s$ under the integral and by taking the upper limit of integration to infinity. It follows that 
\beq\label{eq:BM2}
     \frac{d}{d t}\tilde{\rho}_{\text S}(t)=-\int\limits_{0}^{+\infty}\mathrm{d}s \tr_{\text B}\comm{\tilde{H}_{\text{I}}(t)}{\comm{\tilde{H}_{\text{I}}(t-s)}{\tilde{\rho}_{\text S}(t)\otimes \tilde{\rho}_{\text B}}}.
\eeq

The NAME equation relies on the expansion of the coupling operator $\tilde{A}_{k}(t)=U^{\dagger}_{\text{S}}(t,0)A_{k}U_{\text{S}}(t,0)$, written in the interaction picture, in terms of specific operators acting on the Hilbert space of S. Below, we summarize the main derivation steps leading to  Eq.~\eqref{eq:BMSchro} in the main text.   

The unitary dynamics ruled by $H_{\text S}(t)$ can be studied by adopting the formalism of the Liouville-Hilbert-Schmidt (LHS) operator space \cite{Gyamfi_2020}, that is a linear space of operators endowed with the inner product $\ev{A,B}=\tr[A^{\dagger}B]$, where $A$ and $B$ are two arbitrary operators.~Once a reference basis of operators ${v_{j}}$ of dimension $N_{d}$ has been chosen, the state of the system at any time t can thus be described by a vector $\vec{v}^{\ \!\text H}(t)$ of operators. Each component of this vector is written in the Heisenberg picture with respect to $H_{\text{S}}(t)$. The dynamics of this state vector is ruled by the Heisenberg equation 
\beq\label{eq:Heiseq1} 
    \frac{d\vec{v}^{\ \!\text{H}}(t)}{dt}=U_{\text{S}}^{\dagger}(t,0)\qty[\qty(\frac{i}{\hbar}[H_{\text{S}}(t),\circ] +\frac{\partial}{\partial t})\vec{v}(t)] U_{\text{S}}(t,0),    
\eeq
where the vector $\vec{v}(t)$ allows for the explicit time dependence of its component operators. From the assumption in Sec.~\ref{sec:STA} that $H_{\text{S}}(t)$ belongs to a closed Lie Algebra, it follows that the equations of motion for the operators are also closed under the same algebra. It thus follows that Eq.~\eqref{eq:Heiseq1} can be rewritten as
\beq\label{eq:Heiseq2} 
    \frac{\mathrm{d}}{\mathrm{d}t}\vec{v}^{\ \!\text{H}}(t)=-i\mathcal{G}(t)\vec{v}^{\ \!\text{H}}(t).    
\eeq
where $\mathcal{G}(t)$ is a $N_d\times N_d$ matrix whose elements depend on the properties of the Lie algebra.  
The time evolution operator in the LHS space can be expressed in terms of $\mathcal{G}(t)$ as follows
\beq\label{eq:evol}
     \mathcal{U}(t,t_{0})=\exp{-i\int_{0}^t \mathcal{G}(t^{\prime}) \mathrm{d} t^{\prime}}.
 \eeq

The NAME idea is then linked to the search for an analytical approximation to the dynamics of $\vec{v}^{\ \!\text{H}}(t)$, by writing it in terms of left eigenoperators of the matrix $\mathcal{G}(t)$\cite{dann2018time}.
The special feature of these operators is that, if a suitable choice of basis vectors is performed, their time evolution is known. Thus, in the same spirit of more conventional master equation approaches, the operator $\tilde{A}_{k}(t)$ can be expanded in terms of these eigenoperators so that, after performing a secular approximation, an LGKS equation can be written down.

To derive such eigenoperators, a wise choice of the reference basis operators $v_{j}$ can be made, relying on the properties of the Lie algebra. Indeed, it can be shown that a basis of operators exists such that
\beq\label{eq:matrix1}
    \mathcal{G}(t)=-i\Omega(t)\mathcal{B}(\mu(t)),
\eeq       
where $\Omega(t)$ is a function of time-related to the details of the driven system, while $\mathcal{B}(\mu(t))$ is a matrix that depends on the single parameter $\mu(t)$. As anticipated in the main text, $\mu(t)$ stands for the adiabaticity parameter of the time-dependent protocol ruled by $H_{\text S}(t)$, i.e.,
\beq\label{eq:muparapp}
\mu(t)=\sum_{n\neq m} \frac{\bra{\epsilon_{n}(t)}\dot{H}_{\text{S}}(t)\ket{\epsilon_{m}(t)}}{(\epsilon_{n}(t)-\epsilon_{m}(t))^2}, 
\eeq
where $\ket{\epsilon_{m}(t)}$ are the instantaneous eigenstates of $H_{\text{S}}(t)$ with energy $\epsilon_{m}(t)$. 

Let's first focus on the instance of $\mu=\text{const}$. By defining $U(\mu)$ such that $U^{-1}(\mu)\mathcal{B}(\mu)U(\mu)=\text{diag}(\lambda_{1},\lambda_{2},...,\lambda_{N_{d}})$, then the solution for the vector $\vec{v}^{\ \!\text{H}}(t)$ reads
\beq\label{eq:solutv}
    \vec{v}^{\ \!\text{H}}(t)=U(\mu) e^{-i(\int_{0}^t \Omega(t^{\prime})\mathrm{d}t^{\prime})\text{diag}(\lambda_{1},\dots,\lambda_{N_{d}})}U^{-1}(\mu)\vec{v}(0),
\eeq
where we denoted $\vec{v}^{\ \!\text{H}}(0)=\vec{v}(0)$. By considering each of the components ${v_{j}(0)}$ of the vector $\vec{v}(0)$, we define the operators $F_{k}(\mu)$ as
\beq\label{eq:eigenop}
  		 F_{k}(\mu)=\sum_{j}U^{-1}_{kj}(\mu)v_{j}(0).
\eeq   
Notice that if the components of $\vec{v}^{\ \!\text{H}}(t)$ are chosen as the same operators belonging to the reference basis, then $F_{k}(\mu)$ can be identified with the instantaneous left eigenoperators of the evolution operator. From Eq.~\eqref{eq:solutv} it follows that        
\beq\label{eq:solutcomp}
    v^{\text{H}}_{i}(t)=\sum_{k}U_{ik}(\mu)F_{k}(\mu) e^{-i\lambda_{k}\int_{0}^{t} \Omega(t^{\prime})\mathrm{d}t^{\prime}}.
\eeq
As a result, the dynamics of each operator of $\vec{v}(t)$ is expressed in terms of time-independent operators $F_{k}$, while their time dependence is ruled by simple phase factors.     
Combining Eq.~\eqref{eq:solutcomp} with Eq.~\eqref{eq:BM2}, after performing the secular approximation, a LGKS equation where the role of the jump operator is played by the operators $F_{k}$ can be derived.

Here however, since the driving fields linked to the invariant operator in Eq.~\eqref{eq:diffEq} do not necessarily have a constant  $\mu$, we need to adopt an expansion similar to Eq~.\eqref{eq:solutcomp} that accounts for at least a weak change in the $\mu(t)$. Thus, we choose to adopt the \emph{inertial approximation} \cite{dann2020fast,dann2021inertial}. It is based on the hypothesis of slow driving acceleration, i.e., $(\dot{\mu}(t)/\Omega(t))^2\ll 1$, under which Eq.~\eqref{eq:solutcomp} can be replaced by
\beq\label{eq:solinert}
    v^{\text{H}}_{i}(t)=\sum_{k}U_{ik}(\mu(t))e^{-i\int_{0}^{t} \lambda_{k}(t^{\prime}) \Omega(t^{\prime})\mathrm{d}t^{\prime}} F_{k,t},
\eeq
where $\lambda_k(t)$ are the instantaneous eigenvalues of the matrix $\mathcal{B}(\mu(t))$ and $F_{k,t}\equiv F_{k}(\mu(t),0)$ are the static eigenoperators, where the constant $\mu$ is replaced by its instantaneous value, i.e., $\mu\rightarrow \mu(t)$.~This approach is equivalent to neglecting all the phase contributions in the eigenoperators due to the change of $\mu(t)$ over time.~We now assume the operators $v_{i}(t)$ to be the time-evolved of our reference basis w.r.t $H_{\text S}(t)$.~For the sake of simplicity, we also drop the index $k$ in the operator $A$, i.e., from Eq.~\eqref{eq:Hint} in the main text we consider $H_{\text{I}}=A\otimes\sum_{k}c_{k}B_{k}$.~Then, by dropping the superscript H as well, the operator $\tilde{A}(t)$ can be expanded in terms of the reference operators in Eq.~\eqref{eq:solinert} as $\tilde{A}(t)=\sum_{j} s_{j}(t)v_{j}(t)$, where $s_{j}(t)$ are the expansion coefficients. Inserting this expansion into the expression for $\tilde{H}_{\text{I}}(t)$, we obtain  
\begin{multline}\label{eq:intereigen}
       	\tilde{H}_{\text{I}}(t)=  \sum_{j}s_{j}(t)v_{j}(t)\otimes \sum_{k} c_{k} \tilde{B}_{k}(t)=\\
    	=\sum_{jl}s_{j}(t)U_{jl}(\mu(t))e^{-i\int_{0}^{t}\lambda_{l}(t^{\prime})\Omega(t^{\prime})\mathrm{d}t^{\prime}}F_{l,t} \otimes \sum_{k} c_{k} \tilde{B}_{k}(t).
\end{multline}
By defining $\sum_{j}s_{j}(t)U_{jl}(\mu(t))=\xi_{l}(t)e^{-i\eta_{l}(t)}$ and $\Lambda_{l}(t)=\int_{0}^{t}\lambda_{l}(t^{\prime}) \Omega(t^{\prime}) \mathrm{d} t^{\prime} + \eta_{l}(t)$, inserting Eq.~\eqref{eq:intereigen} in Eq.~\eqref{eq:BM2}, we find
    \begin{multline}\label{eq:BM3}
     \frac{\mathrm{d}}{\mathrm{d} t}\tilde{\rho}_{\text S}(t)= \sum_{lmkk^{\prime}} c_{k} c_{k^{\prime}}\int\limits_{0}^{+\infty}\mathrm{d}s \xi_{l}(t)\xi_{m}(t-s) e^{-i\Lambda_{l}(t)}\cdot\\e^{-i\Lambda_{m}(t-s)}
     \cdot\qty(F_{m,(t-s)}\tilde{\rho}_{\text S}(t)F_{l,t}-F_{l,t}F_{m,(t-s)}\tilde{\rho}_{\text S}(t))\cdot\\\cdot\tr_{\text B}[\tilde{B}_{k}(t)\tilde{B}_{k^{\prime}}(t-s)\rho_{\text B}] +\text{h.c.}, 
    \end{multline}
where h.c. stands for the Hermitean conjugate. In order to reduce Eq.~\eqref{eq:BM3} to the LGKS form, further approximations are needed on the time dependence of the functions $\xi_{l}(t)$ and $\Lambda_{l}(t)$.~If $\dot{\mu}(t)$ is sufficiently small, we can neglect the time retardation in the amplitudes and in the instantaneous eigenoperators,~i.e., $\xi_{l}(t-s)\approx \xi_{l}(t),\ F_{m,(t-s)}\approx F_{m,t}$.~Furthermore, the change in the phases over time can be considered small as compared with the instantaneous values, so that we take only the first-order contribution to the phase, i.e.,  
\beq\label{eq:phases}
     \Lambda_{l}(t-s)\approx \Lambda_{l}(t) -\qty(\frac{\mathrm{d}}{\mathrm{d} t^{\prime}}\Lambda_{l}(t^{\prime})|_{t^{\prime}=t}) s + o(s^2),
\eeq
and neglect higher-order ones.~In this limit, all the time-dependent contributions in Eq.~\eqref{eq:BM3} become local in time. It is thus possible to perform the secular approximation, and achieve the final form of the NAME equation in the interaction picture as follows
\beq\label{eq:BMfinal}
     \frac{d}{d t}\tilde{\rho}_{\text S}(t)= -i\comm{H_{\text{LS}}(t)}{\tilde{\rho}_{\text S}(t)} + \mathcal{D}_{t}\tilde{\rho}_{\text S}(t),
\eeq     
where the instantaneous dissipator $\mathcal{D}_{t}$ reads
\begin{multline}\label{eq:Dissipator}
\mathcal{D}_{t}= \sum_{l} \xi^2_{l}(t) \gamma(\dot{\Lambda}(t))\qty(F_{l,t}\circ F^{\dagger}_{l,t}-\frac{1}{2}\qty{F^{\dagger}_{l,t}F_{l,t},\circ}), 
\end{multline}
$\gamma(\omega)$ are the decay rates and $H_{\text{LS}}(t)$ is the Lamb-shift contribution.  Notice that, under the previous assumptions, the decay rates can be computed in the usual way from the function 
\beq\label{eq:gammakk}
\Gamma_{kk^{\prime}}(\omega)=\int\limits_{0}^{+\infty}\mathrm{d}s e^{i\omega s}\tr_{\text B}[\tilde{B}_{k}(t)\tilde{B}_{k^{\prime}}(t-s)\rho_{\text B}].
\eeq
Namely, $\Gamma_{kk^{\prime}}(\omega)=(1/2)\gamma_{kk^{\prime}}(\omega) + iS_{kk^{\prime}}(\omega)$, so that $\gamma(\omega)=\sum_{kk^{\prime}}c_{k}c_{k^{\prime}}\gamma_{kk^{\prime}}(\omega)$ and the Lamb shift reads $H_{\text{LS}}(t)=\sum_{lkk^{\prime}}c_{k}c_{k^{\prime}}S_{kk^{\prime}}(\dot{\Lambda}(t))\xi_{l}^2(t)F^{\dagger}_{l,t}F_{l,t}$. 

It is useful to derive the explicit form of the function $\Gamma_{kk^{\prime}}(\omega)$ for a bath modeled with a set of harmonic oscillators. Indeed, as shown in Secs.~\ref{sec:appCqubit} and ~\ref{sec:appCosc}, we assume the system-bath coupling to take place via the bath's oscillator position operator $B_{k}=\sqrt{1/(2 m_{k}\omega_{k})}(a_{k}+a^{\dagger}_{k})$.~From Eq.~\eqref{eq:gammakk}, we find
\begin{multline}\label{eq:gammakk2}
\Gamma_{kk^{\prime}}(\omega)=\int\limits_{0}^{+\infty}\mathrm{d}s e^{i\omega s}\tr_{\text B}[\tilde{B}_{k}(s)\tilde{B}_{k^{\prime}}(0)\rho_{\text B}]=\\
=\frac{\delta_{kk^{\prime}}}{2m_{k}\omega_{k}}\int\limits_{0}^{+\infty}\mathrm{d}s\qty((\bar{n}_{k} + 1)e^{i(\omega-\omega_{k})s} + \bar{n}_{k}e^{i(\omega + \omega_{k})s})=\\
=\frac{\delta_{kk^{\prime}}}{2m_{k}\omega_{k}}\Bigg[(\bar{n}_{k} + 1)\qty(\pi \delta(\omega_{k}-\omega) -i\text{P.V.}\frac{1}{\omega_{k}-\omega} ) +\\ 
+\bar{n}_{k}\qty(\pi \delta(\omega_{k}+\omega) +i\text{P.V.}\frac{1}{\omega_{k}+\omega} )\Bigg],
\end{multline}
where $\bar{n}_{k}$ is the average number of bosonic excitations in the k-$\text{th}$ mode. From Eq.~\eqref{eq:gammakk2} the expressions of the decay functions as well as the Lamb shift  follow 
\begin{multline}\label{eq:gammakk3}
\gamma(\omega)=2\pi\sum_{k}\nu^2_{k}\qty[(\bar{n}_{k} + 1) \delta(\omega_{k}-\omega) +\bar{n}_{k}\delta(\omega_{k}+\omega) ], \\
S(\omega)=\sum_{k}\nu^2_{k}\Bigg[\bar{n}_{k} \text{P.V.}\qty(\frac{1}{\omega_{k}+\omega}) \\-(\bar{n}_{k} + 1 )\text{P.V.}\qty(\frac{1}{\omega_{k}-\omega}) \Bigg].
\end{multline}
with $\nu_{k}=c_{k}\sqrt{1/(2m_{k}\omega_{k})}$.

\subsection{NAME for the TLS}\label{sec:appCqubit}
%Indeed, in the Liouville-Hilbert space, the state of the system can be described as a vector of operators $\vec{v}^{\text{H}}(t)$, obeying the Heisenberg equation of motion. From the closure of the Lie algebra of S, once a reference basis of operators $\{v_{j}(t),{j=1,\dots,N_{d}}\}$ is found, then the solution of the Heisenberg equation can be written as  
In order to derive Eqs.~\eqref{eq:dissipatorqubit} and \eqref{eq:jumpqubit} from Eq.~\eqref{eq:Hamiltotqubit} in the main text, we start from the matrix $\mathcal{G}(t)$. The latter can be derived once the reduced TLS Hamiltonian in Eq.~\eqref{eq:Hqubit} is known \cite{rezek2006irreversible,uzdin2013effects,dann2020fast}. Indeed, from Eq.~\eqref{eq:Hqubit} it follows that, choosing the basis of operators $\qty(H_{\text{S}}(t),L(t),D(t))=\{H_{\text{S}}(t),(\epsilon(t)/2) \sigma_{z} -(\Delta(t)/2)\sigma_{x},(\Omega(t)/2)\sigma_{y}\}$, the matrix $\mathcal{G}(t)$ can be written as
\beq\label{eq:Gmatrix}
\mathcal{G}(t)=i\frac{\dot{\Omega}(t)}{\Omega(t)}\mathbb{1} + \Omega(t)\mathcal{B}(t),
\eeq
with   
\beq
\mathcal{B}(t)=
\begin{pmatrix}\label{eq:matrB}
0 & i\mu(t) & 0\\
-i\mu(t) & 0 & i\\
0& -i &0
\end{pmatrix}
\eeq
%\beq\label{eq:solut}
%\vec{v}^{\text{H}}(t)=\exp[-i\int_{0}^{t}\mathcal{G}(t^{\prime})\mathrm{d}t^{\prime}]\vec{v}^{\text{H}}(0),
%\eeqIndeed, a special reference basis can be found such that  $\mathcal{G}(t)=\Omega(t)\mathcal{B}(\mu(t))$ 
We can simplify Eqs.~\eqref{eq:Heiseq2} and ~\eqref{eq:Gmatrix} by looking at the vector $\vec{z}(t)=(\Omega(0)/\Omega(t))\vec{v}(t)$, which evolves according to $\mathrm{d}\vec{z}(t)/{\mathrm{d}t}=-i\Omega(t)\mathcal{B}(t)\vec{z}(t)$. Therefore, it is sufficient to study the eigenvalues and eigenvectors of  \eqref{eq:matrB} and then solve for $\vec{v}(t)$ as in \eqref{eq:solutcomp}. The form of  \eqref{eq:matrB} is particularly useful in the case $\mu(t)=\mu=\text{const}$.~By diagonalizing the matrix $\mathcal{B}(\mu)$,~i.e., $U^{-1}(\mu)\mathcal{B}(\mu)U(\mu)=\text{diag}(\lambda_1,\lambda_2,\lambda_{3})$, we find 
\beq\label{eq:staticeig}
\lambda_1=0\mbox{ , }\lambda_2=\sqrt{1+\mu^2}\mbox{ , }\lambda_3=-\sqrt{1+\mu^2}.
\eeq
According to Eq.~\eqref{eq:eigenop}, the eigenoperators $F_{k}(\mu)$ can be found from the rows of the inverse diagonalization matrix $U^{-1}$. In the chosen reference basis, denoting  $k=\sqrt{1+\mu^2}$ they read 
\begin{align}\label{eq:lefteigenconst}
F_1&=\qty(\frac{1}{k},0,\frac{\mu}{k}),\nonumber\\
F_2&=\qty(\frac{\mu}{\sqrt{2}k},\frac{i}{\sqrt{2}},-\frac{1}{\sqrt{2}k}),\nonumber\\
F_3&=\qty(\frac{\mu}{\sqrt{2}k},-\frac{i}{\sqrt{2}},-\frac{1}{\sqrt{2}k}).
\end{align}
It is thus easy to find the expansion of each element of the vector $\vec{v}^{\ \!\text{H}}(t)$ in terms of $F_{l}(\mu)$ as
\beq\label{eq:expandcost}
v_{j}^{\text{H}}(t)=\frac{\Omega(t)}{\Omega(0)}\sum_{l=1}^{3}U_{jl}(\mu)e^{-i\lambda_l\int_{0}^{t}\Omega(t^{\prime})\mathrm{d}t^{\prime}} F_{l}(\mu).
\eeq
%It should be noticed that the expansion is written in terms of the components of $v$ in the reference basis at the initial time, thus coinciding with the operators in the Schr\"odinger picture. I
%The expansion in Eq.~\eqref{eq:expandcost}, once applied to the interaction operator $H_{\text{I}}$ in Eq.~\eqref{eq:Hamiltotqubit}, allows to perform the Born-Markov and secular approximations to derive the time-local LGKS QME when the driving fields in S have $\mu(t)=\text{const}$ \cite{dann2018time}. 
In order to perform the expansion in the case of time-dependent $\mu$, we employ the inertial approximation \cite{dann2020fast,dann2021inertial},~i.e.~, we replace Eq.~\eqref{eq:expandcost} with
\beq\label{eq:expandt}
v_{j}^{\text{H}}(t)=\frac{\Omega(t)}{\Omega(0)}\sum_{l=1}^{3}U_{jl}(\mu(t))e^{-i\int_{0}^{t}\lambda_l(t^{\prime})\Omega(t^{\prime})\mathrm{d}t^{\prime}} F_{l,t},
\eeq
where $F_{l,t}=F_{l}(\mu(t),0)$ are the static eigenoperators in Eq.~\eqref{eq:lefteigenconst} with $\mu\rightarrow\mu(t)$ and the eigenvalues $\lambda_{i}(t)$ are obtained from Eq.~\eqref{eq:staticeig} in similar fashion. When written in terms of the operator reference basis of choice, the $F_{l,t}$ read
\begin{align}\label{eq:jumpqubitapp}
F_{1,t}&=\frac{1}{k(t)}H_{\text{S}}(0) +\frac{\mu(t)}{k(t)}D(0),\nonumber\\
F_{2,t}&=\frac{\mu(t)}{\sqrt{2}k(t)}H_{\text{S}}(0) +\frac{i}{\sqrt{2}}L(0) -\frac{1}{\sqrt{2}k(t)}D(0),\nonumber\\
F_{3,t}&=\frac{\mu(t)}{\sqrt{2}k(t)}H_{\text{S}}(0) -\frac{i}{\sqrt{2}}L(0) -\frac{1}{\sqrt{2}k(t)}D(0).
\end{align}
It should be noticed that this approximation neglects the geometric phase linked to the full time evolution of the operators $F_{l}(t)$ \cite{dann2021inertial}. The operators $F_{l,t}$ can be used to write down the LGKS dissipator as in Eq.~\eqref{eq:Dissipator}. The rates $\kappa_{l}(t)$ corresponding to each operator in Eq.~\eqref{eq:jumpqubitapp} can be computed by looking at the system-bath interaction operator, as written in Eq.~\eqref{eq:Hamiltotqubit}, main text, and in Eq.~\eqref{eq:intereigen}.~We find that $(1/2)\sigma^{\text{H}}_{y}(t)=(1/\Omega(t))v^{\text{H}}_{3}(t)$, and the amplitudes can be found as $\xi_{1}(t)=\mu(t)/(\Omega(0)k(t)),\ \xi_{2}(t)=\xi_{3}(t)=-1/(\Omega(0)\sqrt{2}k(t))$. The dissipation rates can be computed from the decay function $\gamma(\omega)$ in Eq.~\eqref{eq:gammakk3}. Indeed, using Eq.~\eqref{eq:spectral} in the main text, and considering the continuum limit, the sum over the discrete k-indices can be replaced by an integral over the dense frequencies $\omega_{k}$ \cite{weissbook}.  Equation~\eqref{eq:gammakk3} thus reads
\beq\label{eq:continuum1}
\gamma(\omega)=
\begin{cases}
&2\pi \tilde{J}(\omega)(\bar{n}(\omega) + 1), \omega \geq 0\\
&2\pi \tilde{J}(\abs{\omega})\bar{n}(\abs{\omega}),\omega < 0
\end{cases}
\eeq
\beq\label{eq:continuum2}
S(\omega)=\int\limits_{0}^{+\infty}\mathrm{d}\omega^{\prime}\tilde{J}(\omega^{\prime})\text{P.V.}\qty(\frac{\bar{n}(\omega^{\prime}) + 1 }{\omega-\omega^{\prime}} +\frac{\bar{n}(\omega^{\prime})}{\omega+\omega^{\prime}}),
\eeq
As a result, the rates $\kappa_{l}(t)$ can be computed as follows\begin{align}\label{eq:ratesqubit}
\kappa_{1}(t)&=\gamma(0)\mu^2(t)/(\Omega^2(0) k^{2}(t)),\nonumber\\ \kappa_{2}(t)&=\gamma(\dot{\Lambda}_{2}(t))/(2\Omega^2(0) k^{2}(t)),\nonumber\\
\kappa_{3}(t)&=\gamma(\dot{\Lambda}_{3}(t))/(2\Omega^2(0) k^{2}(t)),
\end{align}
with $\dot{\Lambda}_{2}(t)=\Omega(t)k(t),\ \dot{\Lambda}_{3}(t)=-\Omega(t)k(t)$ and $\gamma(\omega)=\sum_{k}\gamma_{kk}(\omega)$. Notice that, although the method can be employed with arbitrary coupling operators, the choice of $\sigma_{y}$, see Eq.~\eqref{eq:Hamiltotqubit} main text, results in easier expressions for the amplitudes $\xi_{i}(t)$ and phases $\Lambda_i(t)$. Moreover, as a consequence of the NAME approach and the inertial approximation, each LGKS operator is written in terms of a linear combination of $\text{su}(2)$ operators computed at time $t=0$, where the coefficients are simple time-dependent functions. Once the driving fields in Eq.~\eqref{eq:Hqubit} are known, the dissipator in Eq.~\eqref{eq:dissipatorqubit} is analytically computed. 

\subsection{NAME for the oscillator}\label{sec:appCosc}

In the case of the driven harmonic oscillator, starting from Eq.~\eqref{eq:Hamiltotosc}, main text, the choice of the operator basis ${\hat{q}(t)=\sqrt{m\omega(t)}q,\ \hat{p}(t)= p/(m\sqrt{\omega(t)})}$ allows to compute the dynamical matrix $\mathcal{G}(t)$ in Eq.~\eqref{eq:Heiseq2} as follows
\beq\label{eq:Gosct}
\mathcal{M}(t)=-i\mathcal{G}(t)=\omega(t)
\begin{pmatrix}
\frac{\mu(t)}{2} & 1\\
-1 & -\frac{\mu(t)}{2}\\
\end{pmatrix}.
\eeq
where the adiabaticity parameter in Eq.~\eqref{eq:muparapp} reads
\beq
\mu(t)=\frac{\dot{\omega}(t)}{\omega^2(t)}.
\eeq
We can compute the left  eigenvalues and eigenvectors of $\mathcal{M}(t)$ in terms of $(\hat{q}(t),\hat{p}(t))$ as follows 
\beq\label{eq:lefteigen}
\lambda_{\pm}(t)=\pm \frac{i}{2}l(t)\omega(t)\;\; {\rm and} \;\; u_{\pm}=\frac{1}{2}\qty(\mu(t) \pm i l(t))\hat{q}(t) +\hat{p}(t),
\eeq
where $l(t)=\sqrt{4-\mu^2(t)}$ and $\abs{\mu(t)}<2$. In the case of $\mu(t)=\text{const}$, we can define the LGKS operators $F_{\pm}$ as $F_{+}=(i/l)u_{-}(0),F_{-}=F^{\dagger}_{+}$. A direct consequence of their definition, from Eqs.~\eqref{eq:evol} and~\eqref{eq:Gosct} the time evolution of these operators is known analytically, i.e. $F^{\text{H}}_{\pm}(t)=F_{\pm}e^{i\theta_{\pm}(t)}$, $\theta_{\pm}(t)=\mp (l/2)\int_{0}^{t}\omega(t^{\prime}) \mathrm{d}t^{\prime}$, so that from Eq.~\eqref{eq:lefteigen} we can expand the solution of the position operator as follows
\beq\label{eq:posexpansion1}
q^{\text{H}}(t)=q_{0}\sqrt{\frac{\omega(0)}{\omega(t)}}\qty(F_{+}e^{i\theta_{+}(t)} + F_{-}e^{i\theta_{-}(t)} ),
\eeq
where $q_{0}=1/\sqrt{m\omega(0)}$ denotes the length scale of the oscillator with initial frequency $\omega(0)$, which we employ as the length scale of our problem.~By making use of the inertial approximation as explained in Sec.~\ref{sec:appCqubit}, Eq.~\eqref{eq:posexpansion1} can be extended to the case of a time-dependent $\mu(t)$ as follows
\beq\label{eq:posexpansion2}
q^{\text{H}}(t)=q_{0}\sqrt{\frac{\omega(0)}{\omega(t)}}\qty(F_{t,+}e^{i\theta^{\text{in}}_{+}(t)} + F_{t,-}e^{i\theta^{\text{in}}_{-}(t)} ),
\eeq
with $F_{t,\pm}=\pm (i/l(t))u_{\mp}(t)$ and $\theta^{\text{in}}_{\pm}(t)=\mp(1/2)\int_{0}^{t}l(t^{\prime})\omega(t^{\prime})\mathrm{d}t^{\prime}$. From Eq.~\eqref{eq:posexpansion2}, it follows that the amplitudes read $\xi_{+}(t)=\xi_{-}(t)=q_{0}\sqrt{\omega(0)/\omega(t)}$. The final form of the dissipator in Eq.~\eqref{eq:Dissosc} can  be computed once the function $\gamma(\omega)$ is known. It is useful to define the modified spectral density function $\tilde{J}(\omega)$ analogous to  Eq.~\eqref{eq:spectral},
\beq\label{eq:redspectralosc}
\tilde{J}(\omega)= \sum_{k}\bar{\nu}^2_{k} \delta(\omega -\omega_{k})=2\frac{\gamma_{s}}{\omega_{0}}\omega e^{-\omega/\omega_{c}}.
\eeq
Notice that from Eq.~\eqref{eq:Hamiltotosc},  $\bar{\nu}_{k}=q_{0}(c_{k}/\sqrt{2 m_{k}\omega_{k}})$ and $\tilde{J}(\omega)=(q^2_{0}/\pi)J(\omega)$. By taking $\alpha(t)=\dot{\theta}^{\text{in}
}_{-}(t)=(l(t)/2)\omega(t)>0$, using Eqs.~\eqref{eq:Dissipator},~\eqref{eq:gammakk3},~\eqref{eq:posexpansion2} it is then easy to find that 
\begin{align}\label{eq:rateosc}
\kappa_{+}(t)&=2\pi\frac{\omega(0)}{\omega(t)}\tilde{J}(\alpha(t))(1+\bar{n}(\alpha(t))),\nonumber \\
\kappa_{-}(t)&=2\pi\frac{\omega(0)}{\omega(t)}\tilde{J}(\alpha(t))\bar{n}(\alpha(t)).
\end{align}

%\begin{multline}\label{eq:Dissosc}
%\mathcal{D}_{t}=r_{+}(t)\qty(F_{t,+}\circ F_{t,-} -\frac{1}{2}\{F_{t,-}F_{t,+},\circ\}) +\\ + r_{-}(t)\qty(F_{t,-}\circ F_{t,+} -\frac{1}{2}\{F_{t,+}F_{t,-},\circ\}),
%\end{multline}

\section{Details on the optimization}\label{sec:appD}

We can study the behavior of the fidelity in the limit of weak dissipation strength. From Eq.~\eqref{eq:fides}, and from the fact that the target state is a pure state, we can write
\beq\label{eq:fidelitypert1}
\mathcal{F}=\sqrt{\bra{\psi_{\text{tar}}}\rho_{\text{S}}(t_{\text{f}}) \ket{\psi_{\text{tar}}}},
\eeq
with $\ket{\psi_{\text{tar}}}=\sum_{n}\abs{c_{n}}e^{i\varphi_{n}(t_{\text{f}})}\ket{\phi_{n}(t_{\text{f}})}$, where  $\varphi_{n}(t)-\varphi_{n}(t_{0})= \int_{t_{0}}^{t}\mathrm{d}s\bra{\phi_{n}(s)}\qty(i\partial/\partial s-H(s))\ket{\phi_{n}(s)})$. As $\rho_{\text{S}}(t)$ is the solution of the QME in Eq.~\eqref{eq:BMSchro}, from Eq.~\eqref{eq:fidelitypert1} it follows 
\begin{multline}\label{eq:fidelitypert2}
\mathcal{F}^2=\bra{\psi_{\text{tar}}}\rho_{\text{S}}(t_{0}) + \int_{t_{0}}^{t_{\text{f}}} \mathcal{L}_{s}(\rho_{\text{S}}(s)) \mathrm{d}s  \ket{\psi_{\text{tar}}}=\\ \bra{\psi_{\text{tar}}}\rho_{\text{S}}(t_{0}) +\int_{t_{0}}^{t_{\text{f}}}\Big(-i\comm{H_{\text{S}}(s)} {\rho_{\text{S}}(s)} +\mathcal{D}_{s}(\rho_{\text{S}}(s))\Big)\mathrm{d}s\ket{\psi_{\text{tar}}} 
\end{multline}
In order to adopt the weak dissipation limit described in the main text, we write the state as $\rho_{\text{S}}(t)= \rho^{0}_{\text{S}}(t) + \chi(t)$, where $\rho^{0}_{\text{S}}(t)$ is the full solution of the Heisenberg equation of the closed system and $\chi(t)$ is a traceless operator that comprises all the corrections to the state of the reduced system due to the open system map. Due to our reverse-engineering approach, the Hamiltonian fields in $H_{\text{S}}(t)$ are devised to transfer the state from its initial preparation to the target $\rho_{\text{tar}}=\ket{\psi_{\text{tar}}}\bra{\psi_{\text{tar}}}$. It follows that from Eq.~\eqref{eq:fidelitypert2} we can write 
\begin{multline}\label{eq:fidelitypertbis}
\mathcal{F}^2= \bra{\psi_{\text{tar}}}\rho_{\text{S}}(t_{0}) +\int_{t_{0}}^{t_{\text{f}}}\Big(-i\comm{H_{\text{S}}(s)} {\rho^{0}_{\text{S}}(s)+\chi(s)}+\\ +\mathcal{D}_{s}(\rho^{0}_{\text{S}}(s)+\chi(s))\Big)\mathrm{d}s\ket{\psi_{\text{tar}}} =\\ =1 + \int_{t_{0}}^{t_{\text{f}}}\bra{\psi_{\text{tar}}}\mathcal{D}_{s}(\rho^{0}_{\text{S}}(s))\ket{\psi_{\text{tar}}}\mathrm{d}s +\\ +\int_{t_{0}}^{t_{\text{f}}}\bra{\psi_{\text{tar}}}\mathcal{D}_{s}(\chi(s))\ket{\psi_{\text{tar}}}\mathrm{d}s+\\-i\int_{t_{0}}^{t_{\text{f}}}\bra{\psi_{\text{tar}}}\comm{H_{\text{S}}(s)} {\chi(s)}\ket{\psi_{\text{tar}}}\mathrm{d}s
\end{multline}
From Eq.~\eqref{eq:fidelitypertbis}, it is evident that the correction $\chi(t)$ is responsible for contributions arising from the commutator with the Hamiltonian as well as from the dissipative part.~We thus need to provide a systematic approximation to $\chi(t)$ in order to correctly define the functional required in the main text.~It can be done by using a Dyson expansion of the full map $V(t,t_{0})=\mathcal{T}\exp [\int_{t_{0}}^{t}\mathcal{L}_{s} \mathrm{d}s]$ in terms of the closed system map $V^{0}(t,t_{0})=\mathcal{T}\exp [-i\int_{t_{0}}^{t}\comm{H_{\text{S}}(s)}{\circ}\mathrm{d}s]$.~Indeed, denoting with $\mathcal{L}^{0}_{t}=-i\comm{H_{\text{S}}(t)}{\circ}$, $\mathcal{L}^{1}_{t}=\mathcal{D}_{t}$ the two maps obey the equations   
\begin{align} \label{eq:maps}
\frac{\partial}{\partial{t}}V^{0}(t,t_{0})&=\mathcal{L}^{0}_{t} V^{0}(t,t_{0}),\nonumber \\ 
\frac{\partial}{\partial{t}}V(t,t_{0})&=(\mathcal{L}^{0}_{t} +\mathcal{L}^{1}_{t}) V(t,t_{0}), 
\end{align}
with $V^{0}(t_{0},t_{0})=V(t_{0},t_{0})=\mathbb{1}$. It follows that 
\beq\label{eq:maps2}
\frac{\partial}{\partial{t}}(V(t,t_{0})-V^{0}(t,t_{0}))=\mathcal{L}^{0}_{t} (V(t,t_{0})-V^{0}(t,t_{0})) + \mathcal{L}^{1}_{t}V(t,t_{0}).
\eeq
From the integral equation corresponding to Eq.~\eqref{eq:maps2} we find
\beq\label{eq:maps3}
V(t,t_{0})-V^{0}(t,t_{0})=\int_{t_{0}}^{t} V^{0}(t,s)\mathcal{L}^{1}_{s}V(s,t_{0})\mathrm{d}s.
\eeq
Eq.~\eqref{eq:maps3} has the form of a recurrence equation of the kind of those involving Green functions, and they arise in the context of Time Convolutionless QME \cite{breuer,smirne2010nakajima,joye2022adiabatic}.~It allows to rewrite Eq.~\eqref{eq:fidelitypertbis} in terms of a systematic expansion for $V(t,t_{0})$, whose terms can be computed once the closed system map $V^{0}(t,t_{0})$ is known.~Alternatively, the expansion could be also performed in the interaction picture. Thus, at first order in $\mathcal{D}_{t}$, setting $t_{0}=0$, Eq.~\eqref{eq:fidelitypertbis} reads
\begin{multline}\label{eq:maps3}
\mathcal{F}^2=\bra{\psi_{\text{tar}}} V(t_{\text{f}},0)\rho(0) \ket{\psi_{\text{tar}}} \simeq \bra{\psi_{\text{tar}}} \Big(V^{0}(t_{\text{f}},0)+\\ + \int_{0}^{t_{\text{f}}}V^{0}(t_{\text{f}},s)\mathcal{D}_{s}V^{0}(s,0)\mathrm{d}s\Big)\rho_{\text{S}}(0)\ket{\psi_{\text{tar}}}=\\=1 +\bra{\psi_{\text{tar}}}\int_{0}^{t_{\text{f}}}V^{0}(t_{\text{f}},s)(\mathcal{D}_{s}(\rho^{0}_{\text{S}}(s)))\mathrm{d}s\ket{\psi_{\text{tar}}}  
\end{multline}
Equation~\eqref{eq:maps3} includes the corrections linked to the commutators of any order between $\comm{H_{\text{S}}(t)}{\circ}$ and the dissipator $\mathcal{D}_{t}$. Moreover, the map $V^{0}(t^{\prime},t^{\prime \prime})$ for any pair of intermediate times $t^{\prime}>t^{\prime \prime}$ can be written down analytically in terms of the invariant eigenbasis. Indeed, if we consider the evolution operator in the Hilbert space,~i.e., $U(t^{\prime},t^{\prime \prime})=\mathcal{T}\exp[-i\int_{t^{\prime \prime}}^{t^{\prime}}H_{\text{S}}(s)\mathrm{d}s]$, it can be written analytically in terms of the projectors on the invariant eigenstates at different times $\ket{\phi_{n}(t^{\prime})}\bra{\phi_{m}(t^{\prime\prime})}$, i.e.
\beq\label{eq:maps4}
U(t^{\prime},t^{\prime\prime})=\sum_{n} e^{i(\varphi_{n}(t^{\prime}) -\varphi_{n}(t^{\prime\prime}))}\ket{\phi_{n}(t^{\prime})}\bra{\phi_{n}(t^{\prime\prime})}.
\eeq
From the definition of the unitary map $V^{0}(t,t_{0})=U(t,t_{0})\circ U^{\dagger}(t,t_{0})$ and from Eqs.~\eqref{eq:rates}, we rewrite Eq.~\eqref{eq:maps3} as 
\begin{multline}\label{eq:maps5}
\mathcal{F}^2= 1 + \sum_{mnkj} \bra{\psi_{\text{tar}}}\ket{\phi_{n}(t_{\text{f}})}\bra{\phi_{m}(t_{\text{f}})}\ket{\psi_{\text{tar}}}\cdot\\ \cdot \int_{0}^{t_{\text{f}}} \mathrm{d}s e^{i(\alpha_{n}(t_{\text{f}},s)-\alpha_{m}(t_{\text{f}},s))}\mathcal{R}^{(mn)}_{kj}(s)\rho_{\text{S}kj}^{0}(s),
\end{multline}
with $\alpha_{n}(t,t^{\prime})=\varphi_{n}(t)-\varphi_{n}(t^{\prime})$. Notice that, due to the special properties of the invariant operator, Eq.~\eqref{eq:maps5} is a functional of the control fields and of the target state. As a consequence, it can be analytically computed. Moreover, higher-order contributions can be computed as well in terms of the same protocol-related quantities. 
Equations \eqref{eq:maps3}, \eqref{eq:maps5} show that, in the easy instance of a pure target state, due to the concavity of $\mathcal{F}$ and the property $\mathcal{F}\in [0,1]$, we can maximize $\mathcal{F}^2$ by minimizing the magnitude of the corrections to its maximum value. Notice also that, from the properties of the reverse-engineered map $V^{0}(t,t_{0})$, the latter corrections arise from the dissipative character of the map $V(t,t_{0})$. Adding higher order terms in Eq.~\eqref{eq:maps3} allows us to improve the accuracy in the optimal fields. This procedure can be generalized to the case of mixed target states \cite{Liang_2019}. 

\bibliography{STA}

\begin{thebibliography}{120}%
\makeatletter
\providecommand \@ifxundefined [1]{%
 \@ifx{#1\undefined}
}%
\providecommand \@ifnum [1]{%
 \ifnum #1\expandafter \@firstoftwo
 \else \expandafter \@secondoftwo
 \fi
}%
\providecommand \@ifx [1]{%
 \ifx #1\expandafter \@firstoftwo
 \else \expandafter \@secondoftwo
 \fi
}%
\providecommand \natexlab [1]{#1}%
\providecommand \enquote  [1]{``#1''}%
\providecommand \bibnamefont  [1]{#1}%
\providecommand \bibfnamefont [1]{#1}%
\providecommand \citenamefont [1]{#1}%
\providecommand \href@noop [0]{\@secondoftwo}%
\providecommand \href [0]{\begingroup \@sanitize@url \@href}%
\providecommand \@href[1]{\@@startlink{#1}\@@href}%
\providecommand \@@href[1]{\endgroup#1\@@endlink}%
\providecommand \@sanitize@url [0]{\catcode `\\12\catcode `\$12\catcode `\&12\catcode `\#12\catcode `\^12\catcode `\_12\catcode `\%12\relax}%
\providecommand \@@startlink[1]{}%
\providecommand \@@endlink[0]{}%
\providecommand \url  [0]{\begingroup\@sanitize@url \@url }%
\providecommand \@url [1]{\endgroup\@href {#1}{\urlprefix }}%
\providecommand \urlprefix  [0]{URL }%
\providecommand \Eprint [0]{\href }%
\providecommand \doibase [0]{https://doi.org/}%
\providecommand \selectlanguage [0]{\@gobble}%
\providecommand \bibinfo  [0]{\@secondoftwo}%
\providecommand \bibfield  [0]{\@secondoftwo}%
\providecommand \translation [1]{[#1]}%
\providecommand \BibitemOpen [0]{}%
\providecommand \bibitemStop [0]{}%
\providecommand \bibitemNoStop [0]{.\EOS\space}%
\providecommand \EOS [0]{\spacefactor3000\relax}%
\providecommand \BibitemShut  [1]{\csname bibitem#1\endcsname}%
\let\auto@bib@innerbib\@empty
%</preamble>
\bibitem [{\citenamefont {Acín}\ \emph {et~al.}(2018)\citenamefont {Acín} \emph {et~al.}}]{acin2018quantum}%
  \BibitemOpen
  \bibfield  {author} {\bibinfo {author} {\bibfnamefont {A.}~\bibnamefont {Acín}} \emph {et~al.},\ }\bibfield  {title} {\bibinfo {title} {The quantum technologies roadmap: a {E}uropean community view},\ }\href {https://doi.org/10.1088/1367-2630/aad1ea} {\bibfield  {journal} {\bibinfo  {journal} {New J. Phys.}\ }\textbf {\bibinfo {volume} {20}},\ \bibinfo {pages} {080201} (\bibinfo {year} {2018})}\BibitemShut {NoStop}%
\bibitem [{\citenamefont {Preskill}(2018)}]{preskill2018quantum}%
  \BibitemOpen
  \bibfield  {author} {\bibinfo {author} {\bibfnamefont {J.}~\bibnamefont {Preskill}},\ }\bibfield  {title} {\bibinfo {title} {Quantum {C}omputing in the {NISQ} era and beyond},\ }\href {https://doi.org/https://doi.org/10.22331/q-2018-08-06-79} {\bibfield  {journal} {\bibinfo  {journal} {Quantum}\ }\textbf {\bibinfo {volume} {2}},\ \bibinfo {pages} {79} (\bibinfo {year} {2018})}\BibitemShut {NoStop}%
\bibitem [{\citenamefont {Koch}\ \emph {et~al.}(2022)\citenamefont {Koch}, \citenamefont {Boscain}, \citenamefont {Calarco}, \citenamefont {Dirr}, \citenamefont {Filipp}, \citenamefont {Glaser}, \citenamefont {Kosloff}, \citenamefont {Montangero}, \citenamefont {Schulte-Herbrüggen}, \citenamefont {Sugny},\ and\ \citenamefont {Wilhelm}}]{koch2022quantum}%
  \BibitemOpen
  \bibfield  {author} {\bibinfo {author} {\bibfnamefont {C.~P.}\ \bibnamefont {Koch}}, \bibinfo {author} {\bibfnamefont {U.}~\bibnamefont {Boscain}}, \bibinfo {author} {\bibfnamefont {T.}~\bibnamefont {Calarco}}, \bibinfo {author} {\bibfnamefont {G.}~\bibnamefont {Dirr}}, \bibinfo {author} {\bibfnamefont {S.}~\bibnamefont {Filipp}}, \bibinfo {author} {\bibfnamefont {S.~J.}\ \bibnamefont {Glaser}}, \bibinfo {author} {\bibfnamefont {R.}~\bibnamefont {Kosloff}}, \bibinfo {author} {\bibfnamefont {S.}~\bibnamefont {Montangero}}, \bibinfo {author} {\bibfnamefont {T.}~\bibnamefont {Schulte-Herbrüggen}}, \bibinfo {author} {\bibfnamefont {D.}~\bibnamefont {Sugny}},\ and\ \bibinfo {author} {\bibfnamefont {F.~K.}\ \bibnamefont {Wilhelm}},\ }\bibfield  {title} {\bibinfo {title} {Quantum optimal control in quantum technologies. strategic report on current status, visions and goals for research in {E}urope},\ }\href {https://doi.org/10.1140/epjqt/s40507-022-00138-x} {\bibfield  {journal} {\bibinfo  {journal} {EPJ Quantum
  Technol.}\ }\textbf {\bibinfo {volume} {9}},\ \bibinfo {pages} {19} (\bibinfo {year} {2022})}\BibitemShut {NoStop}%
\bibitem [{\citenamefont {d'Alessandro}(2007)}]{d2007introduction}%
  \BibitemOpen
  \bibfield  {author} {\bibinfo {author} {\bibfnamefont {D.}~\bibnamefont {d'Alessandro}},\ }\href@noop {} {\emph {\bibinfo {title} {Introduction to quantum control and dynamics}}}\ (\bibinfo  {publisher} {Chapman and Hall/CRC},\ \bibinfo {year} {2007})\BibitemShut {NoStop}%
\bibitem [{\citenamefont {Glaser}\ \emph {et~al.}(2015)\citenamefont {Glaser}, \citenamefont {Boscain}, \citenamefont {Calarco}, \citenamefont {Koch}, \citenamefont {K\"ockenberger}, \citenamefont {Kosloff}, \citenamefont {Kuprov}, \citenamefont {Luy}, \citenamefont {Schirmer}, \citenamefont {Schulte-Herbr\"uggen}, \citenamefont {Sugny},\ and\ \citenamefont {Wilhelm}}]{glaser2015training}%
  \BibitemOpen
  \bibfield  {author} {\bibinfo {author} {\bibfnamefont {S.~J.}\ \bibnamefont {Glaser}}, \bibinfo {author} {\bibfnamefont {U.}~\bibnamefont {Boscain}}, \bibinfo {author} {\bibfnamefont {T.}~\bibnamefont {Calarco}}, \bibinfo {author} {\bibfnamefont {C.~P.}\ \bibnamefont {Koch}}, \bibinfo {author} {\bibfnamefont {W.}~\bibnamefont {K\"ockenberger}}, \bibinfo {author} {\bibfnamefont {R.}~\bibnamefont {Kosloff}}, \bibinfo {author} {\bibfnamefont {I.}~\bibnamefont {Kuprov}}, \bibinfo {author} {\bibfnamefont {B.}~\bibnamefont {Luy}}, \bibinfo {author} {\bibfnamefont {S.}~\bibnamefont {Schirmer}}, \bibinfo {author} {\bibfnamefont {T.}~\bibnamefont {Schulte-Herbr\"uggen}}, \bibinfo {author} {\bibfnamefont {D.}~\bibnamefont {Sugny}},\ and\ \bibinfo {author} {\bibfnamefont {F.~K.}\ \bibnamefont {Wilhelm}},\ }\bibfield  {title} {\bibinfo {title} {Training {S}chr{\"o}dinger’s cat: Quantum optimal control},\ }\href {https://doi.org/10.1140/epjd/e2015-60464-1} {\bibfield  {journal} {\bibinfo  {journal} {Eur. Phys. J. D}\
  }\textbf {\bibinfo {volume} {69}},\ \bibinfo {pages} {1} (\bibinfo {year} {2015})}\BibitemShut {NoStop}%
\bibitem [{\citenamefont {Skinner}\ \emph {et~al.}(2003)\citenamefont {Skinner}, \citenamefont {Reiss}, \citenamefont {Luy}, \citenamefont {Khaneja},\ and\ \citenamefont {Glaser}}]{skinner2003application}%
  \BibitemOpen
  \bibfield  {author} {\bibinfo {author} {\bibfnamefont {T.~E.}\ \bibnamefont {Skinner}}, \bibinfo {author} {\bibfnamefont {T.~O.}\ \bibnamefont {Reiss}}, \bibinfo {author} {\bibfnamefont {B.}~\bibnamefont {Luy}}, \bibinfo {author} {\bibfnamefont {N.}~\bibnamefont {Khaneja}},\ and\ \bibinfo {author} {\bibfnamefont {S.~J.}\ \bibnamefont {Glaser}},\ }\bibfield  {title} {\bibinfo {title} {Application of optimal control theory to the design of broadband excitation pulses for high-resolution {NMR}},\ }\href {https://doi.org/https://doi.org/10.1016/S1090-7807(03)00153-8} {\bibfield  {journal} {\bibinfo  {journal} {J. Magn. Reson.}\ }\textbf {\bibinfo {volume} {163}},\ \bibinfo {pages} {8} (\bibinfo {year} {2003})}\BibitemShut {NoStop}%
\bibitem [{\citenamefont {Khaneja}\ \emph {et~al.}(2005)\citenamefont {Khaneja}, \citenamefont {Reiss}, \citenamefont {Kehlet}, \citenamefont {Schulte-Herbr{\"u}ggen},\ and\ \citenamefont {Glaser}}]{khaneja2005optimal}%
  \BibitemOpen
  \bibfield  {author} {\bibinfo {author} {\bibfnamefont {N.}~\bibnamefont {Khaneja}}, \bibinfo {author} {\bibfnamefont {T.}~\bibnamefont {Reiss}}, \bibinfo {author} {\bibfnamefont {C.}~\bibnamefont {Kehlet}}, \bibinfo {author} {\bibfnamefont {T.}~\bibnamefont {Schulte-Herbr{\"u}ggen}},\ and\ \bibinfo {author} {\bibfnamefont {S.~J.}\ \bibnamefont {Glaser}},\ }\bibfield  {title} {\bibinfo {title} {Optimal control of coupled spin dynamics: design of {NMR} pulse sequences by gradient ascent algorithms},\ }\href {https://doi.org/https://doi.org/10.1016/j.jmr.2004.11.004} {\bibfield  {journal} {\bibinfo  {journal} {J. Magn. Reson.}\ }\textbf {\bibinfo {volume} {172}},\ \bibinfo {pages} {296} (\bibinfo {year} {2005})}\BibitemShut {NoStop}%
\bibitem [{\citenamefont {Koch}\ \emph {et~al.}(2004)\citenamefont {Koch}, \citenamefont {Palao}, \citenamefont {Kosloff},\ and\ \citenamefont {Masnou-Seeuws}}]{koch2004stabilization}%
  \BibitemOpen
  \bibfield  {author} {\bibinfo {author} {\bibfnamefont {C.~P.}\ \bibnamefont {Koch}}, \bibinfo {author} {\bibfnamefont {J.~P.}\ \bibnamefont {Palao}}, \bibinfo {author} {\bibfnamefont {R.}~\bibnamefont {Kosloff}},\ and\ \bibinfo {author} {\bibfnamefont {F.}~\bibnamefont {Masnou-Seeuws}},\ }\bibfield  {title} {\bibinfo {title} {Stabilization of ultracold molecules using optimal control theory},\ }\href {https://doi.org/10.1103/PhysRevA.70.013402} {\bibfield  {journal} {\bibinfo  {journal} {Phys. Rev. A}\ }\textbf {\bibinfo {volume} {70}},\ \bibinfo {pages} {013402} (\bibinfo {year} {2004})}\BibitemShut {NoStop}%
\bibitem [{\citenamefont {Reich}\ and\ \citenamefont {Koch}(2013)}]{Reich_2013}%
  \BibitemOpen
  \bibfield  {author} {\bibinfo {author} {\bibfnamefont {D.~M.}\ \bibnamefont {Reich}}\ and\ \bibinfo {author} {\bibfnamefont {C.~P.}\ \bibnamefont {Koch}},\ }\bibfield  {title} {\bibinfo {title} {Cooling molecular vibrations with shaped laser pulses: optimal control theory exploiting the timescale separation between coherent excitation and spontaneous emission},\ }\href {https://doi.org/10.1088/1367-2630/15/12/125028} {\bibfield  {journal} {\bibinfo  {journal} {New Journal of Physics}\ }\textbf {\bibinfo {volume} {15}},\ \bibinfo {pages} {125028} (\bibinfo {year} {2013})}\BibitemShut {NoStop}%
\bibitem [{\citenamefont {Nielsen}\ and\ \citenamefont {Chuang}(2010)}]{nielsenchuang2010}%
  \BibitemOpen
  \bibfield  {author} {\bibinfo {author} {\bibfnamefont {M.}~\bibnamefont {Nielsen}}\ and\ \bibinfo {author} {\bibfnamefont {I.}~\bibnamefont {Chuang}},\ }\href@noop {} {\emph {\bibinfo {title} {Quantum Computation and Quantum Information: 10th Anniversary Edition}}}\ (\bibinfo  {publisher} {Cambridge University Press},\ \bibinfo {year} {2010})\BibitemShut {NoStop}%
\bibitem [{\citenamefont {Schulte-Herbr{\"u}ggen}\ \emph {et~al.}(2005)\citenamefont {Schulte-Herbr{\"u}ggen}, \citenamefont {Sp{\"o}rl}, \citenamefont {Khaneja},\ and\ \citenamefont {Glaser}}]{schulte2005optimal}%
  \BibitemOpen
  \bibfield  {author} {\bibinfo {author} {\bibfnamefont {T.}~\bibnamefont {Schulte-Herbr{\"u}ggen}}, \bibinfo {author} {\bibfnamefont {A.}~\bibnamefont {Sp{\"o}rl}}, \bibinfo {author} {\bibfnamefont {N.}~\bibnamefont {Khaneja}},\ and\ \bibinfo {author} {\bibfnamefont {S.~J.}\ \bibnamefont {Glaser}},\ }\bibfield  {title} {\bibinfo {title} {Optimal control-based efficient synthesis of building blocks of quantum algorithms: A perspective from network complexity towards time complexity},\ }\href {https://doi.org/10.1103/PhysRevA.72.042331} {\bibfield  {journal} {\bibinfo  {journal} {Phys. Rev. A}\ }\textbf {\bibinfo {volume} {72}},\ \bibinfo {pages} {042331} (\bibinfo {year} {2005})}\BibitemShut {NoStop}%
\bibitem [{\citenamefont {Waldherr}\ \emph {et~al.}(2014)\citenamefont {Waldherr}, \citenamefont {Wang}, \citenamefont {Zaiser}, \citenamefont {Jamali}, \citenamefont {Schulte-Herbr\"uggen}, \citenamefont {Abe}, \citenamefont {Ohshima}, \citenamefont {Isoya}, \citenamefont {Du}, \citenamefont {Neumann},\ and\ \citenamefont {Wrachtrup}}]{Waldherr_2014_NV}%
  \BibitemOpen
  \bibfield  {author} {\bibinfo {author} {\bibfnamefont {G.}~\bibnamefont {Waldherr}}, \bibinfo {author} {\bibfnamefont {Y.}~\bibnamefont {Wang}}, \bibinfo {author} {\bibfnamefont {S.}~\bibnamefont {Zaiser}}, \bibinfo {author} {\bibfnamefont {M.}~\bibnamefont {Jamali}}, \bibinfo {author} {\bibfnamefont {T.}~\bibnamefont {Schulte-Herbr\"uggen}}, \bibinfo {author} {\bibfnamefont {H.}~\bibnamefont {Abe}}, \bibinfo {author} {\bibfnamefont {T.}~\bibnamefont {Ohshima}}, \bibinfo {author} {\bibfnamefont {J.}~\bibnamefont {Isoya}}, \bibinfo {author} {\bibfnamefont {J.~F.}\ \bibnamefont {Du}}, \bibinfo {author} {\bibfnamefont {P.}~\bibnamefont {Neumann}},\ and\ \bibinfo {author} {\bibfnamefont {J.}~\bibnamefont {Wrachtrup}},\ }\bibfield  {title} {\bibinfo {title} {Quantum error correction in a solid-state hybrid spin register},\ }\href {https://doi.org/10.1038/nature12919} {\bibfield  {journal} {\bibinfo  {journal} {Nature}\ }\textbf {\bibinfo {volume} {506}},\ \bibinfo {pages} {204} (\bibinfo {year}
  {2014})}\BibitemShut {NoStop}%
\bibitem [{\citenamefont {Garc\'{\i}a-Ripoll}\ \emph {et~al.}(2003)\citenamefont {Garc\'{\i}a-Ripoll}, \citenamefont {Zoller},\ and\ \citenamefont {Cirac}}]{garcia2003speed}%
  \BibitemOpen
  \bibfield  {author} {\bibinfo {author} {\bibfnamefont {J.~J.}\ \bibnamefont {Garc\'{\i}a-Ripoll}}, \bibinfo {author} {\bibfnamefont {P.}~\bibnamefont {Zoller}},\ and\ \bibinfo {author} {\bibfnamefont {J.~I.}\ \bibnamefont {Cirac}},\ }\bibfield  {title} {\bibinfo {title} {Speed optimized two-qubit gates with laser coherent control techniques for ion trap quantum computing},\ }\href {https://doi.org/10.1103/PhysRevLett.91.157901} {\bibfield  {journal} {\bibinfo  {journal} {Phys. Rev. Lett.}\ }\textbf {\bibinfo {volume} {91}},\ \bibinfo {pages} {157901} (\bibinfo {year} {2003})}\BibitemShut {NoStop}%
\bibitem [{\citenamefont {Garc\'{\i}a-Ripoll}\ \emph {et~al.}(2005)\citenamefont {Garc\'{\i}a-Ripoll}, \citenamefont {Zoller},\ and\ \citenamefont {Cirac}}]{garcia2005coherent}%
  \BibitemOpen
  \bibfield  {author} {\bibinfo {author} {\bibfnamefont {J.~J.}\ \bibnamefont {Garc\'{\i}a-Ripoll}}, \bibinfo {author} {\bibfnamefont {P.}~\bibnamefont {Zoller}},\ and\ \bibinfo {author} {\bibfnamefont {J.~I.}\ \bibnamefont {Cirac}},\ }\bibfield  {title} {\bibinfo {title} {Coherent control of trapped ions using off-resonant lasers},\ }\href {https://doi.org/10.1103/PhysRevA.71.062309} {\bibfield  {journal} {\bibinfo  {journal} {Phys. Rev. A}\ }\textbf {\bibinfo {volume} {71}},\ \bibinfo {pages} {062309} (\bibinfo {year} {2005})}\BibitemShut {NoStop}%
\bibitem [{\citenamefont {Choi}\ \emph {et~al.}(2014)\citenamefont {Choi}, \citenamefont {Debnath}, \citenamefont {Manning}, \citenamefont {Figgatt}, \citenamefont {Gong}, \citenamefont {Duan},\ and\ \citenamefont {Monroe}}]{manning2014optimal}%
  \BibitemOpen
  \bibfield  {author} {\bibinfo {author} {\bibfnamefont {T.}~\bibnamefont {Choi}}, \bibinfo {author} {\bibfnamefont {S.}~\bibnamefont {Debnath}}, \bibinfo {author} {\bibfnamefont {T.~A.}\ \bibnamefont {Manning}}, \bibinfo {author} {\bibfnamefont {C.}~\bibnamefont {Figgatt}}, \bibinfo {author} {\bibfnamefont {Z.-X.}\ \bibnamefont {Gong}}, \bibinfo {author} {\bibfnamefont {L.-M.}\ \bibnamefont {Duan}},\ and\ \bibinfo {author} {\bibfnamefont {C.}~\bibnamefont {Monroe}},\ }\bibfield  {title} {\bibinfo {title} {Optimal quantum control of multimode couplings between trapped ion qubits for scalable entanglement},\ }\href {https://doi.org/10.1103/PhysRevLett.112.190502} {\bibfield  {journal} {\bibinfo  {journal} {Phys. Rev. Lett.}\ }\textbf {\bibinfo {volume} {112}},\ \bibinfo {pages} {190502} (\bibinfo {year} {2014})}\BibitemShut {NoStop}%
\bibitem [{\citenamefont {Leibfried}\ \emph {et~al.}(2005)\citenamefont {Leibfried}, \citenamefont {Knill}, \citenamefont {Seidelin}, \citenamefont {Britton}, \citenamefont {Blakestad}, \citenamefont {Chiaverini}, \citenamefont {Hume}, \citenamefont {Itano}, \citenamefont {Jost}, \citenamefont {Langer}, \citenamefont {Ozeri}, \citenamefont {Reichle},\ and\ \citenamefont {Wineland}}]{leibfried2005creation}%
  \BibitemOpen
  \bibfield  {author} {\bibinfo {author} {\bibfnamefont {D.}~\bibnamefont {Leibfried}}, \bibinfo {author} {\bibfnamefont {E.}~\bibnamefont {Knill}}, \bibinfo {author} {\bibfnamefont {S.}~\bibnamefont {Seidelin}}, \bibinfo {author} {\bibfnamefont {J.}~\bibnamefont {Britton}}, \bibinfo {author} {\bibfnamefont {R.~B.}\ \bibnamefont {Blakestad}}, \bibinfo {author} {\bibfnamefont {J.}~\bibnamefont {Chiaverini}}, \bibinfo {author} {\bibfnamefont {D.~B.}\ \bibnamefont {Hume}}, \bibinfo {author} {\bibfnamefont {W.~M.}\ \bibnamefont {Itano}}, \bibinfo {author} {\bibfnamefont {J.~D.}\ \bibnamefont {Jost}}, \bibinfo {author} {\bibfnamefont {C.}~\bibnamefont {Langer}}, \bibinfo {author} {\bibfnamefont {R.}~\bibnamefont {Ozeri}}, \bibinfo {author} {\bibfnamefont {R.}~\bibnamefont {Reichle}},\ and\ \bibinfo {author} {\bibfnamefont {D.~J.}\ \bibnamefont {Wineland}},\ }\bibfield  {title} {\bibinfo {title} {Creation of a six-atom ‘{S}chr{\"o}dinger cat’ state},\ }\href {https://doi.org/10.1038/nature04251} {\bibfield
  {journal} {\bibinfo  {journal} {Nature}\ }\textbf {\bibinfo {volume} {438}},\ \bibinfo {pages} {639} (\bibinfo {year} {2005})}\BibitemShut {NoStop}%
\bibitem [{\citenamefont {Goerz}\ \emph {et~al.}(2014)\citenamefont {Goerz}, \citenamefont {Halperin}, \citenamefont {Aytac}, \citenamefont {Koch},\ and\ \citenamefont {Whaley}}]{goerz2014robustness}%
  \BibitemOpen
  \bibfield  {author} {\bibinfo {author} {\bibfnamefont {M.~H.}\ \bibnamefont {Goerz}}, \bibinfo {author} {\bibfnamefont {E.~J.}\ \bibnamefont {Halperin}}, \bibinfo {author} {\bibfnamefont {J.~M.}\ \bibnamefont {Aytac}}, \bibinfo {author} {\bibfnamefont {C.~P.}\ \bibnamefont {Koch}},\ and\ \bibinfo {author} {\bibfnamefont {K.~B.}\ \bibnamefont {Whaley}},\ }\bibfield  {title} {\bibinfo {title} {Robustness of high-fidelity {R}ydberg gates with single-site addressability},\ }\href {https://doi.org/10.1103/PhysRevA.90.032329} {\bibfield  {journal} {\bibinfo  {journal} {Phys. Rev. A}\ }\textbf {\bibinfo {volume} {90}},\ \bibinfo {pages} {032329} (\bibinfo {year} {2014})}\BibitemShut {NoStop}%
\bibitem [{\citenamefont {Omran}\ \emph {et~al.}(2019)\citenamefont {Omran} \emph {et~al.}}]{omran2019generation}%
  \BibitemOpen
  \bibfield  {author} {\bibinfo {author} {\bibfnamefont {A.}~\bibnamefont {Omran}} \emph {et~al.},\ }\bibfield  {title} {\bibinfo {title} {Generation and manipulation of {S}chr{\"o}dinger cat states in {R}ydberg atom arrays},\ }\href {https://doi.org/10.1126/science.aax9743} {\bibfield  {journal} {\bibinfo  {journal} {Science}\ }\textbf {\bibinfo {volume} {365}},\ \bibinfo {pages} {570} (\bibinfo {year} {2019})}\BibitemShut {NoStop}%
\bibitem [{\citenamefont {Rojan}\ \emph {et~al.}(2014)\citenamefont {Rojan}, \citenamefont {Reich}, \citenamefont {Dotsenko}, \citenamefont {Raimond}, \citenamefont {Koch},\ and\ \citenamefont {Morigi}}]{rojan_stateprep_2014}%
  \BibitemOpen
  \bibfield  {author} {\bibinfo {author} {\bibfnamefont {K.}~\bibnamefont {Rojan}}, \bibinfo {author} {\bibfnamefont {D.~M.}\ \bibnamefont {Reich}}, \bibinfo {author} {\bibfnamefont {I.}~\bibnamefont {Dotsenko}}, \bibinfo {author} {\bibfnamefont {J.-M.}\ \bibnamefont {Raimond}}, \bibinfo {author} {\bibfnamefont {C.~P.}\ \bibnamefont {Koch}},\ and\ \bibinfo {author} {\bibfnamefont {G.}~\bibnamefont {Morigi}},\ }\bibfield  {title} {\bibinfo {title} {Arbitrary-quantum-state preparation of a harmonic oscillator via optimal control},\ }\href {https://doi.org/10.1103/PhysRevA.90.023824} {\bibfield  {journal} {\bibinfo  {journal} {Phys. Rev. A}\ }\textbf {\bibinfo {volume} {90}},\ \bibinfo {pages} {023824} (\bibinfo {year} {2014})}\BibitemShut {NoStop}%
\bibitem [{\citenamefont {Moll}\ \emph {et~al.}(2018)\citenamefont {Moll} \emph {et~al.}}]{moll2018quantum}%
  \BibitemOpen
  \bibfield  {author} {\bibinfo {author} {\bibfnamefont {N.}~\bibnamefont {Moll}} \emph {et~al.},\ }\bibfield  {title} {\bibinfo {title} {Quantum optimization using variational algorithms on near-term quantum devices},\ }\href {https://doi.org/10.1088/2058-9565/aab822} {\bibfield  {journal} {\bibinfo  {journal} {Quantum Sci. Technol.}\ }\textbf {\bibinfo {volume} {3}},\ \bibinfo {pages} {030503} (\bibinfo {year} {2018})}\BibitemShut {NoStop}%
\bibitem [{\citenamefont {Choquette}\ \emph {et~al.}(2021)\citenamefont {Choquette}, \citenamefont {Di~Paolo}, \citenamefont {Barkoutsos}, \citenamefont {S{\'e}n{\'e}chal}, \citenamefont {Tavernelli},\ and\ \citenamefont {Blais}}]{choquette2021quantum}%
  \BibitemOpen
  \bibfield  {author} {\bibinfo {author} {\bibfnamefont {A.}~\bibnamefont {Choquette}}, \bibinfo {author} {\bibfnamefont {A.}~\bibnamefont {Di~Paolo}}, \bibinfo {author} {\bibfnamefont {P.~K.}\ \bibnamefont {Barkoutsos}}, \bibinfo {author} {\bibfnamefont {D.}~\bibnamefont {S{\'e}n{\'e}chal}}, \bibinfo {author} {\bibfnamefont {I.}~\bibnamefont {Tavernelli}},\ and\ \bibinfo {author} {\bibfnamefont {A.}~\bibnamefont {Blais}},\ }\bibfield  {title} {\bibinfo {title} {Quantum-optimal-control-inspired ansatz for variational quantum algorithms},\ }\href {https://doi.org/10.1103/PhysRevResearch.3.023092} {\bibfield  {journal} {\bibinfo  {journal} {Phys. Rev. Res.}\ }\textbf {\bibinfo {volume} {3}},\ \bibinfo {pages} {023092} (\bibinfo {year} {2021})}\BibitemShut {NoStop}%
\bibitem [{\citenamefont {Albash}\ and\ \citenamefont {Lidar}(2018)}]{albash2018adiabatic}%
  \BibitemOpen
  \bibfield  {author} {\bibinfo {author} {\bibfnamefont {T.}~\bibnamefont {Albash}}\ and\ \bibinfo {author} {\bibfnamefont {D.~A.}\ \bibnamefont {Lidar}},\ }\bibfield  {title} {\bibinfo {title} {Adiabatic quantum computation},\ }\href {https://doi.org/10.1103/RevModPhys.90.015002} {\bibfield  {journal} {\bibinfo  {journal} {Rev. Mod. Phys.}\ }\textbf {\bibinfo {volume} {90}},\ \bibinfo {pages} {015002} (\bibinfo {year} {2018})}\BibitemShut {NoStop}%
\bibitem [{\citenamefont {Hauke}\ \emph {et~al.}(2020)\citenamefont {Hauke}, \citenamefont {Katzgraber}, \citenamefont {Lechner}, \citenamefont {Nishimori},\ and\ \citenamefont {Oliver}}]{hauke2020perspectives}%
  \BibitemOpen
  \bibfield  {author} {\bibinfo {author} {\bibfnamefont {P.}~\bibnamefont {Hauke}}, \bibinfo {author} {\bibfnamefont {H.~G.}\ \bibnamefont {Katzgraber}}, \bibinfo {author} {\bibfnamefont {W.}~\bibnamefont {Lechner}}, \bibinfo {author} {\bibfnamefont {H.}~\bibnamefont {Nishimori}},\ and\ \bibinfo {author} {\bibfnamefont {W.~D.}\ \bibnamefont {Oliver}},\ }\bibfield  {title} {\bibinfo {title} {Perspectives of quantum annealing: Methods and implementations},\ }\href {https://doi.org/10.1088/1361-6633/ab85b8} {\bibfield  {journal} {\bibinfo  {journal} {Rep. Progr. Phys.}\ }\textbf {\bibinfo {volume} {83}},\ \bibinfo {pages} {054401} (\bibinfo {year} {2020})}\BibitemShut {NoStop}%
\bibitem [{\citenamefont {Blais}\ \emph {et~al.}(2021)\citenamefont {Blais}, \citenamefont {Grimsmo}, \citenamefont {Girvin},\ and\ \citenamefont {Wallraff}}]{blais2021circuit}%
  \BibitemOpen
  \bibfield  {author} {\bibinfo {author} {\bibfnamefont {A.}~\bibnamefont {Blais}}, \bibinfo {author} {\bibfnamefont {A.~L.}\ \bibnamefont {Grimsmo}}, \bibinfo {author} {\bibfnamefont {S.~M.}\ \bibnamefont {Girvin}},\ and\ \bibinfo {author} {\bibfnamefont {A.}~\bibnamefont {Wallraff}},\ }\bibfield  {title} {\bibinfo {title} {Circuit quantum electrodynamics},\ }\href {https://doi.org/10.1103/RevModPhys.93.025005} {\bibfield  {journal} {\bibinfo  {journal} {Rev. Mod. Phys}\ }\textbf {\bibinfo {volume} {93}},\ \bibinfo {pages} {025005} (\bibinfo {year} {2021})}\BibitemShut {NoStop}%
\bibitem [{\citenamefont {Goerz}\ \emph {et~al.}(2017)\citenamefont {Goerz}, \citenamefont {Motzoi}, \citenamefont {Whaley},\ and\ \citenamefont {Koch}}]{goerz2017charting}%
  \BibitemOpen
  \bibfield  {author} {\bibinfo {author} {\bibfnamefont {M.~H.}\ \bibnamefont {Goerz}}, \bibinfo {author} {\bibfnamefont {F.}~\bibnamefont {Motzoi}}, \bibinfo {author} {\bibfnamefont {K.~B.}\ \bibnamefont {Whaley}},\ and\ \bibinfo {author} {\bibfnamefont {C.~P.}\ \bibnamefont {Koch}},\ }\bibfield  {title} {\bibinfo {title} {Charting the circuit {QED} design landscape using optimal control theory},\ }\href {https://doi.org/10.1038/s41534-017-0036-0} {\bibfield  {journal} {\bibinfo  {journal} {Npj Quantum Inf.}\ }\textbf {\bibinfo {volume} {3}},\ \bibinfo {pages} {37} (\bibinfo {year} {2017})}\BibitemShut {NoStop}%
\bibitem [{\citenamefont {Setiawan}\ \emph {et~al.}(2021)\citenamefont {Setiawan}, \citenamefont {Groszkowski}, \citenamefont {Ribeiro},\ and\ \citenamefont {Clerk}}]{setiawan_2021}%
  \BibitemOpen
  \bibfield  {author} {\bibinfo {author} {\bibfnamefont {F.}~\bibnamefont {Setiawan}}, \bibinfo {author} {\bibfnamefont {P.}~\bibnamefont {Groszkowski}}, \bibinfo {author} {\bibfnamefont {H.}~\bibnamefont {Ribeiro}},\ and\ \bibinfo {author} {\bibfnamefont {A.~A.}\ \bibnamefont {Clerk}},\ }\bibfield  {title} {\bibinfo {title} {Analytic design of accelerated adiabatic gates in realistic qubits: General theory and applications to superconducting circuits},\ }\href {https://doi.org/10.1103/PRXQuantum.2.030306} {\bibfield  {journal} {\bibinfo  {journal} {PRX Quantum}\ }\textbf {\bibinfo {volume} {2}},\ \bibinfo {pages} {030306} (\bibinfo {year} {2021})}\BibitemShut {NoStop}%
\bibitem [{\citenamefont {Werninghaus}\ \emph {et~al.}(2021)\citenamefont {Werninghaus}, \citenamefont {Egger}, \citenamefont {Roy}, \citenamefont {Machnes}, \citenamefont {Wilhelm},\ and\ \citenamefont {Filipp}}]{werninghaus2021leakage}%
  \BibitemOpen
  \bibfield  {author} {\bibinfo {author} {\bibfnamefont {M.}~\bibnamefont {Werninghaus}}, \bibinfo {author} {\bibfnamefont {D.~J.}\ \bibnamefont {Egger}}, \bibinfo {author} {\bibfnamefont {F.}~\bibnamefont {Roy}}, \bibinfo {author} {\bibfnamefont {S.}~\bibnamefont {Machnes}}, \bibinfo {author} {\bibfnamefont {F.~K.}\ \bibnamefont {Wilhelm}},\ and\ \bibinfo {author} {\bibfnamefont {S.}~\bibnamefont {Filipp}},\ }\bibfield  {title} {\bibinfo {title} {Leakage reduction in fast superconducting qubit gates via optimal control},\ }\href {https://doi.org/10.1038/s41534-020-00346-2} {\bibfield  {journal} {\bibinfo  {journal} {Npj Quantum Inf.}\ }\textbf {\bibinfo {volume} {7}},\ \bibinfo {pages} {14} (\bibinfo {year} {2021})}\BibitemShut {NoStop}%
\bibitem [{\citenamefont {Doria}\ \emph {et~al.}(2011)\citenamefont {Doria}, \citenamefont {Calarco},\ and\ \citenamefont {Montangero}}]{doria2011optimal}%
  \BibitemOpen
  \bibfield  {author} {\bibinfo {author} {\bibfnamefont {P.}~\bibnamefont {Doria}}, \bibinfo {author} {\bibfnamefont {T.}~\bibnamefont {Calarco}},\ and\ \bibinfo {author} {\bibfnamefont {S.}~\bibnamefont {Montangero}},\ }\bibfield  {title} {\bibinfo {title} {Optimal control technique for many-body quantum dynamics},\ }\href {https://doi.org/10.1103/PhysRevLett.106.190501} {\bibfield  {journal} {\bibinfo  {journal} {Phys. Rev. Lett.}\ }\textbf {\bibinfo {volume} {106}},\ \bibinfo {pages} {190501} (\bibinfo {year} {2011})}\BibitemShut {NoStop}%
\bibitem [{\citenamefont {Caneva}\ \emph {et~al.}(2011)\citenamefont {Caneva}, \citenamefont {Calarco},\ and\ \citenamefont {Montangero}}]{caneva2011chopped}%
  \BibitemOpen
  \bibfield  {author} {\bibinfo {author} {\bibfnamefont {T.}~\bibnamefont {Caneva}}, \bibinfo {author} {\bibfnamefont {T.}~\bibnamefont {Calarco}},\ and\ \bibinfo {author} {\bibfnamefont {S.}~\bibnamefont {Montangero}},\ }\bibfield  {title} {\bibinfo {title} {Chopped random-basis quantum optimization},\ }\href {https://doi.org/10.1103/PhysRevA.84.022326} {\bibfield  {journal} {\bibinfo  {journal} {Phys. Rev. A}\ }\textbf {\bibinfo {volume} {84}},\ \bibinfo {pages} {022326} (\bibinfo {year} {2011})}\BibitemShut {NoStop}%
\bibitem [{\citenamefont {M{\"u}ller}\ \emph {et~al.}(2022)\citenamefont {M{\"u}ller}, \citenamefont {Said}, \citenamefont {Jelezko}, \citenamefont {Calarco},\ and\ \citenamefont {Montangero}}]{muller2022one}%
  \BibitemOpen
  \bibfield  {author} {\bibinfo {author} {\bibfnamefont {M.~M.}\ \bibnamefont {M{\"u}ller}}, \bibinfo {author} {\bibfnamefont {R.~S.}\ \bibnamefont {Said}}, \bibinfo {author} {\bibfnamefont {F.}~\bibnamefont {Jelezko}}, \bibinfo {author} {\bibfnamefont {T.}~\bibnamefont {Calarco}},\ and\ \bibinfo {author} {\bibfnamefont {S.}~\bibnamefont {Montangero}},\ }\bibfield  {title} {\bibinfo {title} {One decade of quantum optimal control in the chopped random basis},\ }\href {https://doi.org/10.1088/1361-6633/ac723c} {\bibfield  {journal} {\bibinfo  {journal} {Rep. Prog. Phys.}\ }\textbf {\bibinfo {volume} {85}},\ \bibinfo {pages} {076001} (\bibinfo {year} {2022})}\BibitemShut {NoStop}%
\bibitem [{\citenamefont {Torrontegui}\ \emph {et~al.}(2013)\citenamefont {Torrontegui}, \citenamefont {Ibáñez}, \citenamefont {Martínez-Garaot}, \citenamefont {Modugno}, \citenamefont {{del Campo}}, \citenamefont {Guéry-Odelin}, \citenamefont {Ruschhaupt}, \citenamefont {Chen},\ and\ \citenamefont {Muga}}]{torrontegui2013shortcuts}%
  \BibitemOpen
  \bibfield  {author} {\bibinfo {author} {\bibfnamefont {E.}~\bibnamefont {Torrontegui}}, \bibinfo {author} {\bibfnamefont {S.}~\bibnamefont {Ibáñez}}, \bibinfo {author} {\bibfnamefont {S.}~\bibnamefont {Martínez-Garaot}}, \bibinfo {author} {\bibfnamefont {M.}~\bibnamefont {Modugno}}, \bibinfo {author} {\bibfnamefont {A.}~\bibnamefont {{del Campo}}}, \bibinfo {author} {\bibfnamefont {D.}~\bibnamefont {Guéry-Odelin}}, \bibinfo {author} {\bibfnamefont {A.}~\bibnamefont {Ruschhaupt}}, \bibinfo {author} {\bibfnamefont {X.}~\bibnamefont {Chen}},\ and\ \bibinfo {author} {\bibfnamefont {J.~G.}\ \bibnamefont {Muga}},\ }\bibfield  {title} {\bibinfo {title} {Shortcuts to adiabaticity},\ }in\ \href {https://doi.org/https://doi.org/10.1016/B978-0-12-408090-4.00002-5} {\emph {\bibinfo {booktitle} {Adv. At. Mol. Opt.}}},\ Vol.~\bibinfo {volume} {62}\ (\bibinfo  {publisher} {Elsevier},\ \bibinfo {year} {2013})\ pp.\ \bibinfo {pages} {117--169}\BibitemShut {NoStop}%
\bibitem [{\citenamefont {Gu\'ery-Odelin}\ \emph {et~al.}(2019)\citenamefont {Gu\'ery-Odelin}, \citenamefont {Ruschhaupt}, \citenamefont {Kiely}, \citenamefont {Torrontegui}, \citenamefont {Mart\'{\i}nez-Garaot},\ and\ \citenamefont {Muga}}]{guery2019shortcuts}%
  \BibitemOpen
  \bibfield  {author} {\bibinfo {author} {\bibfnamefont {D.}~\bibnamefont {Gu\'ery-Odelin}}, \bibinfo {author} {\bibfnamefont {A.}~\bibnamefont {Ruschhaupt}}, \bibinfo {author} {\bibfnamefont {A.}~\bibnamefont {Kiely}}, \bibinfo {author} {\bibfnamefont {E.}~\bibnamefont {Torrontegui}}, \bibinfo {author} {\bibfnamefont {S.}~\bibnamefont {Mart\'{\i}nez-Garaot}},\ and\ \bibinfo {author} {\bibfnamefont {J.~G.}\ \bibnamefont {Muga}},\ }\bibfield  {title} {\bibinfo {title} {Shortcuts to adiabaticity: Concepts, methods, and applications},\ }\href {https://doi.org/10.1103/RevModPhys.91.045001} {\bibfield  {journal} {\bibinfo  {journal} {Rev. Mod. Phys.}\ }\textbf {\bibinfo {volume} {91}},\ \bibinfo {pages} {045001} (\bibinfo {year} {2019})}\BibitemShut {NoStop}%
\bibitem [{\citenamefont {Demirplak}\ and\ \citenamefont {Rice}(2003)}]{demirplak2003adiabatic}%
  \BibitemOpen
  \bibfield  {author} {\bibinfo {author} {\bibfnamefont {M.}~\bibnamefont {Demirplak}}\ and\ \bibinfo {author} {\bibfnamefont {S.~A.}\ \bibnamefont {Rice}},\ }\bibfield  {title} {\bibinfo {title} {Adiabatic population transfer with control fields},\ }\href {https://doi.org/doi: 10.1021/jp030708a} {\bibfield  {journal} {\bibinfo  {journal} {J. Phys. Chem. A}\ }\textbf {\bibinfo {volume} {107}},\ \bibinfo {pages} {9937} (\bibinfo {year} {2003})}\BibitemShut {NoStop}%
\bibitem [{\citenamefont {Demirplak}\ and\ \citenamefont {Rice}(2005)}]{demirplak2005assisted}%
  \BibitemOpen
  \bibfield  {author} {\bibinfo {author} {\bibfnamefont {M.}~\bibnamefont {Demirplak}}\ and\ \bibinfo {author} {\bibfnamefont {S.~A.}\ \bibnamefont {Rice}},\ }\bibfield  {title} {\bibinfo {title} {Assisted adiabatic passage revisited},\ }\href {https://doi.org/https://doi.org/10.1021/jp040647w} {\bibfield  {journal} {\bibinfo  {journal} {J. Phys. Chem. B}\ }\textbf {\bibinfo {volume} {109}},\ \bibinfo {pages} {6838} (\bibinfo {year} {2005})}\BibitemShut {NoStop}%
\bibitem [{\citenamefont {Berry}(2009)}]{berry2009transitionless}%
  \BibitemOpen
  \bibfield  {author} {\bibinfo {author} {\bibfnamefont {M.~V.}\ \bibnamefont {Berry}},\ }\bibfield  {title} {\bibinfo {title} {Transitionless quantum driving},\ }\href {https://doi.org/10.1088/1751-8113/42/36/365303} {\bibfield  {journal} {\bibinfo  {journal} {J. Phys. A Math. Theor.}\ }\textbf {\bibinfo {volume} {42}},\ \bibinfo {pages} {365303} (\bibinfo {year} {2009})}\BibitemShut {NoStop}%
\bibitem [{\citenamefont {del Campo}(2013)}]{del2013shortcuts}%
  \BibitemOpen
  \bibfield  {author} {\bibinfo {author} {\bibfnamefont {A.}~\bibnamefont {del Campo}},\ }\bibfield  {title} {\bibinfo {title} {Shortcuts to adiabaticity by counterdiabatic driving},\ }\href {https://doi.org/10.1103/PhysRevLett.111.100502} {\bibfield  {journal} {\bibinfo  {journal} {Phys. Rev. Lett.}\ }\textbf {\bibinfo {volume} {111}},\ \bibinfo {pages} {100502} (\bibinfo {year} {2013})}\BibitemShut {NoStop}%
\bibitem [{\citenamefont {An}\ \emph {et~al.}(2016)\citenamefont {An}, \citenamefont {Lv}, \citenamefont {Del~Campo},\ and\ \citenamefont {Kim}}]{an2016shortcuts}%
  \BibitemOpen
  \bibfield  {author} {\bibinfo {author} {\bibfnamefont {S.}~\bibnamefont {An}}, \bibinfo {author} {\bibfnamefont {D.}~\bibnamefont {Lv}}, \bibinfo {author} {\bibfnamefont {A.}~\bibnamefont {Del~Campo}},\ and\ \bibinfo {author} {\bibfnamefont {K.}~\bibnamefont {Kim}},\ }\bibfield  {title} {\bibinfo {title} {Shortcuts to adiabaticity by counterdiabatic driving for trapped-ion displacement in phase space},\ }\href {https://doi.org/10.1038/ncomms12999} {\bibfield  {journal} {\bibinfo  {journal} {Nat. Commun.}\ }\textbf {\bibinfo {volume} {7}},\ \bibinfo {pages} {12999} (\bibinfo {year} {2016})}\BibitemShut {NoStop}%
\bibitem [{\citenamefont {Chen}\ \emph {et~al.}(2011)\citenamefont {Chen}, \citenamefont {Torrontegui},\ and\ \citenamefont {Muga}}]{chen2011lewis}%
  \BibitemOpen
  \bibfield  {author} {\bibinfo {author} {\bibfnamefont {X.}~\bibnamefont {Chen}}, \bibinfo {author} {\bibfnamefont {E.}~\bibnamefont {Torrontegui}},\ and\ \bibinfo {author} {\bibfnamefont {J.~G.}\ \bibnamefont {Muga}},\ }\bibfield  {title} {\bibinfo {title} {Lewis-{R}iesenfeld invariants and transitionless quantum driving},\ }\href {https://doi.org/10.1103/PhysRevA.83.062116} {\bibfield  {journal} {\bibinfo  {journal} {Phys. Rev. A}\ }\textbf {\bibinfo {volume} {83}},\ \bibinfo {pages} {062116} (\bibinfo {year} {2011})}\BibitemShut {NoStop}%
\bibitem [{\citenamefont {Masuda}\ and\ \citenamefont {Nakamura}(2008)}]{masuda2008fast}%
  \BibitemOpen
  \bibfield  {author} {\bibinfo {author} {\bibfnamefont {S.}~\bibnamefont {Masuda}}\ and\ \bibinfo {author} {\bibfnamefont {K.}~\bibnamefont {Nakamura}},\ }\bibfield  {title} {\bibinfo {title} {Fast-forward problem in quantum mechanics},\ }\href {https://doi.org/10.1103/PhysRevA.78.062108} {\bibfield  {journal} {\bibinfo  {journal} {Phys. Rev. A}\ }\textbf {\bibinfo {volume} {78}},\ \bibinfo {pages} {062108} (\bibinfo {year} {2008})}\BibitemShut {NoStop}%
\bibitem [{\citenamefont {Masuda}\ and\ \citenamefont {Nakamura}(2010)}]{masuda2010fast}%
  \BibitemOpen
  \bibfield  {author} {\bibinfo {author} {\bibfnamefont {S.}~\bibnamefont {Masuda}}\ and\ \bibinfo {author} {\bibfnamefont {K.}~\bibnamefont {Nakamura}},\ }\bibfield  {title} {\bibinfo {title} {Fast-forward of adiabatic dynamics in quantum mechanics},\ }\href {https://doi.org/http://doi.org/10.1098/rspa.2009.0446} {\bibfield  {journal} {\bibinfo  {journal} {Proc. R. Soc. A.}\ }\textbf {\bibinfo {volume} {466}},\ \bibinfo {pages} {1135} (\bibinfo {year} {2010})}\BibitemShut {NoStop}%
\bibitem [{\citenamefont {Kolodrubetz}\ \emph {et~al.}(2017)\citenamefont {Kolodrubetz}, \citenamefont {Sels}, \citenamefont {Mehta},\ and\ \citenamefont {Polkovnikov}}]{kolodrubetz2017geometry}%
  \BibitemOpen
  \bibfield  {author} {\bibinfo {author} {\bibfnamefont {M.}~\bibnamefont {Kolodrubetz}}, \bibinfo {author} {\bibfnamefont {D.}~\bibnamefont {Sels}}, \bibinfo {author} {\bibfnamefont {P.}~\bibnamefont {Mehta}},\ and\ \bibinfo {author} {\bibfnamefont {A.}~\bibnamefont {Polkovnikov}},\ }\bibfield  {title} {\bibinfo {title} {Geometry and non-adiabatic response in quantum and classical systems},\ }\href {https://doi.org/https://doi.org/10.1016/j.physrep.2017.07.001} {\bibfield  {journal} {\bibinfo  {journal} {Phys. Rep.}\ }\textbf {\bibinfo {volume} {697}},\ \bibinfo {pages} {1} (\bibinfo {year} {2017})}\BibitemShut {NoStop}%
\bibitem [{\citenamefont {Vitanov}\ \emph {et~al.}(2017)\citenamefont {Vitanov}, \citenamefont {Rangelov}, \citenamefont {Shore},\ and\ \citenamefont {Bergmann}}]{vitanov2017stimulated}%
  \BibitemOpen
  \bibfield  {author} {\bibinfo {author} {\bibfnamefont {N.~V.}\ \bibnamefont {Vitanov}}, \bibinfo {author} {\bibfnamefont {A.~A.}\ \bibnamefont {Rangelov}}, \bibinfo {author} {\bibfnamefont {B.~W.}\ \bibnamefont {Shore}},\ and\ \bibinfo {author} {\bibfnamefont {K.}~\bibnamefont {Bergmann}},\ }\bibfield  {title} {\bibinfo {title} {Stimulated {R}aman adiabatic passage in physics, chemistry, and beyond},\ }\href {https://doi.org/10.1103/RevModPhys.89.015006} {\bibfield  {journal} {\bibinfo  {journal} {Rev. Mod. Phys.}\ }\textbf {\bibinfo {volume} {89}},\ \bibinfo {pages} {015006} (\bibinfo {year} {2017})}\BibitemShut {NoStop}%
\bibitem [{\citenamefont {Du}\ \emph {et~al.}(2016)\citenamefont {Du}, \citenamefont {Liang}, \citenamefont {Li}, \citenamefont {Yue}, \citenamefont {Lv}, \citenamefont {Huang}, \citenamefont {Chen}, \citenamefont {Yan},\ and\ \citenamefont {Zhu}}]{Du_fastSTIRAP_2016}%
  \BibitemOpen
  \bibfield  {author} {\bibinfo {author} {\bibfnamefont {Y.-X.}\ \bibnamefont {Du}}, \bibinfo {author} {\bibfnamefont {Z.-T.}\ \bibnamefont {Liang}}, \bibinfo {author} {\bibfnamefont {Y.-C.}\ \bibnamefont {Li}}, \bibinfo {author} {\bibfnamefont {X.-X.}\ \bibnamefont {Yue}}, \bibinfo {author} {\bibfnamefont {Q.-X.}\ \bibnamefont {Lv}}, \bibinfo {author} {\bibfnamefont {W.}~\bibnamefont {Huang}}, \bibinfo {author} {\bibfnamefont {X.}~\bibnamefont {Chen}}, \bibinfo {author} {\bibfnamefont {H.}~\bibnamefont {Yan}},\ and\ \bibinfo {author} {\bibfnamefont {S.-L.}\ \bibnamefont {Zhu}},\ }\bibfield  {title} {\bibinfo {title} {Experimental realization of stimulated raman shortcut-to-adiabatic passage with cold atoms},\ }\href {https://doi.org/10.1038/ncomms12479} {\bibfield  {journal} {\bibinfo  {journal} {Nat. Comm.}\ }\textbf {\bibinfo {volume} {7}},\ \bibinfo {pages} {12479} (\bibinfo {year} {2016})}\BibitemShut {NoStop}%
\bibitem [{\citenamefont {Vepsäläinen}\ \emph {et~al.}(2019)\citenamefont {Vepsäläinen}, \citenamefont {Danilin},\ and\ \citenamefont {Paraoanu}}]{Vespsalsuper_2019}%
  \BibitemOpen
  \bibfield  {author} {\bibinfo {author} {\bibfnamefont {A.}~\bibnamefont {Vepsäläinen}}, \bibinfo {author} {\bibfnamefont {S.}~\bibnamefont {Danilin}},\ and\ \bibinfo {author} {\bibfnamefont {G.~S.}\ \bibnamefont {Paraoanu}},\ }\bibfield  {title} {\bibinfo {title} {Superadiabatic population transfer in a three-level superconducting circuit},\ }\href {https://doi.org/10.1126/sciadv.aau5999} {\bibfield  {journal} {\bibinfo  {journal} {Science Advances}\ }\textbf {\bibinfo {volume} {5}},\ \bibinfo {pages} {eaau5999} (\bibinfo {year} {2019})}\BibitemShut {NoStop}%
\bibitem [{\citenamefont {del Campo}\ \emph {et~al.}(2012)\citenamefont {del Campo}, \citenamefont {Rams},\ and\ \citenamefont {Zurek}}]{del2012assisted}%
  \BibitemOpen
  \bibfield  {author} {\bibinfo {author} {\bibfnamefont {A.}~\bibnamefont {del Campo}}, \bibinfo {author} {\bibfnamefont {M.~M.}\ \bibnamefont {Rams}},\ and\ \bibinfo {author} {\bibfnamefont {W.~H.}\ \bibnamefont {Zurek}},\ }\bibfield  {title} {\bibinfo {title} {Assisted finite-rate adiabatic passage across a quantum critical point: exact solution for the quantum {I}sing model},\ }\href {https://doi.org/10.1103/PhysRevLett.109.115703} {\bibfield  {journal} {\bibinfo  {journal} {Phys. Rev. Lett.}\ }\textbf {\bibinfo {volume} {109}},\ \bibinfo {pages} {115703} (\bibinfo {year} {2012})}\BibitemShut {NoStop}%
\bibitem [{\citenamefont {Sels}\ and\ \citenamefont {Polkovnikov}(2017)}]{sels2017minimizing}%
  \BibitemOpen
  \bibfield  {author} {\bibinfo {author} {\bibfnamefont {D.}~\bibnamefont {Sels}}\ and\ \bibinfo {author} {\bibfnamefont {A.}~\bibnamefont {Polkovnikov}},\ }\bibfield  {title} {\bibinfo {title} {Minimizing irreversible losses in quantum systems by local counterdiabatic driving},\ }\href {https://doi.org/10.1073/pnas.1619826114} {\bibfield  {journal} {\bibinfo  {journal} {PNAS}\ }\textbf {\bibinfo {volume} {114}},\ \bibinfo {pages} {E3909} (\bibinfo {year} {2017})}\BibitemShut {NoStop}%
\bibitem [{\citenamefont {Claeys}\ \emph {et~al.}(2019)\citenamefont {Claeys}, \citenamefont {Pandey}, \citenamefont {Sels},\ and\ \citenamefont {Polkovnikov}}]{claeys2019floquet}%
  \BibitemOpen
  \bibfield  {author} {\bibinfo {author} {\bibfnamefont {P.~W.}\ \bibnamefont {Claeys}}, \bibinfo {author} {\bibfnamefont {M.}~\bibnamefont {Pandey}}, \bibinfo {author} {\bibfnamefont {D.}~\bibnamefont {Sels}},\ and\ \bibinfo {author} {\bibfnamefont {A.}~\bibnamefont {Polkovnikov}},\ }\bibfield  {title} {\bibinfo {title} {Floquet-engineering counterdiabatic protocols in quantum many-body systems},\ }\href {https://doi.org/10.1103/PhysRevLett.123.090602} {\bibfield  {journal} {\bibinfo  {journal} {Phys. Rev. Lett.}\ }\textbf {\bibinfo {volume} {123}},\ \bibinfo {pages} {090602} (\bibinfo {year} {2019})}\BibitemShut {NoStop}%
\bibitem [{\citenamefont {{\v{C}}epait{\.e}}\ \emph {et~al.}(2023)\citenamefont {{\v{C}}epait{\.e}}, \citenamefont {Polkovnikov}, \citenamefont {Daley},\ and\ \citenamefont {Duncan}}]{vcepaite2023counterdiabatic}%
  \BibitemOpen
  \bibfield  {author} {\bibinfo {author} {\bibfnamefont {I.}~\bibnamefont {{\v{C}}epait{\.e}}}, \bibinfo {author} {\bibfnamefont {A.}~\bibnamefont {Polkovnikov}}, \bibinfo {author} {\bibfnamefont {A.~J.}\ \bibnamefont {Daley}},\ and\ \bibinfo {author} {\bibfnamefont {C.~W.}\ \bibnamefont {Duncan}},\ }\bibfield  {title} {\bibinfo {title} {Counterdiabatic optimized local driving},\ }\href {https://doi.org/10.1103/PRXQuantum.4.010312} {\bibfield  {journal} {\bibinfo  {journal} {PRX Quantum}\ }\textbf {\bibinfo {volume} {4}},\ \bibinfo {pages} {010312} (\bibinfo {year} {2023})}\BibitemShut {NoStop}%
\bibitem [{\citenamefont {Sun}\ \emph {et~al.}(2022)\citenamefont {Sun}, \citenamefont {Chandarana}, \citenamefont {Xin},\ and\ \citenamefont {Chen}}]{sun2022optimizing}%
  \BibitemOpen
  \bibfield  {author} {\bibinfo {author} {\bibfnamefont {D.}~\bibnamefont {Sun}}, \bibinfo {author} {\bibfnamefont {P.}~\bibnamefont {Chandarana}}, \bibinfo {author} {\bibfnamefont {Z.}~\bibnamefont {Xin}},\ and\ \bibinfo {author} {\bibfnamefont {X.}~\bibnamefont {Chen}},\ }\bibfield  {title} {\bibinfo {title} {Optimizing counterdiabaticity by variational quantum circuits},\ }\href {http://doi.org/10.1098/rsta.2021.0282} {\bibfield  {journal} {\bibinfo  {journal} {Philos. Trans. R. Soc. A}\ }\textbf {\bibinfo {volume} {380}} (\bibinfo {year} {2022})}\BibitemShut {NoStop}%
\bibitem [{\citenamefont {Chandarana}\ \emph {et~al.}(2022)\citenamefont {Chandarana}, \citenamefont {Hegade}, \citenamefont {Paul}, \citenamefont {Albarr\'an-Arriagada}, \citenamefont {Solano}, \citenamefont {del Campo},\ and\ \citenamefont {Chen}}]{chen2022digitized}%
  \BibitemOpen
  \bibfield  {author} {\bibinfo {author} {\bibfnamefont {P.}~\bibnamefont {Chandarana}}, \bibinfo {author} {\bibfnamefont {N.~N.}\ \bibnamefont {Hegade}}, \bibinfo {author} {\bibfnamefont {K.}~\bibnamefont {Paul}}, \bibinfo {author} {\bibfnamefont {F.}~\bibnamefont {Albarr\'an-Arriagada}}, \bibinfo {author} {\bibfnamefont {E.}~\bibnamefont {Solano}}, \bibinfo {author} {\bibfnamefont {A.}~\bibnamefont {del Campo}},\ and\ \bibinfo {author} {\bibfnamefont {X.}~\bibnamefont {Chen}},\ }\bibfield  {title} {\bibinfo {title} {Digitized-counterdiabatic quantum approximate optimization algorithm},\ }\href {https://doi.org/10.1103/PhysRevResearch.4.013141} {\bibfield  {journal} {\bibinfo  {journal} {Phys. Rev. Res.}\ }\textbf {\bibinfo {volume} {4}},\ \bibinfo {pages} {013141} (\bibinfo {year} {2022})}\BibitemShut {NoStop}%
\bibitem [{\citenamefont {Espinós}\ \emph {et~al.}(2023{\natexlab{a}})\citenamefont {Espinós}, \citenamefont {Cangemi}, \citenamefont {Levy}, \citenamefont {Puebla},\ and\ \citenamefont {Torrontegui}}]{espinós2023invariantbased}%
  \BibitemOpen
  \bibfield  {author} {\bibinfo {author} {\bibfnamefont {H.}~\bibnamefont {Espinós}}, \bibinfo {author} {\bibfnamefont {L.~M.}\ \bibnamefont {Cangemi}}, \bibinfo {author} {\bibfnamefont {A.}~\bibnamefont {Levy}}, \bibinfo {author} {\bibfnamefont {R.}~\bibnamefont {Puebla}},\ and\ \bibinfo {author} {\bibfnamefont {E.}~\bibnamefont {Torrontegui}},\ }\bibfield  {title} {\bibinfo {title} {Invariant-based control of quantum many-body systems across critical points},\ }\href {https://doi.org/10.48550/arXiv.2309.05469} {\bibfield  {journal} {\bibinfo  {journal} {arXiv:2309.05469}\ } (\bibinfo {year} {2023}{\natexlab{a}})}\BibitemShut {NoStop}%
\bibitem [{\citenamefont {Zhou}\ \emph {et~al.}(2020)\citenamefont {Zhou}, \citenamefont {Ji}, \citenamefont {Nie}, \citenamefont {Yang}, \citenamefont {Chen}, \citenamefont {Bian},\ and\ \citenamefont {Peng}}]{Zhoulocalcounter2020}%
  \BibitemOpen
  \bibfield  {author} {\bibinfo {author} {\bibfnamefont {H.}~\bibnamefont {Zhou}}, \bibinfo {author} {\bibfnamefont {Y.}~\bibnamefont {Ji}}, \bibinfo {author} {\bibfnamefont {X.}~\bibnamefont {Nie}}, \bibinfo {author} {\bibfnamefont {X.}~\bibnamefont {Yang}}, \bibinfo {author} {\bibfnamefont {X.}~\bibnamefont {Chen}}, \bibinfo {author} {\bibfnamefont {J.}~\bibnamefont {Bian}},\ and\ \bibinfo {author} {\bibfnamefont {X.}~\bibnamefont {Peng}},\ }\bibfield  {title} {\bibinfo {title} {Experimental realization of shortcuts to adiabaticity in a nonintegrable spin chain by local counterdiabatic driving},\ }\href {https://doi.org/10.1103/PhysRevApplied.13.044059} {\bibfield  {journal} {\bibinfo  {journal} {Phys. Rev. Appl.}\ }\textbf {\bibinfo {volume} {13}},\ \bibinfo {pages} {044059} (\bibinfo {year} {2020})}\BibitemShut {NoStop}%
\bibitem [{\citenamefont {Hegade}\ \emph {et~al.}(2021)\citenamefont {Hegade}, \citenamefont {Paul}, \citenamefont {Ding}, \citenamefont {Sanz}, \citenamefont {Albarr\'an-Arriagada}, \citenamefont {Solano},\ and\ \citenamefont {Chen}}]{hegade2021shortcuts}%
  \BibitemOpen
  \bibfield  {author} {\bibinfo {author} {\bibfnamefont {N.~N.}\ \bibnamefont {Hegade}}, \bibinfo {author} {\bibfnamefont {K.}~\bibnamefont {Paul}}, \bibinfo {author} {\bibfnamefont {Y.}~\bibnamefont {Ding}}, \bibinfo {author} {\bibfnamefont {M.}~\bibnamefont {Sanz}}, \bibinfo {author} {\bibfnamefont {F.}~\bibnamefont {Albarr\'an-Arriagada}}, \bibinfo {author} {\bibfnamefont {E.}~\bibnamefont {Solano}},\ and\ \bibinfo {author} {\bibfnamefont {X.}~\bibnamefont {Chen}},\ }\bibfield  {title} {\bibinfo {title} {Shortcuts to adiabaticity in digitized adiabatic quantum computing},\ }\href {https://doi.org/10.1103/PhysRevApplied.15.024038} {\bibfield  {journal} {\bibinfo  {journal} {Phys. Rev. Appl.}\ }\textbf {\bibinfo {volume} {15}},\ \bibinfo {pages} {024038} (\bibinfo {year} {2021})}\BibitemShut {NoStop}%
\bibitem [{\citenamefont {C{\'a}rdenas-L{\'o}pez}\ \emph {et~al.}(2023)\citenamefont {C{\'a}rdenas-L{\'o}pez}, \citenamefont {Retamal},\ and\ \citenamefont {Chen}}]{cardenas2023shortcuts}%
  \BibitemOpen
  \bibfield  {author} {\bibinfo {author} {\bibfnamefont {F.~A.}\ \bibnamefont {C{\'a}rdenas-L{\'o}pez}}, \bibinfo {author} {\bibfnamefont {J.~C.}\ \bibnamefont {Retamal}},\ and\ \bibinfo {author} {\bibfnamefont {X.}~\bibnamefont {Chen}},\ }\bibfield  {title} {\bibinfo {title} {Shortcuts to adiabaticity in superconducting circuits for fast multi-partite state generation},\ }\href {https://doi.org/10.1038/s42005-023-01283-0} {\bibfield  {journal} {\bibinfo  {journal} {Commun. Phys.}\ }\textbf {\bibinfo {volume} {6}},\ \bibinfo {pages} {167} (\bibinfo {year} {2023})}\BibitemShut {NoStop}%
\bibitem [{\citenamefont {Yin}\ \emph {et~al.}(2022)\citenamefont {Yin}, \citenamefont {Li}, \citenamefont {Allcock}, \citenamefont {Zheng}, \citenamefont {Gu}, \citenamefont {Dai}, \citenamefont {Zhang},\ and\ \citenamefont {An}}]{yin2022shortcuts}%
  \BibitemOpen
  \bibfield  {author} {\bibinfo {author} {\bibfnamefont {Z.}~\bibnamefont {Yin}}, \bibinfo {author} {\bibfnamefont {C.}~\bibnamefont {Li}}, \bibinfo {author} {\bibfnamefont {J.}~\bibnamefont {Allcock}}, \bibinfo {author} {\bibfnamefont {Y.}~\bibnamefont {Zheng}}, \bibinfo {author} {\bibfnamefont {X.}~\bibnamefont {Gu}}, \bibinfo {author} {\bibfnamefont {M.}~\bibnamefont {Dai}}, \bibinfo {author} {\bibfnamefont {S.}~\bibnamefont {Zhang}},\ and\ \bibinfo {author} {\bibfnamefont {S.}~\bibnamefont {An}},\ }\bibfield  {title} {\bibinfo {title} {Shortcuts to adiabaticity for open systems in circuit quantum electrodynamics},\ }\href {https://doi.org/10.1038/s41467-021-27900-6} {\bibfield  {journal} {\bibinfo  {journal} {Nat. Commun.}\ }\textbf {\bibinfo {volume} {13}},\ \bibinfo {pages} {188} (\bibinfo {year} {2022})}\BibitemShut {NoStop}%
\bibitem [{\citenamefont {Koch}(2016)}]{koch2016controlling}%
  \BibitemOpen
  \bibfield  {author} {\bibinfo {author} {\bibfnamefont {C.~P.}\ \bibnamefont {Koch}},\ }\bibfield  {title} {\bibinfo {title} {Controlling open quantum systems: tools, achievements, and limitations},\ }\href {https://doi.org/10.1088/0953-8984/28/21/213001} {\bibfield  {journal} {\bibinfo  {journal} {J. Condens. Matter Phys}\ }\textbf {\bibinfo {volume} {28}},\ \bibinfo {pages} {213001} (\bibinfo {year} {2016})}\BibitemShut {NoStop}%
\bibitem [{\citenamefont {Levy}\ \emph {et~al.}(2017)\citenamefont {Levy}, \citenamefont {Torrontegui},\ and\ \citenamefont {Kosloff}}]{levy2017action}%
  \BibitemOpen
  \bibfield  {author} {\bibinfo {author} {\bibfnamefont {A.}~\bibnamefont {Levy}}, \bibinfo {author} {\bibfnamefont {E.}~\bibnamefont {Torrontegui}},\ and\ \bibinfo {author} {\bibfnamefont {R.}~\bibnamefont {Kosloff}},\ }\bibfield  {title} {\bibinfo {title} {Action-noise-assisted quantum control},\ }\href {https://doi.org/10.1103/PhysRevA.96.033417} {\bibfield  {journal} {\bibinfo  {journal} {Physical Review A}\ }\textbf {\bibinfo {volume} {96}},\ \bibinfo {pages} {033417} (\bibinfo {year} {2017})}\BibitemShut {NoStop}%
\bibitem [{\citenamefont {Levy}\ \emph {et~al.}(2018)\citenamefont {Levy}, \citenamefont {Kiely}, \citenamefont {Muga}, \citenamefont {Kosloff},\ and\ \citenamefont {Torrontegui}}]{levy2018noise}%
  \BibitemOpen
  \bibfield  {author} {\bibinfo {author} {\bibfnamefont {A.}~\bibnamefont {Levy}}, \bibinfo {author} {\bibfnamefont {A.}~\bibnamefont {Kiely}}, \bibinfo {author} {\bibfnamefont {J.}~\bibnamefont {Muga}}, \bibinfo {author} {\bibfnamefont {R.}~\bibnamefont {Kosloff}},\ and\ \bibinfo {author} {\bibfnamefont {E.}~\bibnamefont {Torrontegui}},\ }\bibfield  {title} {\bibinfo {title} {Noise resistant quantum control using dynamical invariants},\ }\href {https://doi.org/10.1088/1367-2630/aaa9e5} {\bibfield  {journal} {\bibinfo  {journal} {New J. Phys.}\ }\textbf {\bibinfo {volume} {20}},\ \bibinfo {pages} {025006} (\bibinfo {year} {2018})}\BibitemShut {NoStop}%
\bibitem [{\citenamefont {Kallush}\ \emph {et~al.}(2022)\citenamefont {Kallush}, \citenamefont {Dann},\ and\ \citenamefont {Kosloff}}]{kallush2022controlling}%
  \BibitemOpen
  \bibfield  {author} {\bibinfo {author} {\bibfnamefont {S.}~\bibnamefont {Kallush}}, \bibinfo {author} {\bibfnamefont {R.}~\bibnamefont {Dann}},\ and\ \bibinfo {author} {\bibfnamefont {R.}~\bibnamefont {Kosloff}},\ }\bibfield  {title} {\bibinfo {title} {Controlling the uncontrollable: Quantum control of open-system dynamics},\ }\href {https://www.science.org/doi/abs/10.1126/sciadv.add0828} {\bibfield  {journal} {\bibinfo  {journal} {Sci. Adv.}\ }\textbf {\bibinfo {volume} {8}},\ \bibinfo {pages} {eadd0828} (\bibinfo {year} {2022})}\BibitemShut {NoStop}%
\bibitem [{\citenamefont {Venuti}\ \emph {et~al.}(2021)\citenamefont {Venuti}, \citenamefont {D’Alessandro},\ and\ \citenamefont {Lidar}}]{venuti2021optimal}%
  \BibitemOpen
  \bibfield  {author} {\bibinfo {author} {\bibfnamefont {L.~C.}\ \bibnamefont {Venuti}}, \bibinfo {author} {\bibfnamefont {D.}~\bibnamefont {D’Alessandro}},\ and\ \bibinfo {author} {\bibfnamefont {D.~A.}\ \bibnamefont {Lidar}},\ }\bibfield  {title} {\bibinfo {title} {Optimal control for quantum optimization of closed and open systems},\ }\href {https://doi.org/10.1103/PhysRevApplied.16.054023} {\bibfield  {journal} {\bibinfo  {journal} {Phys. Rev. Appl.}\ }\textbf {\bibinfo {volume} {16}},\ \bibinfo {pages} {054023} (\bibinfo {year} {2021})}\BibitemShut {NoStop}%
\bibitem [{\citenamefont {Vacanti}\ \emph {et~al.}(2014)\citenamefont {Vacanti}, \citenamefont {Fazio}, \citenamefont {Montangero}, \citenamefont {Palma}, \citenamefont {Paternostro},\ and\ \citenamefont {Vedral}}]{vacanti2014transitionless}%
  \BibitemOpen
  \bibfield  {author} {\bibinfo {author} {\bibfnamefont {G.}~\bibnamefont {Vacanti}}, \bibinfo {author} {\bibfnamefont {R.}~\bibnamefont {Fazio}}, \bibinfo {author} {\bibfnamefont {S.}~\bibnamefont {Montangero}}, \bibinfo {author} {\bibfnamefont {G.}~\bibnamefont {Palma}}, \bibinfo {author} {\bibfnamefont {M.}~\bibnamefont {Paternostro}},\ and\ \bibinfo {author} {\bibfnamefont {V.}~\bibnamefont {Vedral}},\ }\bibfield  {title} {\bibinfo {title} {Transitionless quantum driving in open quantum systems},\ }\href {https://doi.org/10.1088/1367-2630/16/5/053017} {\bibfield  {journal} {\bibinfo  {journal} {New J. Phys.}\ }\textbf {\bibinfo {volume} {16}},\ \bibinfo {pages} {053017} (\bibinfo {year} {2014})}\BibitemShut {NoStop}%
\bibitem [{\citenamefont {Villazon}\ \emph {et~al.}(2019)\citenamefont {Villazon}, \citenamefont {Polkovnikov},\ and\ \citenamefont {Chandran}}]{villazon2019swift}%
  \BibitemOpen
  \bibfield  {author} {\bibinfo {author} {\bibfnamefont {T.}~\bibnamefont {Villazon}}, \bibinfo {author} {\bibfnamefont {A.}~\bibnamefont {Polkovnikov}},\ and\ \bibinfo {author} {\bibfnamefont {A.}~\bibnamefont {Chandran}},\ }\bibfield  {title} {\bibinfo {title} {Swift heat transfer by fast-forward driving in open quantum systems},\ }\href {https://doi.org/10.1103/PhysRevA.100.012126} {\bibfield  {journal} {\bibinfo  {journal} {Phys. Rev. A}\ }\textbf {\bibinfo {volume} {100}},\ \bibinfo {pages} {012126} (\bibinfo {year} {2019})}\BibitemShut {NoStop}%
\bibitem [{\citenamefont {Alipour}\ \emph {et~al.}(2020)\citenamefont {Alipour}, \citenamefont {Chenu}, \citenamefont {Rezakhani},\ and\ \citenamefont {del Campo}}]{alipour2020shortcuts}%
  \BibitemOpen
  \bibfield  {author} {\bibinfo {author} {\bibfnamefont {S.}~\bibnamefont {Alipour}}, \bibinfo {author} {\bibfnamefont {A.}~\bibnamefont {Chenu}}, \bibinfo {author} {\bibfnamefont {A.~T.}\ \bibnamefont {Rezakhani}},\ and\ \bibinfo {author} {\bibfnamefont {A.}~\bibnamefont {del Campo}},\ }\bibfield  {title} {\bibinfo {title} {Shortcuts to adiabaticity in driven open quantum systems: Balanced gain and loss and non-{M}arkovian evolution},\ }\href {https://doi.org/https://doi.org/10.22331/q-2020-09-28-336} {\bibfield  {journal} {\bibinfo  {journal} {Quantum}\ }\textbf {\bibinfo {volume} {4}},\ \bibinfo {pages} {336} (\bibinfo {year} {2020})}\BibitemShut {NoStop}%
\bibitem [{\citenamefont {Funo}\ \emph {et~al.}(2021)\citenamefont {Funo}, \citenamefont {Lambert},\ and\ \citenamefont {Nori}}]{funo2021general}%
  \BibitemOpen
  \bibfield  {author} {\bibinfo {author} {\bibfnamefont {K.}~\bibnamefont {Funo}}, \bibinfo {author} {\bibfnamefont {N.}~\bibnamefont {Lambert}},\ and\ \bibinfo {author} {\bibfnamefont {F.}~\bibnamefont {Nori}},\ }\bibfield  {title} {\bibinfo {title} {General bound on the performance of counter-diabatic driving acting on dissipative spin systems},\ }\href {https://doi.org/10.1103/PhysRevLett.127.150401} {\bibfield  {journal} {\bibinfo  {journal} {Phys. Rev. Lett.}\ }\textbf {\bibinfo {volume} {127}},\ \bibinfo {pages} {150401} (\bibinfo {year} {2021})}\BibitemShut {NoStop}%
\bibitem [{\citenamefont {Lewis}\ and\ \citenamefont {Riesenfeld}(1969)}]{lewis69}%
  \BibitemOpen
  \bibfield  {author} {\bibinfo {author} {\bibfnamefont {H.}~\bibnamefont {Lewis}}\ and\ \bibinfo {author} {\bibfnamefont {W.~B.}\ \bibnamefont {Riesenfeld}},\ }\bibfield  {title} {\bibinfo {title} {An exact quantum theory of the time-dependent harmonic oscillator and of a charged particle in a time-dependent electromagnetic field},\ }\href {https://doi.org/10.1063/1.1664991} {\bibfield  {journal} {\bibinfo  {journal} {J. Math. Phys.}\ }\textbf {\bibinfo {volume} {10}},\ \bibinfo {pages} {1458} (\bibinfo {year} {1969})}\BibitemShut {NoStop}%
\bibitem [{\citenamefont {Dodonov}\ and\ \citenamefont {Man'ko}(1979)}]{Dodonov79}%
  \BibitemOpen
  \bibfield  {author} {\bibinfo {author} {\bibfnamefont {V.~V.}\ \bibnamefont {Dodonov}}\ and\ \bibinfo {author} {\bibfnamefont {V.~I.}\ \bibnamefont {Man'ko}},\ }\bibfield  {title} {\bibinfo {title} {Coherent states and the resonance of a quantum damped oscillator},\ }\href {https://doi.org/10.1103/PhysRevA.20.550} {\bibfield  {journal} {\bibinfo  {journal} {Phys. Rev. A}\ }\textbf {\bibinfo {volume} {20}},\ \bibinfo {pages} {550} (\bibinfo {year} {1979})}\BibitemShut {NoStop}%
\bibitem [{\citenamefont {Dodonov}\ and\ \citenamefont {Man'ko}(1989)}]{Dodonov89}%
  \BibitemOpen
  \bibfield  {author} {\bibinfo {author} {\bibfnamefont {V.~V.}\ \bibnamefont {Dodonov}}\ and\ \bibinfo {author} {\bibfnamefont {V.~I.}\ \bibnamefont {Man'ko}},\ }in\ \href@noop {} {\emph {\bibinfo {booktitle} {Invariants and the evolution of nonstationary quantum system}}}\ (\bibinfo  {publisher} {Nova Science Publishers, Inc},\ \bibinfo {year} {1989})\BibitemShut {NoStop}%
\bibitem [{\citenamefont {Dann}\ \emph {et~al.}(2018)\citenamefont {Dann}, \citenamefont {Levy},\ and\ \citenamefont {Kosloff}}]{dann2018time}%
  \BibitemOpen
  \bibfield  {author} {\bibinfo {author} {\bibfnamefont {R.}~\bibnamefont {Dann}}, \bibinfo {author} {\bibfnamefont {A.}~\bibnamefont {Levy}},\ and\ \bibinfo {author} {\bibfnamefont {R.}~\bibnamefont {Kosloff}},\ }\bibfield  {title} {\bibinfo {title} {Time-dependent {M}arkovian quantum master equation},\ }\href {https://doi.org/10.1103/PhysRevA.98.052129} {\bibfield  {journal} {\bibinfo  {journal} {Phys. Rev. A}\ }\textbf {\bibinfo {volume} {98}},\ \bibinfo {pages} {052129} (\bibinfo {year} {2018})}\BibitemShut {NoStop}%
\bibitem [{\citenamefont {Espinós}\ \emph {et~al.}(2023{\natexlab{b}})\citenamefont {Espinós}, \citenamefont {Panadero}, \citenamefont {García-Ripoll},\ and\ \citenamefont {Torrontegui}}]{Espinós_2023}%
  \BibitemOpen
  \bibfield  {author} {\bibinfo {author} {\bibfnamefont {H.}~\bibnamefont {Espinós}}, \bibinfo {author} {\bibfnamefont {I.}~\bibnamefont {Panadero}}, \bibinfo {author} {\bibfnamefont {J.~J.}\ \bibnamefont {García-Ripoll}},\ and\ \bibinfo {author} {\bibfnamefont {E.}~\bibnamefont {Torrontegui}},\ }\bibfield  {title} {\bibinfo {title} {Quantum control of tunable-coupling transmons using dynamical invariants of motion},\ }\href {https://doi.org/10.1088/2058-9565/acbed7} {\bibfield  {journal} {\bibinfo  {journal} {Quantum Sci. Technol.}\ }\textbf {\bibinfo {volume} {8}},\ \bibinfo {pages} {025017} (\bibinfo {year} {2023}{\natexlab{b}})}\BibitemShut {NoStop}%
\bibitem [{\citenamefont {Chen}\ \emph {et~al.}(2010)\citenamefont {Chen}, \citenamefont {Ruschhaupt}, \citenamefont {Schmidt}, \citenamefont {del Campo}, \citenamefont {Gu\'ery-Odelin},\ and\ \citenamefont {Muga}}]{chen2010fast}%
  \BibitemOpen
  \bibfield  {author} {\bibinfo {author} {\bibfnamefont {X.}~\bibnamefont {Chen}}, \bibinfo {author} {\bibfnamefont {A.}~\bibnamefont {Ruschhaupt}}, \bibinfo {author} {\bibfnamefont {S.}~\bibnamefont {Schmidt}}, \bibinfo {author} {\bibfnamefont {A.}~\bibnamefont {del Campo}}, \bibinfo {author} {\bibfnamefont {D.}~\bibnamefont {Gu\'ery-Odelin}},\ and\ \bibinfo {author} {\bibfnamefont {J.~G.}\ \bibnamefont {Muga}},\ }\bibfield  {title} {\bibinfo {title} {Fast optimal frictionless atom cooling in harmonic traps: Shortcut to adiabaticity},\ }\href {https://doi.org/10.1103/PhysRevLett.104.063002} {\bibfield  {journal} {\bibinfo  {journal} {Phys. Rev. Lett.}\ }\textbf {\bibinfo {volume} {104}},\ \bibinfo {pages} {063002} (\bibinfo {year} {2010})}\BibitemShut {NoStop}%
\bibitem [{\citenamefont {Uzdin}\ \emph {et~al.}(2013)\citenamefont {Uzdin}, \citenamefont {Dalla~Torre}, \citenamefont {Kosloff},\ and\ \citenamefont {Moiseyev}}]{uzdin2013effects}%
  \BibitemOpen
  \bibfield  {author} {\bibinfo {author} {\bibfnamefont {R.}~\bibnamefont {Uzdin}}, \bibinfo {author} {\bibfnamefont {E.~G.}\ \bibnamefont {Dalla~Torre}}, \bibinfo {author} {\bibfnamefont {R.}~\bibnamefont {Kosloff}},\ and\ \bibinfo {author} {\bibfnamefont {N.}~\bibnamefont {Moiseyev}},\ }\bibfield  {title} {\bibinfo {title} {Effects of an exceptional point on the dynamics of a single particle in a time-dependent harmonic trap},\ }\href {https://doi.org/10.1103/PhysRevA.88.022505} {\bibfield  {journal} {\bibinfo  {journal} {Phys. Rev. A}\ }\textbf {\bibinfo {volume} {88}},\ \bibinfo {pages} {022505} (\bibinfo {year} {2013})}\BibitemShut {NoStop}%
\bibitem [{\citenamefont {Dann}\ and\ \citenamefont {Kosloff}(2021{\natexlab{a}})}]{dann2021inertial}%
  \BibitemOpen
  \bibfield  {author} {\bibinfo {author} {\bibfnamefont {R.}~\bibnamefont {Dann}}\ and\ \bibinfo {author} {\bibfnamefont {R.}~\bibnamefont {Kosloff}},\ }\bibfield  {title} {\bibinfo {title} {Inertial theorem: overcoming the quantum adiabatic limit},\ }\href {https://doi.org/10.1103/PhysRevResearch.3.013064} {\bibfield  {journal} {\bibinfo  {journal} {Phys. Rev. Res.}\ }\textbf {\bibinfo {volume} {3}},\ \bibinfo {pages} {013064} (\bibinfo {year} {2021}{\natexlab{a}})}\BibitemShut {NoStop}%
\bibitem [{\citenamefont {Gilmore}(2008)}]{gilmore_2008}%
  \BibitemOpen
  \bibfield  {author} {\bibinfo {author} {\bibfnamefont {R.}~\bibnamefont {Gilmore}},\ }\href@noop {} {\emph {\bibinfo {title} {Lie Groups, Physics, and Geometry: An Introduction for Physicists, Engineers and Chemists}}}\ (\bibinfo  {publisher} {Cambridge University Press},\ \bibinfo {year} {2008})\BibitemShut {NoStop}%
\bibitem [{\citenamefont {{Weiss, U.}}(1998)}]{weissbook}%
  \BibitemOpen
  \bibfield  {author} {\bibinfo {author} {\bibnamefont {{Weiss, U.}}},\ }\href@noop {} {\emph {\bibinfo {title} {Quantum Dissipative Systems, 2nd edition}}}\ (\bibinfo  {publisher} {Worls Scientific},\ \bibinfo {year} {1998})\BibitemShut {NoStop}%
\bibitem [{\citenamefont {{H.-P. Breuer and F. Petruccione}}(2002)}]{breuer}%
  \BibitemOpen
  \bibfield  {author} {\bibinfo {author} {\bibnamefont {{H.-P. Breuer and F. Petruccione}}},\ }\href@noop {} {\emph {\bibinfo {title} {Open quantum systems}}}\ (\bibinfo  {publisher} {Oxford university press},\ \bibinfo {year} {2002})\BibitemShut {NoStop}%
\bibitem [{\citenamefont {Rivas}\ and\ \citenamefont {Huelga}(2012)}]{rivas2012open}%
  \BibitemOpen
  \bibfield  {author} {\bibinfo {author} {\bibfnamefont {A.}~\bibnamefont {Rivas}}\ and\ \bibinfo {author} {\bibfnamefont {S.~F.}\ \bibnamefont {Huelga}},\ }\href@noop {} {\emph {\bibinfo {title} {Open quantum systems}}},\ Vol.~\bibinfo {volume} {10}\ (\bibinfo  {publisher} {Springer Berlin, Heidelberg},\ \bibinfo {year} {2012})\BibitemShut {NoStop}%
\bibitem [{\citenamefont {Grabert}\ \emph {et~al.}(1988)\citenamefont {Grabert}, \citenamefont {Schramm},\ and\ \citenamefont {Ingold}}]{GRABERT1988115}%
  \BibitemOpen
  \bibfield  {author} {\bibinfo {author} {\bibfnamefont {H.}~\bibnamefont {Grabert}}, \bibinfo {author} {\bibfnamefont {P.}~\bibnamefont {Schramm}},\ and\ \bibinfo {author} {\bibfnamefont {G.-L.}\ \bibnamefont {Ingold}},\ }\bibfield  {title} {\bibinfo {title} {Quantum {B}rownian motion: The functional integral approach},\ }\href {https://doi.org/https://doi.org/10.1016/0370-1573(88)90023-3} {\bibfield  {journal} {\bibinfo  {journal} {Phys. Rep.}\ }\textbf {\bibinfo {volume} {168}},\ \bibinfo {pages} {115} (\bibinfo {year} {1988})}\BibitemShut {NoStop}%
\bibitem [{\citenamefont {Grifoni}\ and\ \citenamefont {H{\"a}nggi}(1998)}]{grifoni1998driven}%
  \BibitemOpen
  \bibfield  {author} {\bibinfo {author} {\bibfnamefont {M.}~\bibnamefont {Grifoni}}\ and\ \bibinfo {author} {\bibfnamefont {P.}~\bibnamefont {H{\"a}nggi}},\ }\bibfield  {title} {\bibinfo {title} {Driven quantum tunneling},\ }\href {https://doi.org/https://doi.org/10.1016/S0370-1573(98)00022-2} {\bibfield  {journal} {\bibinfo  {journal} {Phys. Rep.}\ }\textbf {\bibinfo {volume} {304}},\ \bibinfo {pages} {229} (\bibinfo {year} {1998})}\BibitemShut {NoStop}%
\bibitem [{\citenamefont {Prior}\ \emph {et~al.}(2010)\citenamefont {Prior}, \citenamefont {Chin}, \citenamefont {Huelga},\ and\ \citenamefont {Plenio}}]{prior2010efficient}%
  \BibitemOpen
  \bibfield  {author} {\bibinfo {author} {\bibfnamefont {J.}~\bibnamefont {Prior}}, \bibinfo {author} {\bibfnamefont {A.~W.}\ \bibnamefont {Chin}}, \bibinfo {author} {\bibfnamefont {S.~F.}\ \bibnamefont {Huelga}},\ and\ \bibinfo {author} {\bibfnamefont {M.~B.}\ \bibnamefont {Plenio}},\ }\bibfield  {title} {\bibinfo {title} {Efficient simulation of strong system-environment interactions},\ }\href {https://doi.org/10.1103/PhysRevLett.105.050404} {\bibfield  {journal} {\bibinfo  {journal} {Phys. Rev. Lett.}\ }\textbf {\bibinfo {volume} {105}},\ \bibinfo {pages} {050404} (\bibinfo {year} {2010})}\BibitemShut {NoStop}%
\bibitem [{\citenamefont {Strathearn}\ \emph {et~al.}(2018)\citenamefont {Strathearn}, \citenamefont {Kirton}, \citenamefont {Kilda}, \citenamefont {Keeling},\ and\ \citenamefont {Lovett}}]{strathearn2018efficient}%
  \BibitemOpen
  \bibfield  {author} {\bibinfo {author} {\bibfnamefont {A.}~\bibnamefont {Strathearn}}, \bibinfo {author} {\bibfnamefont {P.}~\bibnamefont {Kirton}}, \bibinfo {author} {\bibfnamefont {D.}~\bibnamefont {Kilda}}, \bibinfo {author} {\bibfnamefont {J.}~\bibnamefont {Keeling}},\ and\ \bibinfo {author} {\bibfnamefont {B.~W.}\ \bibnamefont {Lovett}},\ }\bibfield  {title} {\bibinfo {title} {Efficient non-{M}arkovian quantum dynamics using time-evolving matrix product operators},\ }\href {https://doi.org/10.1038/s41467-018-05617-3} {\bibfield  {journal} {\bibinfo  {journal} {Nat. Commun.}\ }\textbf {\bibinfo {volume} {9}},\ \bibinfo {pages} {3322} (\bibinfo {year} {2018})}\BibitemShut {NoStop}%
\bibitem [{\citenamefont {Alicki}(2007)}]{alicki2007quantum}%
  \BibitemOpen
  \bibfield  {author} {\bibinfo {author} {\bibfnamefont {R.}~\bibnamefont {Alicki}},\ }\href@noop {} {\emph {\bibinfo {title} {Quantum dynamical semigroups and applications}}},\ Vol.\ \bibinfo {volume} {717}\ (\bibinfo  {publisher} {Springer},\ \bibinfo {year} {2007})\BibitemShut {NoStop}%
\bibitem [{\citenamefont {Albash}\ \emph {et~al.}(2012)\citenamefont {Albash}, \citenamefont {Boixo}, \citenamefont {Lidar},\ and\ \citenamefont {Zanardi}}]{albash2012quantum}%
  \BibitemOpen
  \bibfield  {author} {\bibinfo {author} {\bibfnamefont {T.}~\bibnamefont {Albash}}, \bibinfo {author} {\bibfnamefont {S.}~\bibnamefont {Boixo}}, \bibinfo {author} {\bibfnamefont {D.~A.}\ \bibnamefont {Lidar}},\ and\ \bibinfo {author} {\bibfnamefont {P.}~\bibnamefont {Zanardi}},\ }\bibfield  {title} {\bibinfo {title} {Quantum adiabatic {M}arkovian master equations},\ }\href {https://doi.org/10.1088/1367-2630/14/12/123016} {\bibfield  {journal} {\bibinfo  {journal} {New J. Phys.}\ }\textbf {\bibinfo {volume} {14}},\ \bibinfo {pages} {123016} (\bibinfo {year} {2012})}\BibitemShut {NoStop}%
\bibitem [{\citenamefont {Mozgunov}\ and\ \citenamefont {Lidar}(2020)}]{mozgunov2020completely}%
  \BibitemOpen
  \bibfield  {author} {\bibinfo {author} {\bibfnamefont {E.}~\bibnamefont {Mozgunov}}\ and\ \bibinfo {author} {\bibfnamefont {D.}~\bibnamefont {Lidar}},\ }\bibfield  {title} {\bibinfo {title} {Completely positive master equation for arbitrary driving and small level spacing},\ }\href {https://doi.org/https://doi.org/10.22331/q-2020-02-06-227} {\bibfield  {journal} {\bibinfo  {journal} {Quantum}\ }\textbf {\bibinfo {volume} {4}},\ \bibinfo {pages} {227} (\bibinfo {year} {2020})}\BibitemShut {NoStop}%
\bibitem [{\citenamefont {Wu}\ \emph {et~al.}(2022)\citenamefont {Wu}, \citenamefont {Huang},\ and\ \citenamefont {Yi}}]{Wudriven2022}%
  \BibitemOpen
  \bibfield  {author} {\bibinfo {author} {\bibfnamefont {S.~L.}\ \bibnamefont {Wu}}, \bibinfo {author} {\bibfnamefont {X.~L.}\ \bibnamefont {Huang}},\ and\ \bibinfo {author} {\bibfnamefont {X.~X.}\ \bibnamefont {Yi}},\ }\bibfield  {title} {\bibinfo {title} {Driven markovian master equation based on the lewis-riesenfeld-invariant theory},\ }\href {https://doi.org/10.1103/PhysRevA.106.052217} {\bibfield  {journal} {\bibinfo  {journal} {Phys. Rev. A}\ }\textbf {\bibinfo {volume} {106}},\ \bibinfo {pages} {052217} (\bibinfo {year} {2022})}\BibitemShut {NoStop}%
\bibitem [{\citenamefont {Di~Meglio}\ \emph {et~al.}(2023)\citenamefont {Di~Meglio}, \citenamefont {Plenio},\ and\ \citenamefont {Huelga}}]{di2023time}%
  \BibitemOpen
  \bibfield  {author} {\bibinfo {author} {\bibfnamefont {G.}~\bibnamefont {Di~Meglio}}, \bibinfo {author} {\bibfnamefont {M.~B.}\ \bibnamefont {Plenio}},\ and\ \bibinfo {author} {\bibfnamefont {S.~F.}\ \bibnamefont {Huelga}},\ }\bibfield  {title} {\bibinfo {title} {Time dependent {M}arkovian master equation beyond the adiabatic limit},\ }\href {https://doi.org/10.48550/arXiv.2304.06166} {\bibfield  {journal} {\bibinfo  {journal} {arXiv:2304.06166}\ } (\bibinfo {year} {2023})}\BibitemShut {NoStop}%
\bibitem [{\citenamefont {Watad}\ and\ \citenamefont {Lindner}(2023)}]{watad2023variational}%
  \BibitemOpen
  \bibfield  {author} {\bibinfo {author} {\bibfnamefont {T.}~\bibnamefont {Watad}}\ and\ \bibinfo {author} {\bibfnamefont {N.~H.}\ \bibnamefont {Lindner}},\ }\bibfield  {title} {\bibinfo {title} {Variational quantum algorithms for simulation of {L}indblad dynamics},\ }\href {https://doi.org/10.48550/arXiv.2305.02815} {\bibfield  {journal} {\bibinfo  {journal} {arXiv:2305.02815}\ } (\bibinfo {year} {2023})}\BibitemShut {NoStop}%
\bibitem [{\citenamefont {Kosloff}(2013)}]{kosloff2013quantum}%
  \BibitemOpen
  \bibfield  {author} {\bibinfo {author} {\bibfnamefont {R.}~\bibnamefont {Kosloff}},\ }\bibfield  {title} {\bibinfo {title} {Quantum thermodynamics: A dynamical viewpoint},\ }\href {https://doi.org/https://doi.org/10.3390/e15062100} {\bibfield  {journal} {\bibinfo  {journal} {Entropy}\ }\textbf {\bibinfo {volume} {15}},\ \bibinfo {pages} {2100} (\bibinfo {year} {2013})}\BibitemShut {NoStop}%
\bibitem [{\citenamefont {Dann}\ and\ \citenamefont {Kosloff}(2021{\natexlab{b}})}]{dann2021open}%
  \BibitemOpen
  \bibfield  {author} {\bibinfo {author} {\bibfnamefont {R.}~\bibnamefont {Dann}}\ and\ \bibinfo {author} {\bibfnamefont {R.}~\bibnamefont {Kosloff}},\ }\bibfield  {title} {\bibinfo {title} {Open system dynamics from thermodynamic compatibility},\ }\href {https://doi.org/10.1103/PhysRevResearch.3.023006} {\bibfield  {journal} {\bibinfo  {journal} {Phys. Rev. Res.}\ }\textbf {\bibinfo {volume} {3}},\ \bibinfo {pages} {023006} (\bibinfo {year} {2021}{\natexlab{b}})}\BibitemShut {NoStop}%
\bibitem [{\citenamefont {Levy}\ and\ \citenamefont {Kosloff}(2014)}]{levy2014local}%
  \BibitemOpen
  \bibfield  {author} {\bibinfo {author} {\bibfnamefont {A.}~\bibnamefont {Levy}}\ and\ \bibinfo {author} {\bibfnamefont {R.}~\bibnamefont {Kosloff}},\ }\bibfield  {title} {\bibinfo {title} {The local approach to quantum transport may violate the second law of thermodynamics},\ }\href {https://doi.org/10.1209/0295-5075/107/20004} {\bibfield  {journal} {\bibinfo  {journal} {Europhysics Letters}\ }\textbf {\bibinfo {volume} {107}},\ \bibinfo {pages} {20004} (\bibinfo {year} {2014})}\BibitemShut {NoStop}%
\bibitem [{\citenamefont {De~Chiara}\ \emph {et~al.}(2018)\citenamefont {De~Chiara}, \citenamefont {Landi}, \citenamefont {Hewgill}, \citenamefont {Reid}, \citenamefont {Ferraro}, \citenamefont {Roncaglia},\ and\ \citenamefont {Antezza}}]{de2018reconciliation}%
  \BibitemOpen
  \bibfield  {author} {\bibinfo {author} {\bibfnamefont {G.}~\bibnamefont {De~Chiara}}, \bibinfo {author} {\bibfnamefont {G.}~\bibnamefont {Landi}}, \bibinfo {author} {\bibfnamefont {A.}~\bibnamefont {Hewgill}}, \bibinfo {author} {\bibfnamefont {B.}~\bibnamefont {Reid}}, \bibinfo {author} {\bibfnamefont {A.}~\bibnamefont {Ferraro}}, \bibinfo {author} {\bibfnamefont {A.~J.}\ \bibnamefont {Roncaglia}},\ and\ \bibinfo {author} {\bibfnamefont {M.}~\bibnamefont {Antezza}},\ }\bibfield  {title} {\bibinfo {title} {Reconciliation of quantum local master equations with thermodynamics},\ }\href {https://doi.org/10.1088/1367-2630/aaecee} {\bibfield  {journal} {\bibinfo  {journal} {New J. Phys.}\ }\textbf {\bibinfo {volume} {20}},\ \bibinfo {pages} {113024} (\bibinfo {year} {2018})}\BibitemShut {NoStop}%
\bibitem [{\citenamefont {Cangemi}\ \emph {et~al.}(2019)\citenamefont {Cangemi}, \citenamefont {Cataudella}, \citenamefont {Sassetti},\ and\ \citenamefont {De~Filippis}}]{cangemi2019dissipative}%
  \BibitemOpen
  \bibfield  {author} {\bibinfo {author} {\bibfnamefont {L.~M.}\ \bibnamefont {Cangemi}}, \bibinfo {author} {\bibfnamefont {V.}~\bibnamefont {Cataudella}}, \bibinfo {author} {\bibfnamefont {M.}~\bibnamefont {Sassetti}},\ and\ \bibinfo {author} {\bibfnamefont {G.}~\bibnamefont {De~Filippis}},\ }\bibfield  {title} {\bibinfo {title} {Dissipative dynamics of a driven qubit: Interplay between nonadiabatic dynamics and noise effects from the weak to strong coupling regime},\ }\href {https://doi.org/10.1103/PhysRevB.100.014301} {\bibfield  {journal} {\bibinfo  {journal} {Phys. Rev. B}\ }\textbf {\bibinfo {volume} {100}},\ \bibinfo {pages} {014301} (\bibinfo {year} {2019})}\BibitemShut {NoStop}%
\bibitem [{\citenamefont {Chru{\'s}ci{\'n}ski}(2022)}]{chruscinski2022dynamical}%
  \BibitemOpen
  \bibfield  {author} {\bibinfo {author} {\bibfnamefont {D.}~\bibnamefont {Chru{\'s}ci{\'n}ski}},\ }\bibfield  {title} {\bibinfo {title} {Dynamical maps beyond {M}arkovian regime},\ }\href {https://doi.org/https://doi.org/10.1016/j.physrep.2022.09.003} {\bibfield  {journal} {\bibinfo  {journal} {Phys. Rep.}\ }\textbf {\bibinfo {volume} {992}},\ \bibinfo {pages} {1} (\bibinfo {year} {2022})}\BibitemShut {NoStop}%
\bibitem [{\citenamefont {Wiseman}\ and\ \citenamefont {Milburn}(2009)}]{wiseman_2009}%
  \BibitemOpen
  \bibfield  {author} {\bibinfo {author} {\bibfnamefont {H.}~\bibnamefont {Wiseman}}\ and\ \bibinfo {author} {\bibfnamefont {G.}~\bibnamefont {Milburn}},\ }\href@noop {} {\emph {\bibinfo {title} {Quantum Measurement and Control}}}\ (\bibinfo  {publisher} {Cambridge University Press},\ \bibinfo {year} {2009})\BibitemShut {NoStop}%
\bibitem [{\citenamefont {Roy}\ \emph {et~al.}(2020)\citenamefont {Roy}, \citenamefont {Chalker}, \citenamefont {Gornyi},\ and\ \citenamefont {Gefen}}]{roy2020measurement}%
  \BibitemOpen
  \bibfield  {author} {\bibinfo {author} {\bibfnamefont {S.}~\bibnamefont {Roy}}, \bibinfo {author} {\bibfnamefont {J.~T.}\ \bibnamefont {Chalker}}, \bibinfo {author} {\bibfnamefont {I.~V.}\ \bibnamefont {Gornyi}},\ and\ \bibinfo {author} {\bibfnamefont {Y.}~\bibnamefont {Gefen}},\ }\bibfield  {title} {\bibinfo {title} {Measurement-induced steering of quantum systems},\ }\href {https://doi.org/10.1103/PhysRevResearch.2.033347} {\bibfield  {journal} {\bibinfo  {journal} {Phys. Rev. Res.}\ }\textbf {\bibinfo {volume} {2}},\ \bibinfo {pages} {033347} (\bibinfo {year} {2020})}\BibitemShut {NoStop}%
\bibitem [{\citenamefont {Lidar}\ \emph {et~al.}(1998)\citenamefont {Lidar}, \citenamefont {Chuang},\ and\ \citenamefont {Whaley}}]{lidar1998decoherence}%
  \BibitemOpen
  \bibfield  {author} {\bibinfo {author} {\bibfnamefont {D.~A.}\ \bibnamefont {Lidar}}, \bibinfo {author} {\bibfnamefont {I.~L.}\ \bibnamefont {Chuang}},\ and\ \bibinfo {author} {\bibfnamefont {K.~B.}\ \bibnamefont {Whaley}},\ }\bibfield  {title} {\bibinfo {title} {Decoherence-free subspaces for quantum computation},\ }\href {https://doi.org/10.1103/PhysRevLett.81.2594} {\bibfield  {journal} {\bibinfo  {journal} {Phys. Rev. Lett.}\ }\textbf {\bibinfo {volume} {81}},\ \bibinfo {pages} {2594} (\bibinfo {year} {1998})}\BibitemShut {NoStop}%
\bibitem [{\citenamefont {Knill}\ \emph {et~al.}(2000)\citenamefont {Knill}, \citenamefont {Laflamme},\ and\ \citenamefont {Viola}}]{knill2000theory}%
  \BibitemOpen
  \bibfield  {author} {\bibinfo {author} {\bibfnamefont {E.}~\bibnamefont {Knill}}, \bibinfo {author} {\bibfnamefont {R.}~\bibnamefont {Laflamme}},\ and\ \bibinfo {author} {\bibfnamefont {L.}~\bibnamefont {Viola}},\ }\bibfield  {title} {\bibinfo {title} {Theory of quantum error correction for general noise},\ }\href {https://doi.org/10.1103/PhysRevLett.84.2525} {\bibfield  {journal} {\bibinfo  {journal} {Phys. Rev. Lett.}\ }\textbf {\bibinfo {volume} {84}},\ \bibinfo {pages} {2525} (\bibinfo {year} {2000})}\BibitemShut {NoStop}%
\bibitem [{\citenamefont {Lidar}\ and\ \citenamefont {Whaley}(2003)}]{lidar2003decoherence}%
  \BibitemOpen
  \bibfield  {author} {\bibinfo {author} {\bibfnamefont {D.~A.}\ \bibnamefont {Lidar}}\ and\ \bibinfo {author} {\bibfnamefont {K.~B.}\ \bibnamefont {Whaley}},\ }\bibfield  {title} {\bibinfo {title} {Decoherence-free subspaces and subsystems},\ }in\ \href {https://doi.org/https://doi.org/10.1007/3-540-44874-8_5} {\emph {\bibinfo {booktitle} {Irreversible quantum dynamics}}}\ (\bibinfo  {publisher} {Springer},\ \bibinfo {year} {2003})\ pp.\ \bibinfo {pages} {83--120}\BibitemShut {NoStop}%
\bibitem [{\citenamefont {Cavina}\ \emph {et~al.}(2017)\citenamefont {Cavina}, \citenamefont {Mari},\ and\ \citenamefont {Giovannetti}}]{cavina2017slow}%
  \BibitemOpen
  \bibfield  {author} {\bibinfo {author} {\bibfnamefont {V.}~\bibnamefont {Cavina}}, \bibinfo {author} {\bibfnamefont {A.}~\bibnamefont {Mari}},\ and\ \bibinfo {author} {\bibfnamefont {V.}~\bibnamefont {Giovannetti}},\ }\bibfield  {title} {\bibinfo {title} {Slow dynamics and thermodynamics of open quantum systems},\ }\href {https://doi.org/10.1103/PhysRevLett.119.050601} {\bibfield  {journal} {\bibinfo  {journal} {Phys. Rev. Lett.}\ }\textbf {\bibinfo {volume} {119}},\ \bibinfo {pages} {050601} (\bibinfo {year} {2017})}\BibitemShut {NoStop}%
\bibitem [{\citenamefont {Scandi}\ and\ \citenamefont {Perarnau-Llobet}(2019)}]{Scandi2019thermodynamiclength}%
  \BibitemOpen
  \bibfield  {author} {\bibinfo {author} {\bibfnamefont {M.}~\bibnamefont {Scandi}}\ and\ \bibinfo {author} {\bibfnamefont {M.}~\bibnamefont {Perarnau-Llobet}},\ }\bibfield  {title} {\bibinfo {title} {Thermodynamic length in open quantum systems},\ }\href {https://doi.org/10.22331/q-2019-10-24-197} {\bibfield  {journal} {\bibinfo  {journal} {{Quantum}}\ }\textbf {\bibinfo {volume} {3}},\ \bibinfo {pages} {197} (\bibinfo {year} {2019})}\BibitemShut {NoStop}%
\bibitem [{\citenamefont {Scopa}\ \emph {et~al.}(2019)\citenamefont {Scopa}, \citenamefont {Landi}, \citenamefont {Hammoumi},\ and\ \citenamefont {Karevski}}]{scopa2019exact}%
  \BibitemOpen
  \bibfield  {author} {\bibinfo {author} {\bibfnamefont {S.}~\bibnamefont {Scopa}}, \bibinfo {author} {\bibfnamefont {G.~T.}\ \bibnamefont {Landi}}, \bibinfo {author} {\bibfnamefont {A.}~\bibnamefont {Hammoumi}},\ and\ \bibinfo {author} {\bibfnamefont {D.}~\bibnamefont {Karevski}},\ }\bibfield  {title} {\bibinfo {title} {Exact solution of time-dependent {L}indblad equations with closed algebras},\ }\href {https://doi.org/10.1103/PhysRevA.99.022105} {\bibfield  {journal} {\bibinfo  {journal} {Phys. Rev. A}\ }\textbf {\bibinfo {volume} {99}},\ \bibinfo {pages} {022105} (\bibinfo {year} {2019})}\BibitemShut {NoStop}%
\bibitem [{\citenamefont {Liang}\ \emph {et~al.}(2019)\citenamefont {Liang}, \citenamefont {Yeh}, \citenamefont {Mendonça}, \citenamefont {Teh}, \citenamefont {Reid},\ and\ \citenamefont {Drummond}}]{Liang_2019}%
  \BibitemOpen
  \bibfield  {author} {\bibinfo {author} {\bibfnamefont {Y.-C.}\ \bibnamefont {Liang}}, \bibinfo {author} {\bibfnamefont {Y.-H.}\ \bibnamefont {Yeh}}, \bibinfo {author} {\bibfnamefont {P.~E. M.~F.}\ \bibnamefont {Mendonça}}, \bibinfo {author} {\bibfnamefont {R.~Y.}\ \bibnamefont {Teh}}, \bibinfo {author} {\bibfnamefont {M.~D.}\ \bibnamefont {Reid}},\ and\ \bibinfo {author} {\bibfnamefont {P.~D.}\ \bibnamefont {Drummond}},\ }\bibfield  {title} {\bibinfo {title} {Quantum fidelity measures for mixed states},\ }\href {https://doi.org/10.1088/1361-6633/ab1ca4} {\bibfield  {journal} {\bibinfo  {journal} {Reports on Progress in Physics}\ }\textbf {\bibinfo {volume} {82}},\ \bibinfo {pages} {076001} (\bibinfo {year} {2019})}\BibitemShut {NoStop}%
\bibitem [{\citenamefont {Dann}\ \emph {et~al.}(2020)\citenamefont {Dann}, \citenamefont {Tobalina},\ and\ \citenamefont {Kosloff}}]{dann2020fast}%
  \BibitemOpen
  \bibfield  {author} {\bibinfo {author} {\bibfnamefont {R.}~\bibnamefont {Dann}}, \bibinfo {author} {\bibfnamefont {A.}~\bibnamefont {Tobalina}},\ and\ \bibinfo {author} {\bibfnamefont {R.}~\bibnamefont {Kosloff}},\ }\bibfield  {title} {\bibinfo {title} {Fast route to equilibration},\ }\href {https://doi.org/10.1103/PhysRevA.101.052102} {\bibfield  {journal} {\bibinfo  {journal} {Phys. Rev. A}\ }\textbf {\bibinfo {volume} {101}},\ \bibinfo {pages} {052102} (\bibinfo {year} {2020})}\BibitemShut {NoStop}%
\bibitem [{\citenamefont {Turyansky}\ \emph {et~al.}(2024)\citenamefont {Turyansky}, \citenamefont {Ovdat}, \citenamefont {Dann}, \citenamefont {Aqua}, \citenamefont {Kosloff}, \citenamefont {Dayan},\ and\ \citenamefont {Pick}}]{turyansky2023inertial}%
  \BibitemOpen
  \bibfield  {author} {\bibinfo {author} {\bibfnamefont {D.}~\bibnamefont {Turyansky}}, \bibinfo {author} {\bibfnamefont {O.}~\bibnamefont {Ovdat}}, \bibinfo {author} {\bibfnamefont {R.}~\bibnamefont {Dann}}, \bibinfo {author} {\bibfnamefont {Z.}~\bibnamefont {Aqua}}, \bibinfo {author} {\bibfnamefont {R.}~\bibnamefont {Kosloff}}, \bibinfo {author} {\bibfnamefont {B.}~\bibnamefont {Dayan}},\ and\ \bibinfo {author} {\bibfnamefont {A.}~\bibnamefont {Pick}},\ }\bibfield  {title} {\bibinfo {title} {Inertial geometric quantum logic gates},\ }\href {https://doi.org/10.1103/PhysRevApplied.21.054033} {\bibfield  {journal} {\bibinfo  {journal} {Phys. Rev. Appl.}\ }\textbf {\bibinfo {volume} {21}},\ \bibinfo {pages} {054033} (\bibinfo {year} {2024})}\BibitemShut {NoStop}%
\bibitem [{\citenamefont {Caldeira}\ and\ \citenamefont {Leggett}(1983)}]{caldeira1983}%
  \BibitemOpen
  \bibfield  {author} {\bibinfo {author} {\bibfnamefont {A.~O.}\ \bibnamefont {Caldeira}}\ and\ \bibinfo {author} {\bibfnamefont {A.~J.}\ \bibnamefont {Leggett}},\ }\bibfield  {title} {\bibinfo {title} {Quantum tunnelling in a dissipative system},\ }\href {https://doi.org/https://doi.org/10.1016/0003-4916(83)90202-6} {\bibfield  {journal} {\bibinfo  {journal} {Ann. Phys.}\ }\textbf {\bibinfo {volume} {149}},\ \bibinfo {pages} {374} (\bibinfo {year} {1983})}\BibitemShut {NoStop}%
\bibitem [{\citenamefont {Deffner}\ and\ \citenamefont {Campbell}(2017)}]{deffner2017quantum}%
  \BibitemOpen
  \bibfield  {author} {\bibinfo {author} {\bibfnamefont {S.}~\bibnamefont {Deffner}}\ and\ \bibinfo {author} {\bibfnamefont {S.}~\bibnamefont {Campbell}},\ }\bibfield  {title} {\bibinfo {title} {Quantum speed limits: from {H}eisenberg’s uncertainty principle to optimal quantum control},\ }\href {https://doi.org/10.1088/1751-8121/aa86c6} {\bibfield  {journal} {\bibinfo  {journal} {J. Phys. A Math. Theor.}\ }\textbf {\bibinfo {volume} {50}},\ \bibinfo {pages} {453001} (\bibinfo {year} {2017})}\BibitemShut {NoStop}%
\bibitem [{\citenamefont {Taddei}\ \emph {et~al.}(2013)\citenamefont {Taddei}, \citenamefont {Escher}, \citenamefont {Davidovich},\ and\ \citenamefont {de~Matos~Filho}}]{taddei2013quantum}%
  \BibitemOpen
  \bibfield  {author} {\bibinfo {author} {\bibfnamefont {M.~M.}\ \bibnamefont {Taddei}}, \bibinfo {author} {\bibfnamefont {B.~M.}\ \bibnamefont {Escher}}, \bibinfo {author} {\bibfnamefont {L.}~\bibnamefont {Davidovich}},\ and\ \bibinfo {author} {\bibfnamefont {R.~L.}\ \bibnamefont {de~Matos~Filho}},\ }\bibfield  {title} {\bibinfo {title} {Quantum speed limit for physical processes},\ }\href {https://doi.org/10.1103/PhysRevLett.110.050402} {\bibfield  {journal} {\bibinfo  {journal} {Phys. Rev. Lett.}\ }\textbf {\bibinfo {volume} {110}},\ \bibinfo {pages} {050402} (\bibinfo {year} {2013})}\BibitemShut {NoStop}%
\bibitem [{\citenamefont {del Campo}\ \emph {et~al.}(2013)\citenamefont {del Campo}, \citenamefont {Egusquiza}, \citenamefont {Plenio},\ and\ \citenamefont {Huelga}}]{del2013quantum}%
  \BibitemOpen
  \bibfield  {author} {\bibinfo {author} {\bibfnamefont {A.}~\bibnamefont {del Campo}}, \bibinfo {author} {\bibfnamefont {I.~L.}\ \bibnamefont {Egusquiza}}, \bibinfo {author} {\bibfnamefont {M.~B.}\ \bibnamefont {Plenio}},\ and\ \bibinfo {author} {\bibfnamefont {S.~F.}\ \bibnamefont {Huelga}},\ }\bibfield  {title} {\bibinfo {title} {Quantum speed limits in open system dynamics},\ }\href {https://doi.org/10.1103/PhysRevLett.110.050403} {\bibfield  {journal} {\bibinfo  {journal} {Phys. Rev. Lett.}\ }\textbf {\bibinfo {volume} {110}},\ \bibinfo {pages} {050403} (\bibinfo {year} {2013})}\BibitemShut {NoStop}%
\bibitem [{\citenamefont {Deffner}\ and\ \citenamefont {Lutz}(2013)}]{deffner2013quantum}%
  \BibitemOpen
  \bibfield  {author} {\bibinfo {author} {\bibfnamefont {S.}~\bibnamefont {Deffner}}\ and\ \bibinfo {author} {\bibfnamefont {E.}~\bibnamefont {Lutz}},\ }\bibfield  {title} {\bibinfo {title} {Quantum speed limit for non-{M}arkovian dynamics},\ }\href {https://doi.org/10.1103/PhysRevLett.111.010402} {\bibfield  {journal} {\bibinfo  {journal} {Phys. Rev. Lett.}\ }\textbf {\bibinfo {volume} {111}},\ \bibinfo {pages} {010402} (\bibinfo {year} {2013})}\BibitemShut {NoStop}%
\bibitem [{\citenamefont {Pires}\ \emph {et~al.}(2016)\citenamefont {Pires}, \citenamefont {Cianciaruso}, \citenamefont {C{\'e}leri}, \citenamefont {Adesso},\ and\ \citenamefont {Soares-Pinto}}]{pires2016generalized}%
  \BibitemOpen
  \bibfield  {author} {\bibinfo {author} {\bibfnamefont {D.~P.}\ \bibnamefont {Pires}}, \bibinfo {author} {\bibfnamefont {M.}~\bibnamefont {Cianciaruso}}, \bibinfo {author} {\bibfnamefont {L.~C.}\ \bibnamefont {C{\'e}leri}}, \bibinfo {author} {\bibfnamefont {G.}~\bibnamefont {Adesso}},\ and\ \bibinfo {author} {\bibfnamefont {D.~O.}\ \bibnamefont {Soares-Pinto}},\ }\bibfield  {title} {\bibinfo {title} {Generalized geometric quantum speed limits},\ }\href {https://doi.org/10.1103/PhysRevX.6.021031} {\bibfield  {journal} {\bibinfo  {journal} {Phys. Rev. X}\ }\textbf {\bibinfo {volume} {6}},\ \bibinfo {pages} {021031} (\bibinfo {year} {2016})}\BibitemShut {NoStop}%
\bibitem [{\citenamefont {Rezek}\ and\ \citenamefont {Kosloff}(2006)}]{rezek2006irreversible}%
  \BibitemOpen
  \bibfield  {author} {\bibinfo {author} {\bibfnamefont {Y.}~\bibnamefont {Rezek}}\ and\ \bibinfo {author} {\bibfnamefont {R.}~\bibnamefont {Kosloff}},\ }\bibfield  {title} {\bibinfo {title} {Irreversible performance of a quantum harmonic heat engine},\ }\href {https://doi.org/10.1088/1367-2630/8/5/083} {\bibfield  {journal} {\bibinfo  {journal} {New J. Phys.}\ }\textbf {\bibinfo {volume} {8}},\ \bibinfo {pages} {83} (\bibinfo {year} {2006})}\BibitemShut {NoStop}%
\bibitem [{\citenamefont {Bargmann}(1947)}]{bargmann47}%
  \BibitemOpen
  \bibfield  {author} {\bibinfo {author} {\bibfnamefont {V.}~\bibnamefont {Bargmann}},\ }\bibfield  {title} {\bibinfo {title} {Irreducible unitary representations of the {L}orentz group},\ }\href {https://doi.org/https://doi.org/10.2307/1969129} {\bibfield  {journal} {\bibinfo  {journal} {Ann. Math.}\ }\textbf {\bibinfo {volume} {48}},\ \bibinfo {pages} {568} (\bibinfo {year} {1947})}\BibitemShut {NoStop}%
\bibitem [{\citenamefont {Novaes}(2004)}]{novaes2004some}%
  \BibitemOpen
  \bibfield  {author} {\bibinfo {author} {\bibfnamefont {M.}~\bibnamefont {Novaes}},\ }\bibfield  {title} {\bibinfo {title} {Some basics of su(1,1)},\ }\href@noop {} {\bibfield  {journal} {\bibinfo  {journal} {Rev. Bras. de Ensino de Fis.}\ }\textbf {\bibinfo {volume} {26}},\ \bibinfo {pages} {351} (\bibinfo {year} {2004})}\BibitemShut {NoStop}%
\bibitem [{\citenamefont {Torrontegui}\ \emph {et~al.}(2011)\citenamefont {Torrontegui}, \citenamefont {Ib{\'a}{\~n}ez}, \citenamefont {Chen}, \citenamefont {Ruschhaupt}, \citenamefont {Gu{\'e}ry-Odelin},\ and\ \citenamefont {Muga}}]{torrontegui2011fast}%
  \BibitemOpen
  \bibfield  {author} {\bibinfo {author} {\bibfnamefont {E.}~\bibnamefont {Torrontegui}}, \bibinfo {author} {\bibfnamefont {S.}~\bibnamefont {Ib{\'a}{\~n}ez}}, \bibinfo {author} {\bibfnamefont {X.}~\bibnamefont {Chen}}, \bibinfo {author} {\bibfnamefont {A.}~\bibnamefont {Ruschhaupt}}, \bibinfo {author} {\bibfnamefont {D.}~\bibnamefont {Gu{\'e}ry-Odelin}},\ and\ \bibinfo {author} {\bibfnamefont {J.~G.}\ \bibnamefont {Muga}},\ }\bibfield  {title} {\bibinfo {title} {Fast atomic transport without vibrational heating},\ }\href {https://doi.org/10.1103/PhysRevA.83.013415} {\bibfield  {journal} {\bibinfo  {journal} {Phys. Rev. A}\ }\textbf {\bibinfo {volume} {83}},\ \bibinfo {pages} {013415} (\bibinfo {year} {2011})}\BibitemShut {NoStop}%
\bibitem [{\citenamefont {Weedbrook}\ \emph {et~al.}(2012)\citenamefont {Weedbrook}, \citenamefont {Pirandola}, \citenamefont {Garc\'{\i}a-Patr\'on}, \citenamefont {Cerf}, \citenamefont {Ralph}, \citenamefont {Shapiro},\ and\ \citenamefont {Lloyd}}]{weedbrook2012gaussian}%
  \BibitemOpen
  \bibfield  {author} {\bibinfo {author} {\bibfnamefont {C.}~\bibnamefont {Weedbrook}}, \bibinfo {author} {\bibfnamefont {S.}~\bibnamefont {Pirandola}}, \bibinfo {author} {\bibfnamefont {R.}~\bibnamefont {Garc\'{\i}a-Patr\'on}}, \bibinfo {author} {\bibfnamefont {N.~J.}\ \bibnamefont {Cerf}}, \bibinfo {author} {\bibfnamefont {T.~C.}\ \bibnamefont {Ralph}}, \bibinfo {author} {\bibfnamefont {J.~H.}\ \bibnamefont {Shapiro}},\ and\ \bibinfo {author} {\bibfnamefont {S.}~\bibnamefont {Lloyd}},\ }\bibfield  {title} {\bibinfo {title} {Gaussian quantum information},\ }\href {https://doi.org/10.1103/RevModPhys.84.621} {\bibfield  {journal} {\bibinfo  {journal} {Rev. Mod. Phys.}\ }\textbf {\bibinfo {volume} {84}},\ \bibinfo {pages} {621} (\bibinfo {year} {2012})}\BibitemShut {NoStop}%
\bibitem [{\citenamefont {Hillery}\ \emph {et~al.}(1984)\citenamefont {Hillery}, \citenamefont {O'Connell}, \citenamefont {Scully},\ and\ \citenamefont {Wigner}}]{HILLERY1984121}%
  \BibitemOpen
  \bibfield  {author} {\bibinfo {author} {\bibfnamefont {M.}~\bibnamefont {Hillery}}, \bibinfo {author} {\bibfnamefont {R.}~\bibnamefont {O'Connell}}, \bibinfo {author} {\bibfnamefont {M.}~\bibnamefont {Scully}},\ and\ \bibinfo {author} {\bibfnamefont {E.}~\bibnamefont {Wigner}},\ }\bibfield  {title} {\bibinfo {title} {Distribution functions in physics: Fundamentals},\ }\href {https://doi.org/https://doi.org/10.1016/0370-1573(84)90160-1} {\bibfield  {journal} {\bibinfo  {journal} {Phys. Rep.}\ }\textbf {\bibinfo {volume} {106}},\ \bibinfo {pages} {121} (\bibinfo {year} {1984})}\BibitemShut {NoStop}%
\bibitem [{\citenamefont {Lee}(1995)}]{LEE1995147}%
  \BibitemOpen
  \bibfield  {author} {\bibinfo {author} {\bibfnamefont {H.-W.}\ \bibnamefont {Lee}},\ }\bibfield  {title} {\bibinfo {title} {Theory and application of the quantum phase-space distribution functions},\ }\href {https://doi.org/https://doi.org/10.1016/0370-1573(95)00007-4} {\bibfield  {journal} {\bibinfo  {journal} {Phys. Rep.}\ }\textbf {\bibinfo {volume} {259}},\ \bibinfo {pages} {147} (\bibinfo {year} {1995})}\BibitemShut {NoStop}%
\bibitem [{Note1()}]{Note1}%
  \BibitemOpen
  \bibinfo {note} {We emphasize that all these protocols would produce fidelity one in case the dynamics is assumed closed. Maximum and minimum refer to optimization with respect to the parameter $g_6,g_{7}$ in Eq.~(\ref {eq:parametri2}). Better or worse results can be obtained by considering higher-order expansion.}\BibitemShut {Stop}%
\bibitem [{\citenamefont {Gyamfi}(2020)}]{Gyamfi_2020}%
  \BibitemOpen
  \bibfield  {author} {\bibinfo {author} {\bibfnamefont {J.~A.}\ \bibnamefont {Gyamfi}},\ }\bibfield  {title} {\bibinfo {title} {Fundamentals of quantum mechanics in {L}iouville space},\ }\href {https://doi.org/10.1088/1361-6404/ab9fdd} {\bibfield  {journal} {\bibinfo  {journal} {Eur. J. Phys.}\ }\textbf {\bibinfo {volume} {41}},\ \bibinfo {pages} {063002} (\bibinfo {year} {2020})}\BibitemShut {NoStop}%
\bibitem [{\citenamefont {Smirne}\ and\ \citenamefont {Vacchini}(2010)}]{smirne2010nakajima}%
  \BibitemOpen
  \bibfield  {author} {\bibinfo {author} {\bibfnamefont {A.}~\bibnamefont {Smirne}}\ and\ \bibinfo {author} {\bibfnamefont {B.}~\bibnamefont {Vacchini}},\ }\bibfield  {title} {\bibinfo {title} {Nakajima-{Z}wanzig versus time-convolutionless master equation for the non-{M}arkovian dynamics of a two-level system},\ }\href {https://doi.org/10.1103/PhysRevA.82.022110} {\bibfield  {journal} {\bibinfo  {journal} {Phys. Rev. A}\ }\textbf {\bibinfo {volume} {82}},\ \bibinfo {pages} {022110} (\bibinfo {year} {2010})}\BibitemShut {NoStop}%
\bibitem [{\citenamefont {Joye}(2022)}]{joye2022adiabatic}%
  \BibitemOpen
  \bibfield  {author} {\bibinfo {author} {\bibfnamefont {A.}~\bibnamefont {Joye}},\ }\bibfield  {title} {\bibinfo {title} {Adiabatic {L}indbladian evolution with small dissipators},\ }\href {https://doi.org/https://doi.org/10.1007/s00220-021-04306-5} {\bibfield  {journal} {\bibinfo  {journal} {Commun. Math. Phys.}\ }\textbf {\bibinfo {volume} {391}},\ \bibinfo {pages} {223} (\bibinfo {year} {2022})}\BibitemShut {NoStop}%
\end{thebibliography}%

\end{document}